\documentclass[pdflatex,sn-mathphys-num]{sn-jnl}% Math and Physical Sciences Numbered Reference Style
\usepackage{graphicx}%
\usepackage{multirow}%
\usepackage{amsmath,amssymb,amsfonts}%
\usepackage{amsthm}%
\usepackage{mathrsfs}%
\usepackage[title]{appendix}%
\usepackage{xcolor}%
\usepackage{textcomp}%
\usepackage{manyfoot}%
\usepackage{booktabs}%
\usepackage{algorithm}%
\usepackage{algorithmicx}%
\usepackage{algpseudocode}%
\usepackage{listings}%
\usepackage{comment}
\usepackage{siunitx}
\usepackage{makecell}
\usepackage[T1]{fontenc}
\usepackage{pdflscape}

\theoremstyle{thmstyleone}%
\newtheorem{theorem}{Theorem}%  meant for continuous numbers
\newtheorem{proposition}[theorem]{Proposition}% 

\theoremstyle{thmstyletwo}%
\newtheorem{example}{Example}%
\newtheorem{remark}{Remark}%

\theoremstyle{thmstylethree}%
\newtheorem{definition}{Definition}%

\begin{document}

\title[Article Title]{Machine-learned global glacier ice volumes}
%%=============================================================%%
%% GivenName	-> \fnm{Joergen W.}
%% Particle	-> \spfx{van der} -> surname prefix
%% FamilyName	-> \sur{Ploeg}
%% Suffix	-> \sfx{IV}
%% \author*[1,2]{\fnm{Joergen W.} \spfx{van der} \sur{Ploeg} 
%%  \sfx{IV}}\email{iauthor@gmail.com}
%%=============================================================%%

\author*[1,2]{\fnm{Niccolò} \sur{Maffezzoli}}\email{niccolo.maffezzoli@unive.it}

\author[3,4]{\fnm{Eric} \sur{Rignot}}%\email{iiauthor@gmail.com}
%\equalcont{These authors contributed equally to this work.}

\author[1,2]{\fnm{Carlo} \sur{Barbante}}%\email{barbante@unive.it}
%\equalcont{These authors contributed equally to this work.}

\author[5]{\fnm{Mathieu} \sur{Morlighem}}%\email{iiauthor@gmail.com}
%\equalcont{These authors contributed equally to this work.}

\author[6]{\fnm{Troels} \sur{Petersen}}%\email{barbante@unive.it}
%\equalcont{These authors contributed equally to this work.}

\author[1]{\fnm{Sebastiano} \sur{Vascon}}%\email{barbante@unive.it}
%\equalcont{These authors contributed equally to this work.}

\affil*[1]{\orgdiv{Department of Environmental Sciences, Informatics and Statistics}, \orgname{University of Venice}, \orgaddress{\street{Via Torino 155}, \city{Venezia}, \postcode{30172}, \country{Italy}}}

\affil[2]{\orgdiv{Institute of Polar Sciences}, \orgname{National Research Council}, \city{Venezia}, \country{Italy}}

\affil[3]{\orgdiv{Department of Earth System Science}, \orgname{University of California Irvine}, \orgaddress{\street{Croul Hall}, \city{Irvine}, \postcode{92627}, \state{California}, \country{United States}}}

\affil[4]{\orgdiv{Jet Propulsion Laboratory}, \city{Pasadena}, \country{United States}}

\affil[5]{\orgdiv{Department of Earth Sciences}, \orgname{Dartmouth College}, \city{Hanover}, \postcode{03755}, \state{NH}, \country{United States}}

\affil[6]{\orgdiv{Niels Bohr Institute}, \orgname{University of Copenhagen}, \city{Copenhagen}, \country{Denmark}}

%%==================================%%
%% Sample for unstructured abstract %%
%%==================================%%

\abstract{
We present a global dataset of glacier ice thickness modeled with IceBoost v2.0, a gradient-boosted decision tree scheme trained on 7 million ice thickness measurements and informed by physical and geometrical predictors. We model the distributed ice thickness for every glacier in the two latest Randolph Glacier Inventory releases (v6.0 and v7.0), totaling 215,547 and 274,531 glacier outlines, respectively, plus 955 ice masses contiguous with the Greenland Ice Sheet. IceBoost v2.0 represents the third existing global solution alongside Millan et al. (2022, \cite{millan2022}) and Farinotti et al. (2019, \cite{farinotti2019}). We find a global glacier volume of $(149 \pm 38)\times 10^3$ km$^3$, consistent with the previous ensemble estimate of $(147 \pm 28)\times 10^3$ km$^3$. The corresponding sea-level equivalent, $323 \pm 91$ mm, is likewise consistent with the earlier value of $315 \pm 63$ mm. Compared to measurements, IceBoost error is 20–45\% lower than the other solutions in the high Arctic, highlighting the value of machine-learning approaches.
% confidence
We examine major glaciated regions and compare results with the other models. Confidence in our solution is highest at higher latitudes, where abundant training data adequately sample the feature space. Over steep and mountainous terrain, small glaciers, and under-represented lower-latitude regions, confidence is lower. IceBoost v2.0 demonstrates strong generalization at ice sheet margins. On the Geikie Plateau (East Greenland), we find nearly twice as much ice as previously reported, highlighting the method’s potential to infer bed topography in parts of the ice sheets.
% drawbacks
No physical laws are explicitly imposed during training, so sufficient and high-quality training data are crucial. The quality of the solutions depends on the accuracy of the training data, the Digital Elevation Model, ice velocity fields, and glacier geometries, including nunataks.
% Jensen Gap
Using the Jensen Gap, we probe the model’s curvature with respect to input errors and find it is strongly concave over low-slope, thick-ice regions, implying a potential downward bias in predicted thickness under input uncertainty.
% use of the dataset
The released dataset can be used to model future glacier evolution and sea-level rise, inform the design of glaciological surveys and field campaigns, as well as guide policies on freshwater management.
}

\keywords{glaciers, ice thickness, machine learning}

%%\pacs[JEL Classification]{D8, H51}

%%\pacs[MSC Classification]{35A01, 65L10, 65L12, 65L20, 65L70}

\maketitle

\section{Background \& Summary}
\label{sec:background}
% glaciers
Knowledge of the volumes of glaciers and ice caps, as well as their spatially distributed ice thickness, is fundamental for geophysical modeling. Models projecting the future evolution of ice masses must be initialized with and are particularly sensitive to accurate present-day conditions \cite{farinotti2017}. However, this requirement is becoming progressively difficult to satisfy, as accelerated climate warming and glacier shrinkage cause present-day glacier states to evolve faster than our ability to produce updated thickness maps, many of which still reflect conditions from the early 2000s. Glaciers are retreating worldwide, having lost about 5\% of their total mass over the past two decades and up to 39\% in some regions of the world \cite{glambie2025}. They account for approximately 25–30\% of modern sea-level rise \cite{zemp2019, ipcc2021}. The rate of ice loss has accelerated during the last decade \cite{Hugonnet2021, glambie2025}, and projections indicate that glaciers may lose 26–41\% of their total mass by 2100, depending on the undertaken future climate trajectory \cite{rounce2023}. The implications are far-reaching, affecting freshwater availability and management \cite{huss2018, rodell2018, immerzeel2020}, coastal habitability \cite{ipcc2019}, and a wide range of socio-economic systems dependent on glacier-fed environments.

% models
Glacier mass change can be quantified either from surface-elevation change observations \cite{Hugonnet2021, jakob2023} or from gravity variations induced by mass redistribution \cite{ciraci2020}. However, inferring the global distribution of ice thickness remains a major challenge. Only a few global-scale modeling efforts exist to date \cite{millan2022, farinotti2019, huss2012}, each relying on geometrical relationships, mathematical interpolations, physics-based models or mass conservation principles.

% measurements
Over recent decades, millions of in-situ and airborne ice thickness measurements have been collected by international efforts such as the World Glacier Monitoring Service (WGMS; \cite{welty2020}), NASA’s Operation IceBridge, and numerous regional and individual glaciological surveys. The GlaThiDa (Glacier Thickness Database) Consortium \cite{glathida2020} has consolidated most of these measurements into a unified dataset (currently v3.1.0, with v.4 forthcoming). Despite its importance, this resource has remained largely underexploited for global modeling purposes.

% iceboost v2
In this study, we provide a new solution to the distributed ice thickness of the world's glaciers, using a machine learning model. We combine ice thickness observations from GlaThiDa and additional surveys, encompassing 1,661 glaciers worldwide and over seven million measurements, to build a training dataset and train a system of two gradient-boosted decision tree schemes. We then use this system to generate distributed ice thickness map for all glaciers globally. The model, IceBoost v2.0 \cite{maffezzoli2025}, represents an updated version that prioritizes smoothness of the predicted thickness field, over the previous version v1.1. Alongside the global ice thickness maps derived for all glaciers in the Randolph Glacier Inventory (RGI) version 6.2 (hereafter RGI v.62, $n$ = 216,502) and version 7.0 (hereafter RGI v.70, $n$ = 274,531), we provide maps of ice thickness uncertainty, surface elevation, geoid elevation, and Jensen Gap, a metrics of model nonlinearity.

\begin{figure}
    \centering
    \includegraphics[width=1.0\linewidth]{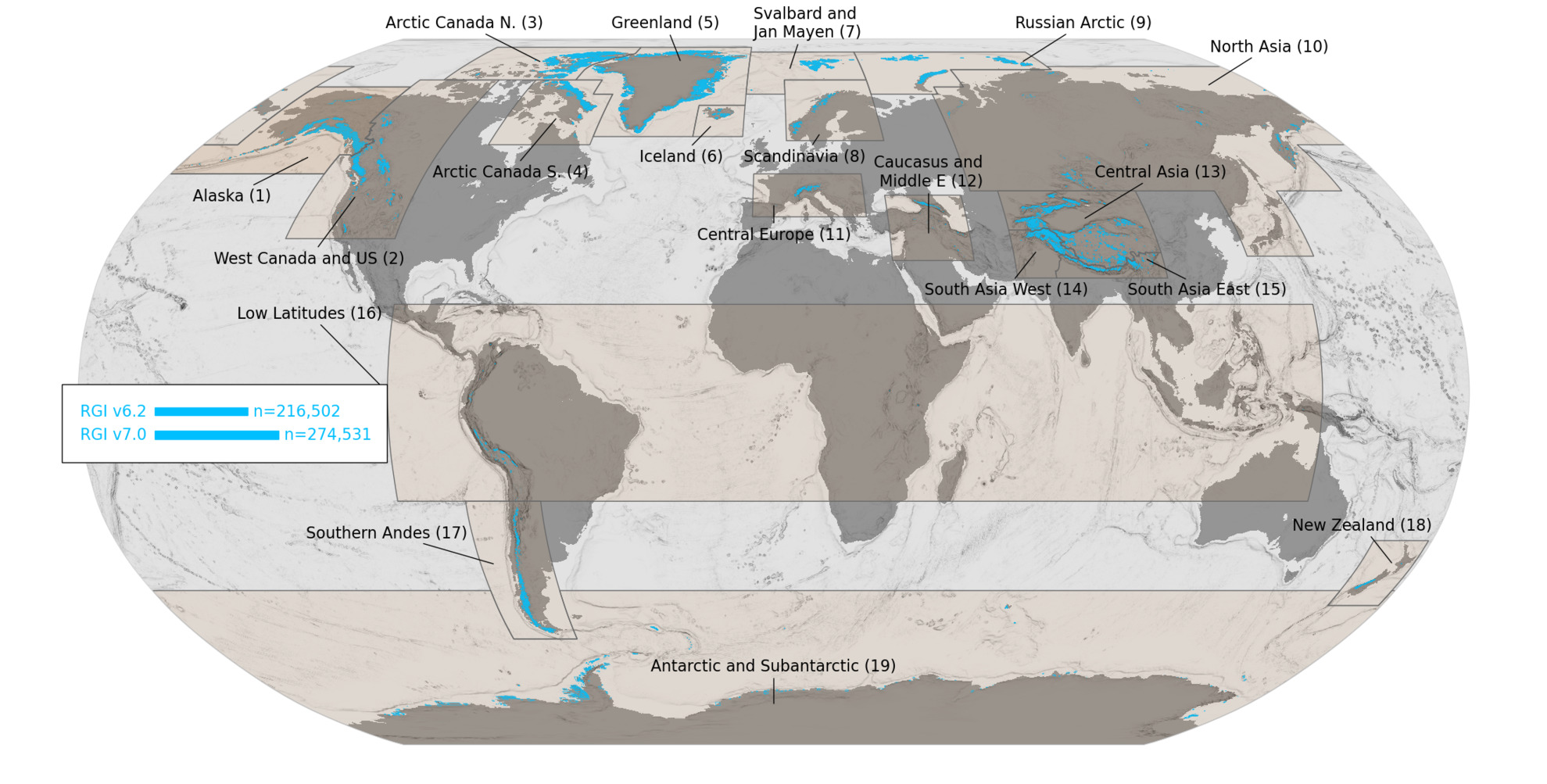}
    \caption{The World's glaciers (cyan), divided into the 19 regions of the Randolph Glacier Inventory. The box indicates the total number of glaciers in versions v.6.2 and v.7}
    \label{fig:fig1}
\end{figure}

\clearpage
\section{Methods}
\subsection{Training data}
\label{sec:updates_model_and_data}

% data updates
IceBoost v2.0 is trained with publicly available ice thickness datasets (Table \ref{table:datasets_training}). GlaThiDa v3.1.0 (n=3,854,279 measurements, \cite{welty2020}) represents the main global dataset. We also incorporate the following regional datasets: the surveys carried out on the Ruth glacier \cite{tober2024} and other glaciers by \cite{tober2025}, totaling n=1,472,965 extra measurements in Alaska. In Scandinavia, we add the survey on the Jostedalsbreen ice cap (n=351,559, \cite{gillespie2024}). Over the two Patagonian ice fields, we add two datasets (\cite{millan2019_patagonia, furst2024}) for a total of n=418,689 extra measurements. At high latitudes we incorporate two main datasets. We use the IceBridge product \cite{icebridge_mcords_l2}, which intercepts glaciers over the Canadian Arctic, coastal Greenland, Svalbard, and the Antarctic periphery. Furthermore, we add the 2002-2023 product that covers glaciers over the Antarctic peninsula acquired by the Center for Remote Sensing of Ice Sheets (CReSIS, \cite{cresis}). The Antarctic peninsula is not part of the official Randolph Glacier Inventory, but have a large number of alpine glaciers with a great number of measurements that offer the opportunity to extend the training dataset and train a model capable of generalizing to the ice sheet periphery. We note that in coastal Greenland, as part of IceBridge, we use multiple surveys taken on the Geikie Plateau, with direct connection to the Greenland ice sheet. Altogether, we train the model with n=2,868,276 additional measurements located at very high latitudes (IceBridge and CReSIS), many of which located in regions proximal to the ice sheets.
Globally, IceBoost v2 is trained with 1.9 times the data of IceBoost v1.1.

\begin{table}[h]
\caption{Ice thickness datasets used as training targets for IceBoost v2.0.}\label{table:datasets_training}%
\begin{tabular}{lllll}
\toprule
\textbf{Dataset names} & \textbf{No. points} & \textbf{Domain} &  \textbf{Data source} & \textbf{Reference} \\
\midrule
GlaThiDa v3.1.0 & 3,854,279 & Global & \href{https://www.gtn-g.ch/glathida/}{GTN-G} & \cite{glathida2020}  \\
\cmidrule(lr){1-5} % fine line spanning all 5 columns

Alaska & 1,472,965 & Alaska and Northwestern Canada & \makecell[l]{\href{https://arizona.figshare.com/articles/dataset/Data_and_Code_for_Thickness_of_Ruth_Glacier_Alaska_and_Depth_of_its_Great_Gorge_from_Ice-Penetrating_Radar_and_Mass_Conservation_/21669824}{University Arizona Data repository} \\
\href{https://nsidc.org/data/iruafhf2/versions/1}{IceBridge}, \href{https://nsidc.org/data/irares2/versions/1}{IceBridge}} & \cite{tober2024, tober2025, tober2025_UAFL2HF, tober2025_ARESL2}  \\
\cmidrule(lr){1-5} % fine line spanning all 5 columns

Patagonia & 418,689 & Patagonian icefields & \makecell[l]{\href{https://qfuego-patagonia.org/}{QFuego-Patagonia}, \\ \href{https://datadryad.org/dataset/doi:10.7280/D11Q17}{UC Irvine Dryad Data Repository}} & \cite{furst2024, millan2019_patagonia}  \\
\cmidrule(lr){1-5} % fine line spanning all 5 columns

Scandinavia & 351,559 & Jostedalsbreen ice cap & \href{https://nva.sikt.no/registration/018fa427d47a-cfa126c6-690b-43f2-967c-05ab9426f41b}{Norwegian Nasjonalt Vitenarkiv} & \cite{gillespie2024}  \\
\cmidrule(lr){1-5} % fine line spanning all 5 columns

Polar & 2,868,276 & \makecell[l]{Greenland, Canadian Arctic, \\ Svalbard, Antarctica} & \href{https://nsidc.org/data/irmcr2/versions/1}{IceBridge}, \href{https://data.cresis.ku.edu/}{CReSIS} & \cite{icebridge_mcords_l2}, \cite{cresis} \\
\cmidrule(lr){1-5} % fine line spanning all 5 columns

\multicolumn{5}{l}{Total points before cleaning \quad 8,965,768 } \\
\multicolumn{5}{l}{Total points after cleaning \quad 7,069,690 (n=1,661 glaciers)} \\
\multicolumn{5}{l}{Total points after encoding \quad 378,373 (n=1,661 glaciers)} \\
\botrule
\end{tabular}
\end{table}

\subsection{Model}
\label{sec:model}
In Section \ref{sec:model_inputs} we briefly describe the model inputs, previously introduced in IceBoost v1.1 \cite{maffezzoli2025}. The differences between the two model versions are presented in Section \ref{sec:model_updates}.

\subsection{Model inputs and training}
\label{sec:model_inputs}

% elevation, slope and curvature
The model is informed by a set of 26 variables (Table \ref{tab_datasets_training}). Elevation, slope, and curvature are calculated from the Tandem-X EDEM v1, an automated edited (filtered, interpolated, and infilled) variant of the global Digital Elevation Model, acquired between 2010 and 2014 under the TanDEM-X mission. As in IceBoost v1.1, we include multiple slope and curvature variables obtained by applying Gaussian kernels of varying sizes to the DEM. Smaller kernels capture fine-scale features, which are important for small glaciers, while larger kernels capture broader-scale variations relevant for extensive ice masses.

\begin{table}[h]
    \caption{IceBoost v2.0 model inputs, their units, resolution and referenced time. tag.}\label{tab_datasets_training}%
    \begin{tabular}{llllllr}
    \toprule
    \textbf{Feature} & \textbf{Variable} & \textbf{Unit}  & \textbf{Resolution} & \textbf{Error} & \textbf{Time tag} & \textbf{Source}\\
    \midrule
    - Curvature & \makecell[l]{$c_{50}$, $c_{100}$, $c_{150}$, \\ $c_{300}$, $c_{450}$, $_{cgfa}$} & \SI{}{\metre\tothe{-1}} & \SI{30}{\metre} & \makecell[l]{$2\sqrt{5}\sigma_z/\Delta x^2$ \\ (Eq. \ref{eq:sigma_curv})} &  2010-2014  & Tandem-X Edited DEM \\
    - \makecell[l]{Distance to glacier \\ margins/ or nunataks} & $d_{noice}$ & \SI{}{\kilo\metre} & 30 m & 100 m & 2000-2010  & RGI polygons v.70, v.62\\
    
    - Distance from ocean & $d_{ocean}$ & \SI{}{\kilo\metre} & 50-500 m & 100 m & $>$2000 & GSHHG \cite{wessel1996}\\
    
    - Surface Elevation & $z$ & \SI{}{\metre} & 30 m & \makecell[l]{2-4 m \\ (Eq. \ref{eq:sigma_z})}&  2010-2014 & Tandem-X Edited DEM \\
    
    - Length & $lmax$ & \SI{}{\metre} & 30 m & 5\% &  2000-2010 & RGI polygons v.70, v.62 \\
    
    - Surface slopes & \makecell[l]{$s_{50}$, $s_{75}$, $s_{100}$, \\ $s_{125}$, $s_{150}$, $s_{300}$,\\ $s_{450}$, $s_{gfa}$} & \SI{1}{} & \SI{30}{\metre} & \makecell[l]{$2\sqrt{2}/\Delta x$ \\ (Eq. \ref{eq:sigma_s})} &  2010-2014  & Tandem-X Edited DEM \\
    
    - Surface Mass balance & $smb$ & kg/(m$^2$$\cdot$yr) & 30-2000 m & 10\% & \makecell[r]{1961-2021 \\ 2000-2019} & \makecell[r]{RACMO2.3p2 \cite{noel2019,noel2023} \\ Hugonnet et al. 2021 \cite{Hugonnet2021}}\\
    
    - Temperature at 2 meters & $t2m$ & K & 9 km & 1 K & 2000-2010 & ERA5, ERA5-Land\\ 
    
    - Velocity & \makecell[l]{$v_{50}$, $v_{100}$, $v_{150}$, \\$v_{300}$, $v_{450}$, $v_{gfa}$} & m/yr & \makecell[l]{50 m\\ 250 m\\ 450 m} & \makecell[l]{10 m/yr\\ 18 m/yr\\ 18 m/yr} & 2017-2018 & \makecell[r]{Millan et al. 2022 \cite{millan2022} \\ Joughin et al. 2016 \cite{joughin2016} \\ Mouginot et al. 2016 \cite{mouginot2019}}\\
    \botrule
    \end{tabular}
\end{table}

% distance from noice and distance from the oceans
The distance to glacier margins or internal rock outcrops (denoted $d_{noice}$) is calculated using glacier polygons from the Randolph Glacier Inventory, v.62 and the most recent v.70. The distance to the ocean is derived from the Global Self-consistent Hierarchical High-resolution Geography (GSHHG) Shorelines product, v2.3.7, used at full ('f') resolution \cite{wessel1996}. These shorelines are based on and updated from the World Vector Shoreline project \cite{soluri1990}.

% mass balance
We use the same distributed surface mass balance products as in IceBoost v1.1 \cite{maffezzoli2025}. For glaciers at the peripheries of Greenland and Antarctica, we use RACMO2.3p2 \cite{noel2018}, downscaled to 1 km \cite{noel2019} and 2 km \cite{noel2023}, and averaged over 1961–1990 and 1979–2021, respectively. Outside the ice sheets, we fit an empirical linear mass balance–elevation lapse rate for each glacier pair in the 19 regions. This involves determining the mass balance at zero elevation and the slope ($\frac{dMB}{dz}$) of the lapse rate using glacier-integrated geodetic mass balance data from Hugonnet et al. (2021) \cite{Hugonnet2021}, along with elevations from Tandem-X EDEM. We impose an inverse squared distance weight to encourage similar parameters for nearby glacier pairs and reduce the influence of distant glaciers.

% t2m
Temperature inputs (2-m air temperature, t2m) are obtained from ERA5-Land (0.1° grid spacing, $\approx$9 km; \cite{era5land}). For pixels missing due to imperfect land masks along coastlines and islands, we supplement with the ERA5 t2m field (0.25° resolution; \cite{era5}), bilinearly interpolated to the ERA5-Land 0.1° grid.

% velocity
Ice velocity inputs are identical to those used in IceBoost v1.1. Specifically, for glaciers in the Greenland periphery and Antarctica (both peripheral and continental), we use the products of Joughin et al. \cite{joughin2016}, and Mouginot et al. \cite{mouginot2019}. For all other glaciers we use the product of Millan et al. \cite{millan2022}.

% Training
IceBoost v2.0 comprises independent XGBoost and CatBoost modules, both trained with a squared loss during a 100-iteration cross-validation pipeline on a random 20 \% subset of the training data. Hyperparameters are reported in the Supplementary Information. At inference time, when the model is tasked to predict the ice thickness of a glacier regular grid, the predictions from the two modules are equally averaged.

\subsection{Model updates}
\label{sec:model_updates}

% model updates

% getting rid of some features
IceBoost v2.0 was refined to prioritize the smoothness of the solution across neighboring glaciers. In IceBoost v1.1, we observed that several input variables introduced discontinuities at glacier boundaries. We removed variables that were prone to DEM artifacts: $z_{min}$ (minimum elevation), $z_{max}$ (maximum elevation), $z - z_{min}$ (elevation above glacier base), $z_{01}$ (normalized elevation), and $\Delta z$ (glacier elevation range). Additional features were excluded due to their sensitivity to imputation (integrated mass balance, MB), or the stochastic nature of glacier delineations (glacier and cluster areas, $A$, $A_{cluster}$, as well as the perimeter). Furthermore, variables found to have limited predictive importance \cite{maffezzoli2025} based on SHAP (Shapley Additive Explanations) analysis but added to the computational burden were removed: glacier-integrated values of aspect, curvature, and slope. 

% velocity features
The velocity product used for continental glaciers (\cite{millan2022}) often contains partially or completely missing data, as well as very low values (<5 m/yr). To address this, we exclude the velocity features when data are missing or when the glacier-wide average velocity falls below 5 m/yr. This approach also removes the need for velocity imputation, which in IceBoost v1.1 was performed using regional averages. Moreover, to enrich the feature set for very small glaciers, we use the variable $l_{max}$ (glacier length, otherwise not used) for glacier clusters smaller than 10 $km^2$ that lack velocity information. For such clusters, IceBoost v2 models all constituent glaciers as a single contiguous ice mass, thereby eliminating the arbitrariness inherent in individual glacier delineations.

% concluding remarks on the model
Though valuable information is discarded in the new model, IceBoost v2.0 becomes a model where all features are continuous across neighboring glaciers, leading to solution that smoother across neighboring glaciers. Tested against ground truth data, its performance is similar to Millan and Farinotti's models in non-polar regions, and up to 20-45\% better over high latitudes (Supplementary Information).

\subsection{Projection, posting and time tag}
\label{sec:posting}

\noindent \textbf{Study domain.}
% discuss here the differences between RGI62, RGI70 and what datasets we will be using here
We model the distributed ice thickness for all n=216,502 glaciers in RGI v6.2 and n=274,531 glaciers in RGI v7.0. We note that, compared to v.6 (n=215,547), the v.62 version includes 955 additional glaciers in the periphery of the Greenland ice sheet, notably the Geikie plateau. They were however retained in this work and in the previous IceBoost v1.1 model version, because they provide a great deal of training measurements in a glacier flow regime, with skillful capabilities in the outer ice sheet. These outlines are not part of version v.7, which excludes all outliers with direct connection to the ice sheets. RGI v7.0 consists of 73\% new or updated glacier outlines compared to version v.6, equivalent to a 42 \% improvement of glacier surface area globally. We also include training data in the Antarctic peninsula. We create n=50 geometries inside the peninsula by simply partitioning the area into randomly-sized Voronoi polygons. The peninsula outer geometry is taken from BedMachine Antarctica v4 (Morlighem et al., in press), truncated at northings=300,000 m ($\simeq$73-75 S). The inner rock outcrops (nunataks) are also taken from BedMachine v4.\\

% updated geometries in the Randolph Glacier Inventory
\noindent \textbf{Updated glacier geometries from RGI v6.0 to RGI v7.0}
The effect of revised glacier geometries in RGI v7.0 can have a significant effect for the modeled thickness. Two main effects can be observed: revised glacier perimeters and modifications of internal nunataks. Regarding the first effect, from RGI v6.0 to RGI v7.0, glaciers have often shrunk. The model running on RGI v6.0 would yield too thick ice near the borders because of the flat surface left from the exposed bedrock (low slope values tend to increase the predicted thickness). Regarding the second effect, the addition of internal nunataks in RGI v7.0 has the effect to reduce (at times significantly) the predicted ice in the neighboring regions to the specified outcrops. For example, a significant more shallow prediction is visible over the South Patagonia ice field (e.g. Pio XI glacier) in v7.0 compared to v6.0, because of a significant increase of nunataks geometries in RGI v7.0, compared to RGI v6.0. 
The quality of the glacier geometry dataset is extremely important. For ice volume estimates, the role of nunataks geometries is larger than any revision of the outside glacier borders, which is marginal, as typically ice at the margins is anyhow shallow. A last effect regarding glacier geometries relates to misplaced geometries, not centered on glaciers. An example for this can be seen over glaciers on the Coronation Islands (Subantarctic Islands, 60.6°S 45.6°W), for which all RGI v6.0 geometries are shifted with respect to the real glacier positions, corrected in RGI v7.0 The distributed ice thickness predicted for RGI v6.0 should be therefore not trusted. We generally observed a very significant improvement in the quality of glacier geometries in the latest RGI v7.0 dataset.\\

% time tag
\noindent \textbf{Time tag and time uncertainty.} 
The TanDEM-X 30m EDEM is produced within 2010-2014, therefore all geodetic features (elevation and its gradients) inherit this time tag \cite{bueso2021, gonzalez2020, martone2018}. The SAR-derived ice velocity product is tagged to 2017-2018 \cite{millan2022}. The glacier polygons used to train the model are those from the Randolph Glacier Inventory v6.0 (\cite{rgi6}). Most glacier polygons in RGI v6.0 and the most recent RGI v7.0 are time tagged between 2000-2010 (Fig 2 in \cite{rgi6}). 

The distance to the ocean is inferred using the ocean vectors digitalized in the GSHHG v2.3.7 product, tagged in June 2017. The glacier-integrated mass balance dataset of Hugonnet et al. (2021, \cite{Hugonnet2021}) are tied to 2000-2019, while the RACMO2 products used in Greenland and Antarctica are used as time averages between 1961–1990 and 1979–2021, respectively. The temperature-above-2-meter input is calculated by combining ERA5 with ERA5-Land. Both products are considered by averaging all monthly maps over 2000–2010 to generate one single global temperature field. Finally, and most importantly, the ice thickness data used to train the model has a lower cutoff at 2005, with much more measurements acquired after 2010.

To conclude, we estimate our product to be tagged to 2010-2018.\\

% ice volumes
%\noindent \textbf{Ice volumes and sea level equivalent.} For each glacier we calculate the total ice volume $V_{ice}$, and the ice volume below sea level $V^{ice}_{bsl}$. 

%\begin{align}
%    V_{ice} &= (A/N) \sum_{i} h_{i} \\
%    V^{ice}_{bsl} &= (A/N) \sum_{i} max(0, h_i -H_i)
%\end{align}

\noindent \textbf{Ice volumes and sea level equivalent.} \\

% ice volume
For each glacier we calculate the ice volume $V_{ice}$ as: 
\begin{align}
    V_{ice} &= \sum_{n} H_{n}A_n \label{eq:vice}\\
\end{align}
where $H_n=f(X_n)$ is the modeled thickness at pixel $n$, $A_n$ is the pixel area and the sum is taken over the $n$ pixels. An upper bound uncertainty on the ice volume is calculated as $\sigma_{V_{ice}}=\sum_{n} \sigma_{H_{n}}A_n$, assuming that all pixel errors are positively correlated with $\rho=1$. The thickness errors are modeled in Section \ref{sec:ice_thickness_uncertainties}.

%where $h_i=f(X_i)$ is the modeled thickness at pixel $i$, $H_i$ is the EIGEN-6C4 geoid orthometric height, N is the glacier pixel count, and $A$ is the glacier area. Both volumes are released in the tif (per-glacier) files as attributes.\\
%An upper bound error on the ice volume is calculated as $\sigma_{V_{ice}}=(A/N)\sum_{i} \sigma_{h_{i}}$, assuming that all pixels errors are perfectly positively correlated. The thickness uncertainties are estimated in Section \ref{sec:ice_thickness_uncertainties}. The error on the ice volume below sea level is calculated in the same way. 

% sea level equivalent
The ice volume sea-level equivalent (SLE, in mm) is calculated from the ice volume above floatation $V_{af}$.

\begin{align}
V_{\mathrm{af}} &= \sum_n \max\!\left[ \left( H_n + \min\!\left(b_n,0\right)\frac{\rho_{\mathrm{ocean}}}{\rho_{\mathrm{ice}}} \right),\; 0 \right] \cdot A_n, \label{eq:vaf}\\[6pt]
\mathrm{SLE} &= \frac{V_{\mathrm{af}}}{A_{\mathrm{ocean}}} \frac{\rho_{\mathrm{ice}}}{\rho_{\mathrm{ocean}}} \cdot 10^6. \label{eq:sle}
\end{align}

In Eq. \ref{eq:vaf}, $b$ is the bed elevation relative to the EIGEN-6C4 geoid, and it is calculated as $b = z - N - H$, where $z$ is the DEM surface elevation with respect to the WGS84 ellipsoid, N is the geoid elevation and H is the ice thickness. We take $\rho_{ice}=917\ kg/m^3$ and $\rho_{ocean}=1027\ kg/m^3$ as the densities of ice and seawater, respectively, and $A_{ocean}=3.618\cdot10^8\ km^2$ as the area of the oceans. In the above calculations, steric and isostatic effects are neglected.

The quantity \[
H + \min(b,0) \frac{\rho_{\mathrm{ocean}}}{\rho_{\mathrm{ice}}}
\] 
is the height above flotation, denoted $H_{\mathrm{af}}$. We set its uncertainty equal to the ice thickness uncertainty:
\begin{align}
\sigma_{H_{\mathrm{af}}} =
\begin{cases}
\sigma_H, & \text{if } H_{\mathrm{af}} > 0,\\
0, & \text{otherwise,}
\end{cases}
\end{align}
and we calculate the uncertainty of the SLE as:
\begin{align}
\sigma_{V_{\mathrm{af}}} &= \sum_n \sigma_{H_{\mathrm{af}}} \cdot A_n, \\
\sigma_{\mathrm{SLE}} &= \frac{\sigma_{V_{\mathrm{af}}}}{A_{\mathrm{ocean}}} \frac{\rho_{\mathrm{ice}}}{\rho_{\mathrm{ocean}}} \cdot 10^6 \,,
\end{align}

In other words, only regions with positive height above floatation contribute to SLE uncertainty.

%The sea-level equivalent (SLE, in mm) of the ice volume above flotation is then:
%\begin{align}
%SLE &= \frac{V_{ice}-V^{ice}_{bsl}}{A_{ocean}} \frac{\rho_{ice}}{\rho_{ocean}}\cdot 10^6, \label{eq:sle}
%SLE &= \frac{V_{af}}{A_{ocean}} \frac{\rho_{ice}}{\rho_{ocean}}\cdot 10^6, \label{eq:sle}
%\end{align}

%where $\rho_{ice}=917\ kg/m^3$ is taken as the density of the ice, and $\rho_{ocean}=1027\ kg/m^3$ as the density of seawater, $A_{ocean}=3.618\cdot10^8\ km^2$ is taken as the area of the oceans. Steric and isostatic effects are neglected.

\subsection{Regional total ice volumes and sea level equivalent}
The regional volumes are calculated by summing all individual glacier volumes (Eq. \ref{eq:vice}). The regional volume error is calculated assuming all individual glacier volumes are correlated (with $\rho=1$, the same assumption is made by \cite{millan2022, farinotti2019}), thereby the error becomes $\sigma_{V_{rgi}} = \sum_i \sigma_{V_i}$, where $i$ indexes regional glaciers. The regional error is to be considered an upper bound, and it is likely too large. The regional SLE errors are calculated in the same way as the regional ice volumes (assuming perfect positive correlation between all glaciers' SLEs in the region).

\begin{table}[h]
    \caption{Global glacier ice volumes and SLE stratified regionally with reference to RGI v7.0. The previous estimate is tied to RGI v6.0 and is computed with a weighted average of values from \cite{millan2022} and \cite{farinotti2019}.
    }\label{table_regional_statistics}%
    \begin{tabular}{@{}lcccccc@{}}
    \toprule
    \makecell{Region} & \makecell{No.\\glaciers} & \makecell{Area\\(\SI{e3}{\kilo\metre\tothe{2}})} &
    \multicolumn{2}{c}{\makecell{Ice volume\\(\SI{e3}{\kilo\metre\tothe{3}})}} &
    \multicolumn{2}{c}{\makecell{SLE\\(\SI{}{\milli\metre})}} \\
    \cmidrule(lr){4-5} \cmidrule(lr){6-7}
    & & & \makecell{This\\work} & \makecell{Prev.\\estimate} & \makecell{This\\work} & \makecell{Prev.\\estimate} \\
    \midrule
    01 Alaska & 27509 & 87 & 16.9 ± 3.2 & 18.5 ± 3.7  & 39.3 ± 7.8 & 43.4 ± 8.8 \\
    02 Western Canada and US  & 18730 & 15 & 1.5 ± 0.4 & 1.1 ± 0.2 & 3.7 ± 1.1 & 2.7 ± 0.6 \\
    03 Arctic Canada North & 5216 & 105 & 24.3 ± 6.2 & 26.8 ± 5.1 & 58.0 ± 15.1 & 62.7 ± 12.6 \\
    04 Arctic Canada South & 11009 & 41 & 7.1 ± 1.8 & 7.8 ± 1.5 & 17.1 ± 4.4 & 19.2 ± 3.9 \\
    05 Greenland Periphery & 19994 & 90 & 13.2 ± 4.6 & 13.6 ± 2.7 & 31.1 ± 11.0 & 30.5 ± 6.4 \\
    06 Iceland & 568 & 11 & 4.6±0.7 & 3.7 ± 0.7 & 11.2 ± 1.8 & 9.2 ± 1.8 \\
    07 Svalbard and Jan Mayen & 1666 & 34 & 6.7 ± 1.5 & 7.3 ± 1.5 & 15.5 ± 3.8 & 16.6 ± 3.5 \\
    08 Scandinavia & 3410 & 2.9 & 0.35 ± 0.09 & 0.30 ± 0.06 & 0.84 ± 0.22 & 0.71 ± 0.17 \\
    09 Russian Arctic & 1069 & 52 & 12.8 ± 3.0 & 15.1 ± 2.7 & 30.3 ± 7.4 & 32.7 ± 6.3 \\
    10 North Asia & 7155 & 2.6 & 0.19 ± 0.07 & 0.12 ± 0.02 & 0.45 ± 0.16 & 0.29 ± 0.07 \\
    11 Central Europe & 4079 & 2.1 & 0.11 ± 0.05 & 0.13 ± 0.03 & 0.26 ± 0.13 & 0.29 ± 0.07 \\
    12 Caucasus and Middle East & 2275 & 1.4 & 0.08 ± 0.03 & 0.06 ± 0.02 & 0.19 ± 0.07 & 0.17 ± 0.07 \\
    13 Central Asia & 75613 & 50 & 3.8 ± 2.2 & 3.4 ± 0.8 & 9.0 ± 5.4 & 8.4 ± 1.9 \\
    14 South Asia West & 37562 & 33 & 3.8 ± 1.5 & 3.0 ± 0.7 & 8.9 ± 3.7 & 7.3 ± 1.7 \\
    15 South Asia East & 18587 & 16 & 1.0 ± 0.5 & 0.9 ± 0.2 & 2.4 ± 1.1 & 2.2 ± 0.5 \\
    16 Low Latitudes & 3695 & 1.9 & 0.09 ± 0.06 & 0.09 ± 0.02 & 0.21 ± 0.13 & 0.19 ± 0.07 \\
    17 Southern Andes & 30634 & 28 & 6.7 ± 1.3 & 5.6 ± 1.1 & 16.2 ± 3.1 & 13.5 ± 2.6 \\
    18 New Zealand & 3018 & 0.9 & 0.08 ± 0.03 & 0.07 ± 0.02 & 0.18 ± 0.06 & 0.19 ± 0.07 \\
    19 Subantarctic and Antarctic Islands & 2742 & 133 & 45.7 ± 10.3 & 39.2 ± 7.3 & 78.2 ± 24.7 & 64.9 ± 12.0 \\
    Global  & 274531 & 707 & 149 ± 38 & 147 ± 28 & 323 ± 91 & 315 ± 63 \\

    \botrule
    \end{tabular}
\end{table}

\subsection{Modeled ice thickness}
\label{sect:ice_thickness_maps}
\begin{figure}[ht!]
    \centering
    \includegraphics[width=.49\linewidth]{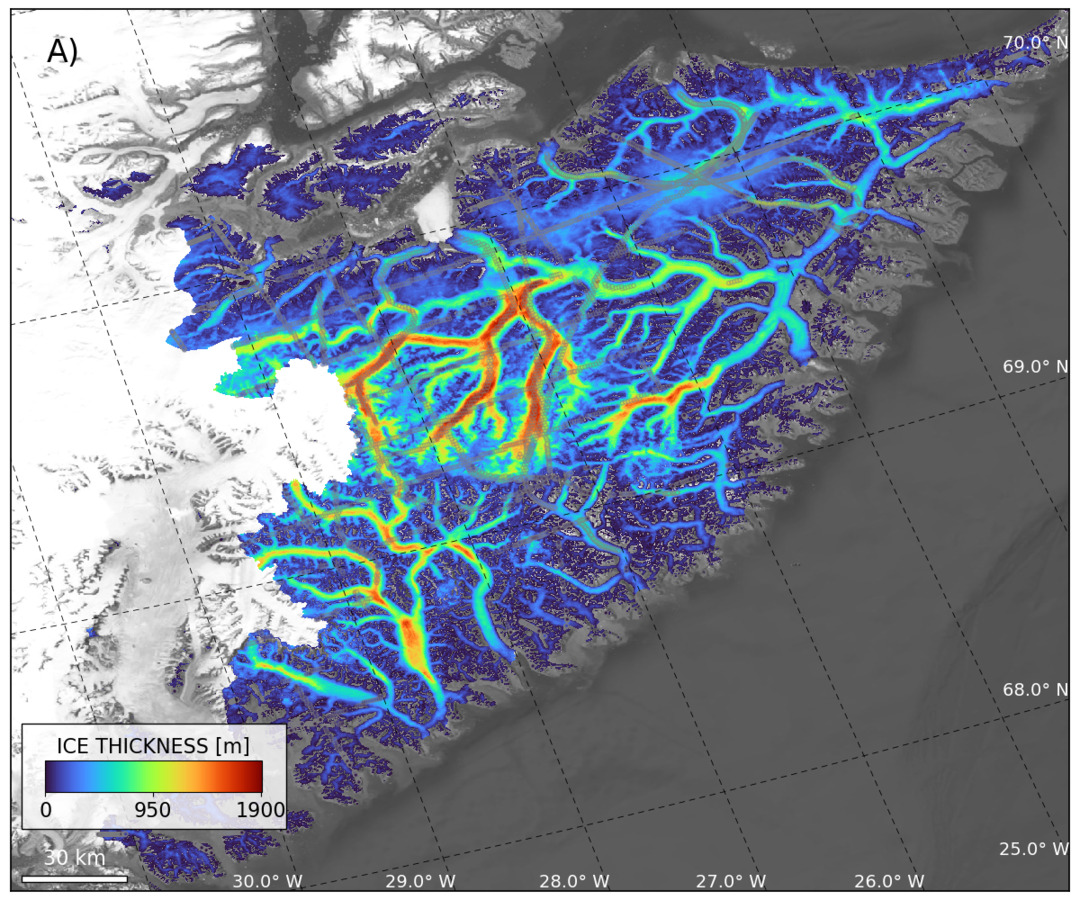}
    \includegraphics[width=.49\linewidth]{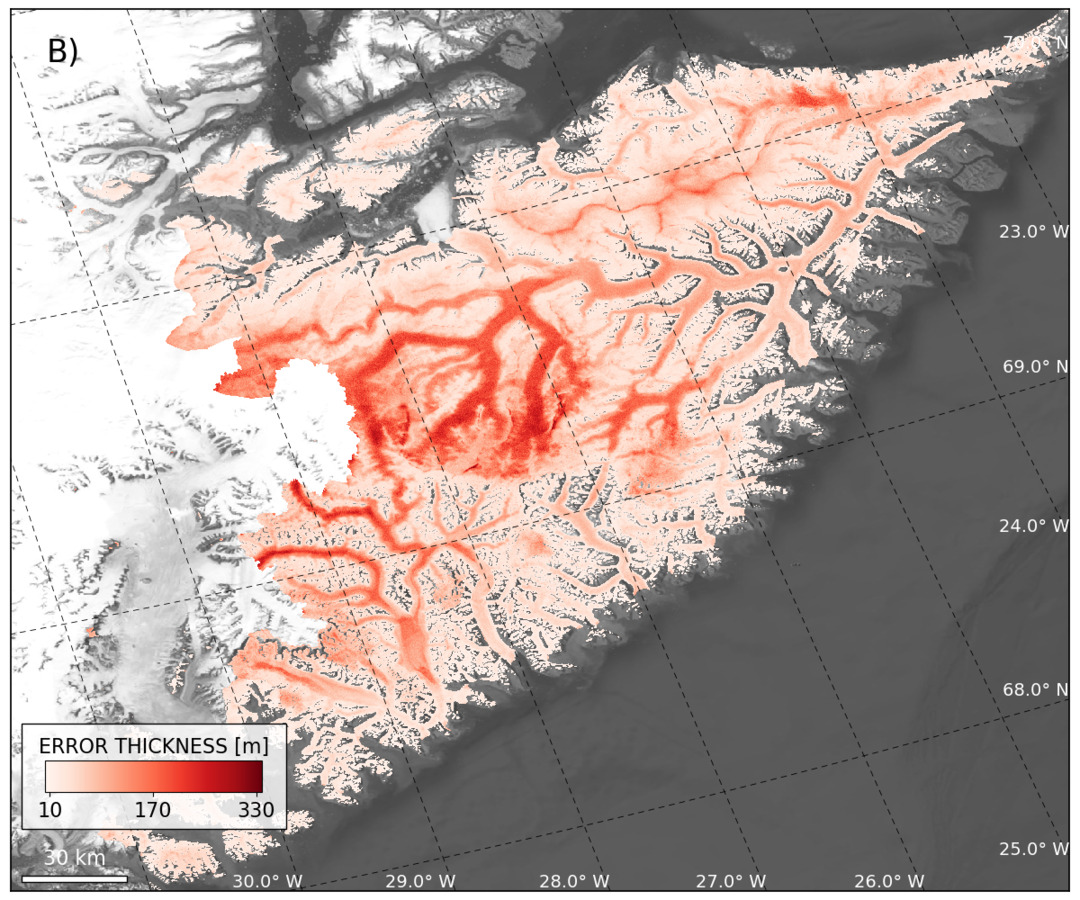}\\
    \includegraphics[width=.49\linewidth]{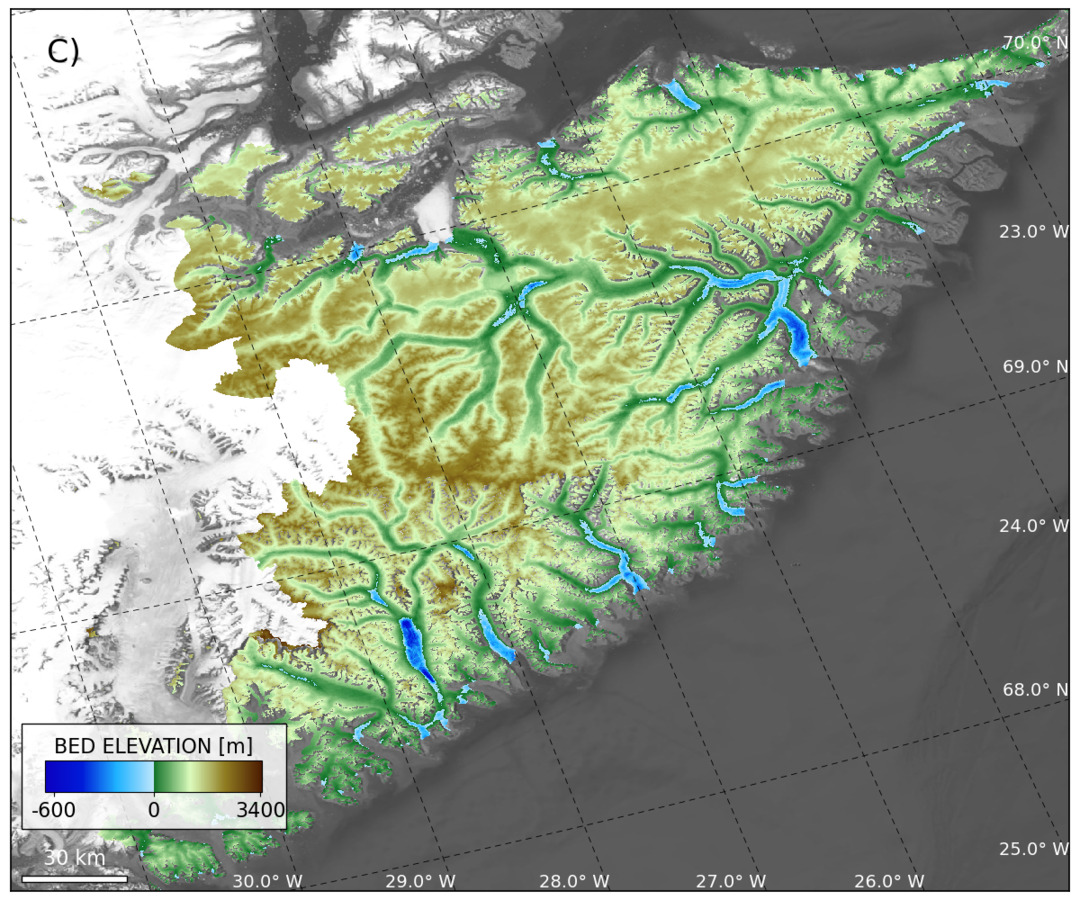}
    \includegraphics[width=.49\linewidth]{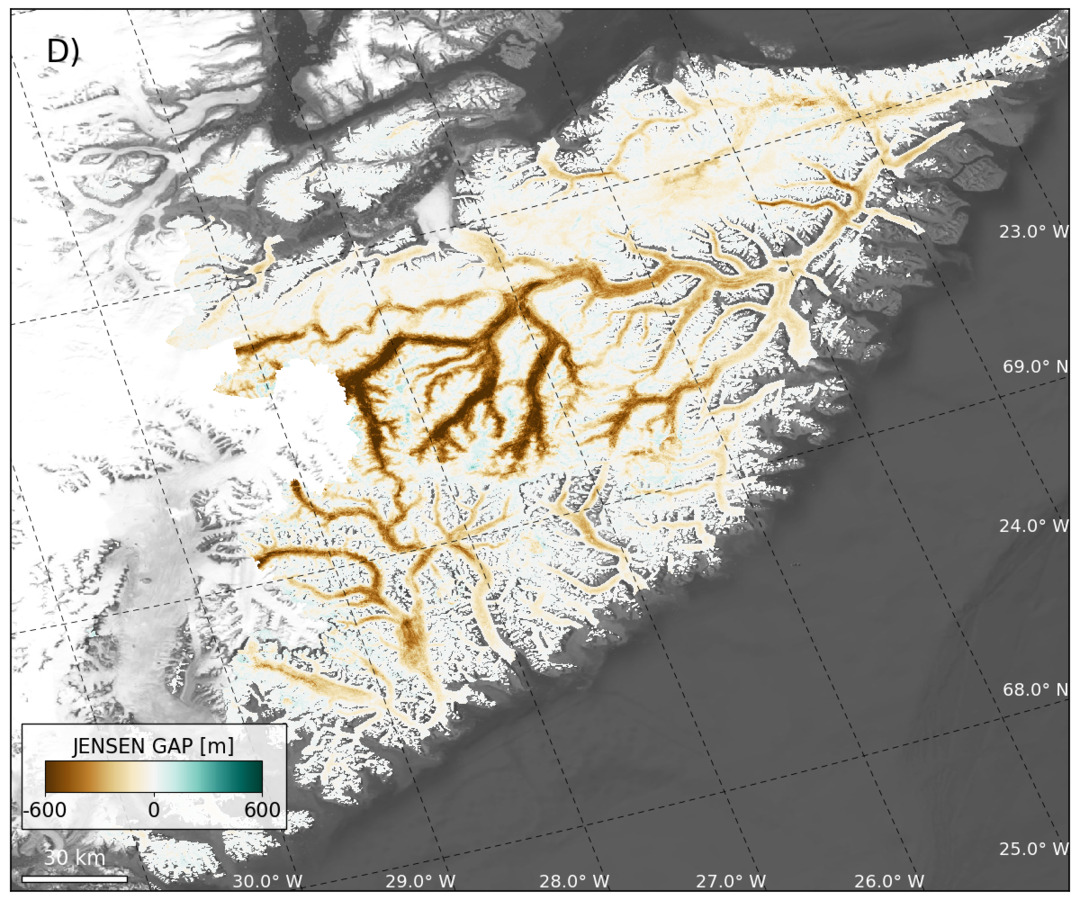}
    \caption{The Geikie Plateau (coastal East Greenland) modeled with IceBoost v.2. A) ice thickness; B) ice thickness error; C) bed elevation; D) Jensen Gap.}
    \label{fig:geikie}
\end{figure}

We model all individual glaciers in RGI v6.2 and RGI v7.0 An example of the data product is shown in Fig. \ref{fig:geikie} for the Geikie Plateau (East Greenland): the modeled ice thickness (A), the modeled ice thickness error (B), the bed elevation (C), and the Jensen Gap (D). The errors and the Jensen gap are discussed in Sections \ref{sec:ice_thickness_uncertainties}-\ref{sect:jensen_gap}.

In Sections \ref{sect:alaska} to \ref{sect:southern_andes}, we present regions of particular interest to the glaciological community: Alaska (Sect. \ref{sect:alaska}), the Canadian Arctic (Sect. \ref{sect:canadian_arctic}), the Russian Arctic (Sect. \ref{sect:russian_arctic}), the Greenland periphery (Sect. \ref{sect:greenland_periphery}), Asia (Sect. \ref{sect:asia}), and the Southern Andes (Sect. \ref{sect:southern_andes}). For each region, we present the most significant glaciated systems and ice caps and discuss differences between IceBoost, the models by Millan et al. \cite{millan2022} and Farinotti et al. \cite{farinotti2019}, and available measurements.

\clearpage

\subsubsection{Alaska (RGI 01)}
\label{sect:alaska}

\begin{table}[h]
    \caption{Alaska ice volumes estimated by different models. All units are \SI{e3}{\kilo\metre\tothe{3}}.}\label{tab1}%
    \begin{tabular}{@{}lccc@{}}
    \toprule
    Alaska (RGI 01) & IceBoost v2 & Millan et al. \cite{millan2022} & Farinotti et al. \cite{farinotti2019}\\
    \midrule
    Total &  16.9 ± 3.2 & 17.8 ± 4.6 & 19.0 ± 5.0 \\
    - Bering-Malaspina-Seward basins & 6.4 ± 1.0 & 6.7 & 7.8 \\
    - Others &  10.5 ± 2.2 &  11.1 & 11.2 \\
    \bottomrule
    \end{tabular}
\end{table}

\begin{figure}[h!]
    \caption{Bering-Malaspina-Seward basin (Alaska). A=IceBoost v2 (with and without overlayed data); B=Millan et al. 2022 \cite{millan2022}; C=Farinotti et al. \cite{farinotti2019}.}
    \includegraphics[width=.49\linewidth]{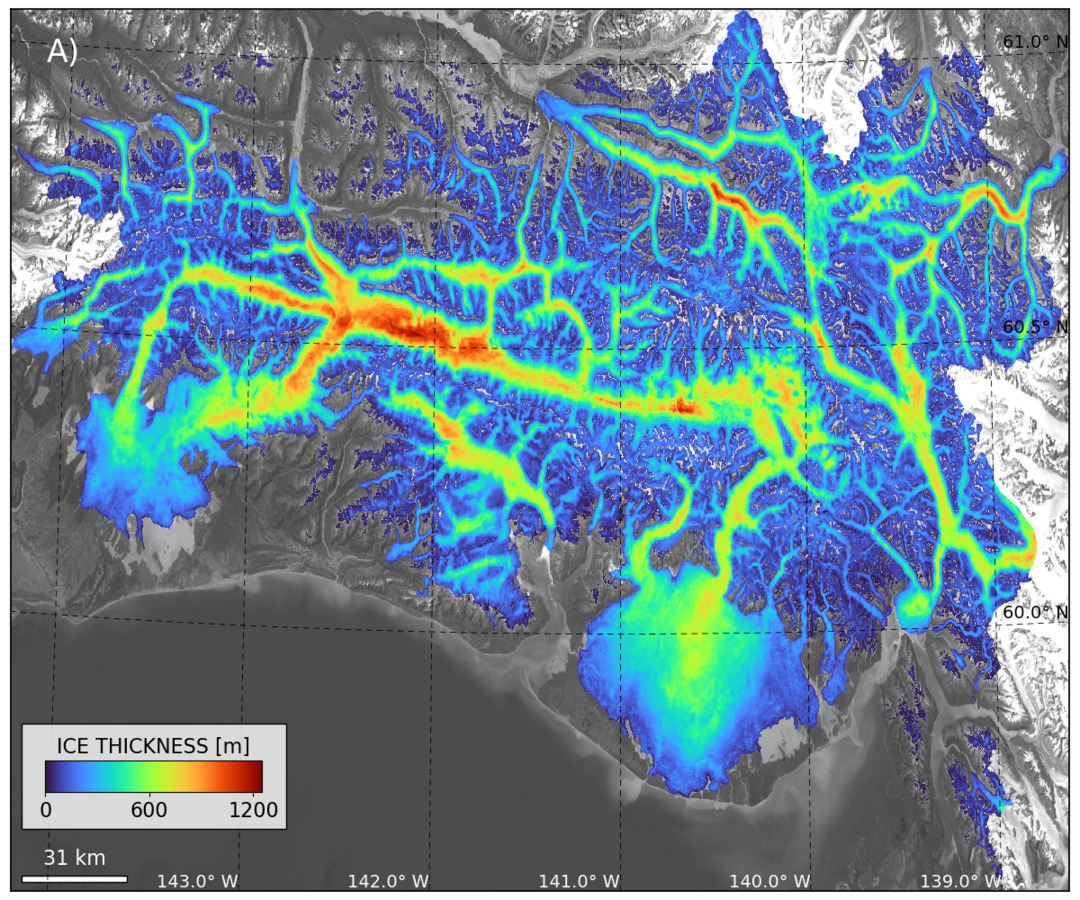}
    \includegraphics[width=.49\linewidth]{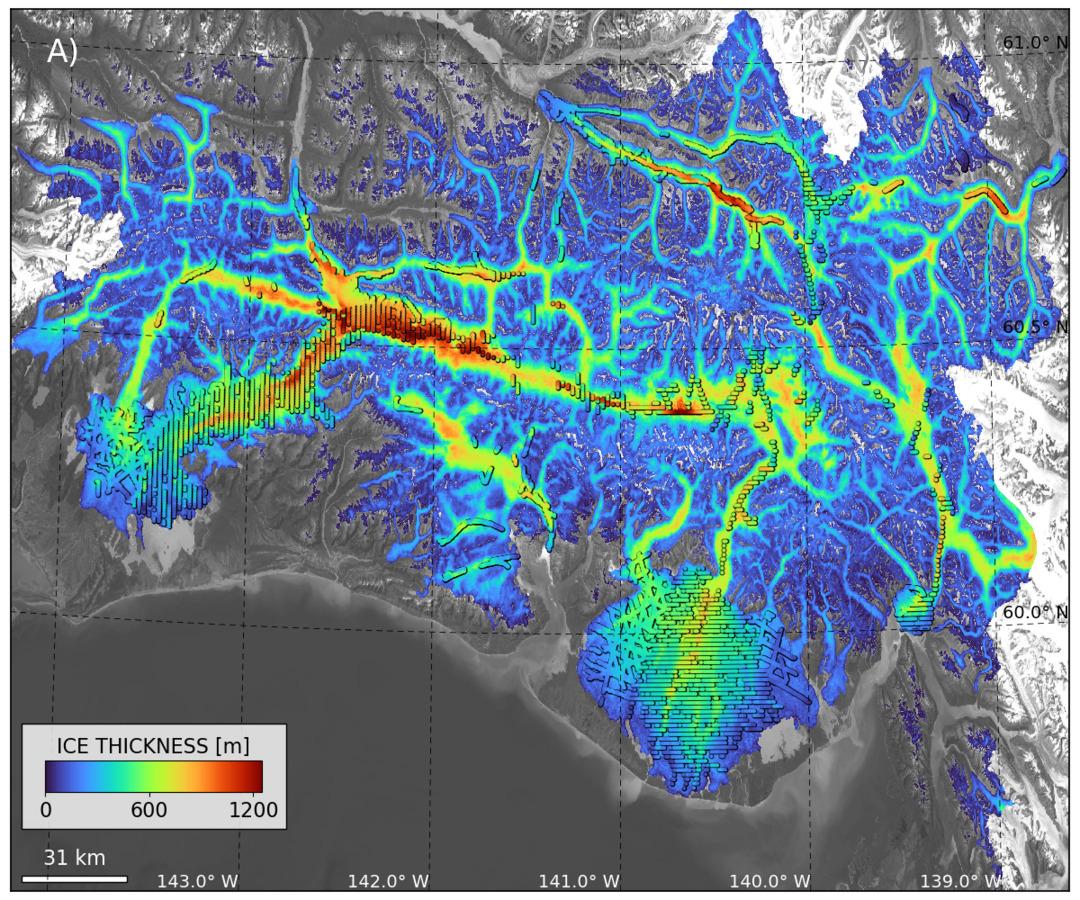}\\
    \includegraphics[width=.49\linewidth]{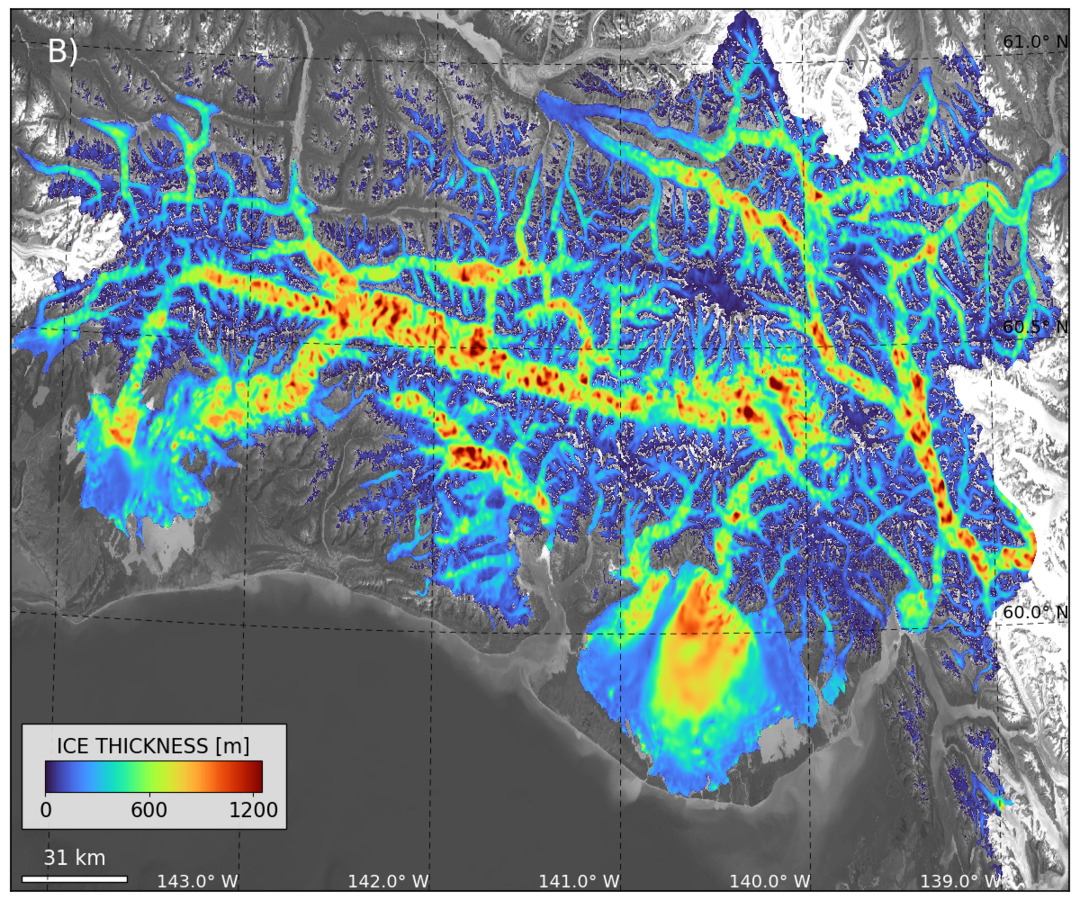}
    \includegraphics[width=.49\linewidth]{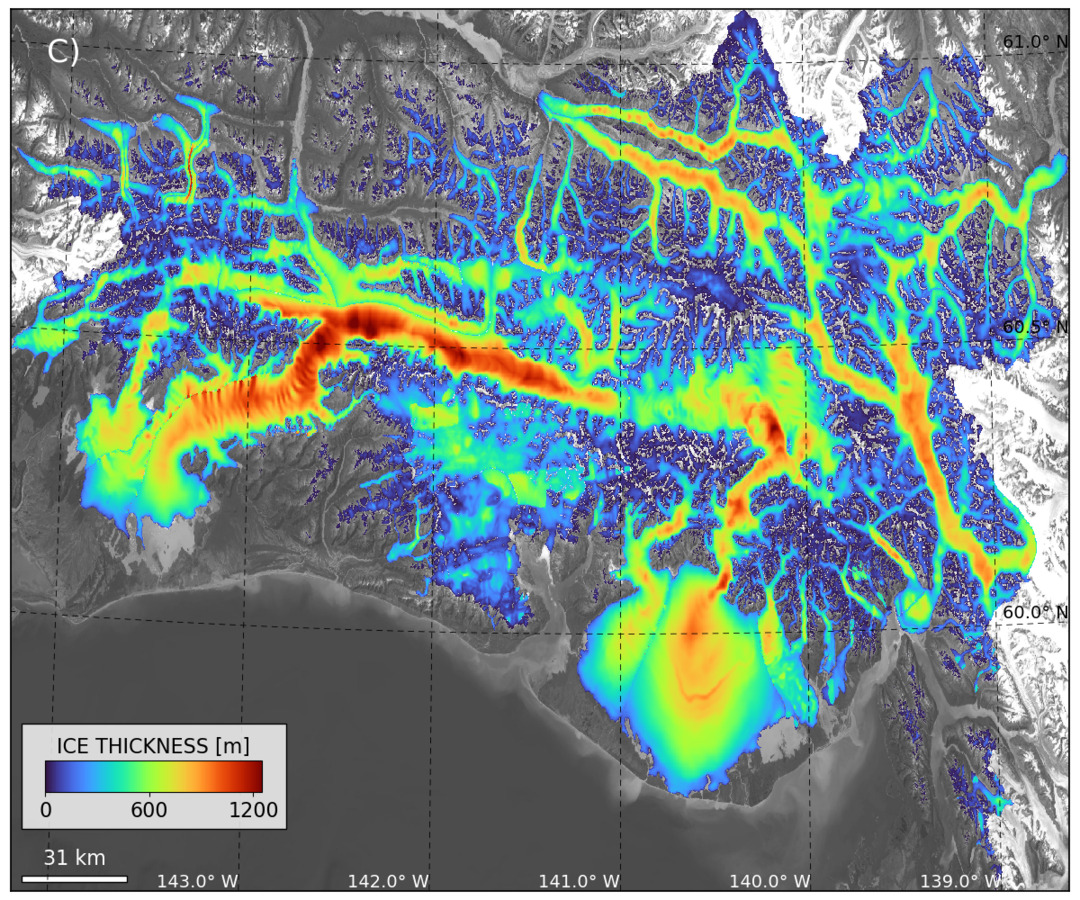}
    \label{fig:alaska_malaspina}
\end{figure}

In the Bering-Malaspina-Seward glacier basins, regions of shallow ice are fairly consistent across all models (Fig. \ref{fig:alaska_malaspina}). Thick-ice areas show differences. Millan's model shows short-scale fluctuations of shallower and deeper ice. IceBoost and Farinotti's models produce smoother fields. Farinotti models ice is too thick compared to data in the deepest parts of the Bering glacier terminus as well as in the Malaspina terminal lobe. When evaluated against ground-truth data collected in Alaska, IceBoost shows errors that are 25\% and 40\% lower (Supp. Info. Fig. S1), indicating that this is the most accurate model for this region. Data and IceBoost indicate that Malaspina, Agassiz, Steller and Bering termini are grounded 100-300 meters below sea level. Yet, IceBoost cannot resolve the bed troughs captured by radar profiles at spatial scales of 100 meters on the Malaspina lobe \cite{tober2023}. All models and data indicate that the Hubbard glacier terminus is grounded up to half a kilometer below sea level in the Disenchantment Bay.

\subsubsection{Canadian Arctic North and South (RGI 03-04)}
\label{sect:canadian_arctic}

\begin{table}[h]
    \caption{Canadian Arctic ice volumes estimated by different models. All units are \SI{e3}{\kilo\metre\tothe{3}}.}\label{tab1}%
    \begin{tabular}{@{}lccc@{}}
    \toprule
    Canadian Arctic North (RGI 03) & IceBoost v2 & Millan et al. \cite{millan2022} & Farinotti et al. \cite{farinotti2019}\\
    \midrule
    Total & 24.3 ± 6.2 & 25.4 ± 7.2  & 28.3 ± 7.4  \\
    - Northern Ellesmere Island & 11.137±3.106 & 10.651 & 12.180 \\
    - Muller icefield & 2.28 ± 0.56 & 2.38 & 2.49 \\
    - Prince Of Wales, SydKap, Manson Icefields & 6.269 ± 1.512 & 6.978 & 7.871 \\
    - Devon ice cap & 4.365 ± 0.943 & 4.87 & 5.421 \\
    - Others & 0.249 ± 0.079 & 0.521 & 0.338 \\
    \midrule
    Canadian Arctic South (RGI 04) & IceBoost v2 & Millan et al. \cite{millan2022} & Farinotti et al. \cite{farinotti2019}\\
    \midrule
    Total & 7.1 ± 1.8 & 7.0 ± 2.1  & 8.6 ± 2.2 \\
    - Baffin Island North & 1.741±0.466 & 1.509 & 2.220 \\
    - Baffin Island Central & 2.893±0.719  & 2.836 & 3.313 \\
    - Baffin Island South & 2.163±0.538 & 2.308 & 2.728 \\
    - Others & 0.303±0.077 & 0.347 & 0.339 \\
    \bottomrule
    \end{tabular}
\end{table}

Comparison with measurements suggests that IceBoost performs best among all models in the Canadian Arctic (RGI 3 and 4), with errors lower by 32\% in the Canadian Arctic North and by 46–53\% in the Canadian Arctic South (Supp. Info. Fig. S1).\\

\begin{landscape}
\begin{figure}[h!]
    \caption{Canadian Arctic North. Top: Northern Ellesmere Island. Bottom: Devon Ice Cap. A=IceBoost v2; B=Millan et al. 2022 \cite{millan2022}; C=Farinotti et al. \cite{farinotti2019}.}
    \includegraphics[width=.32\linewidth]{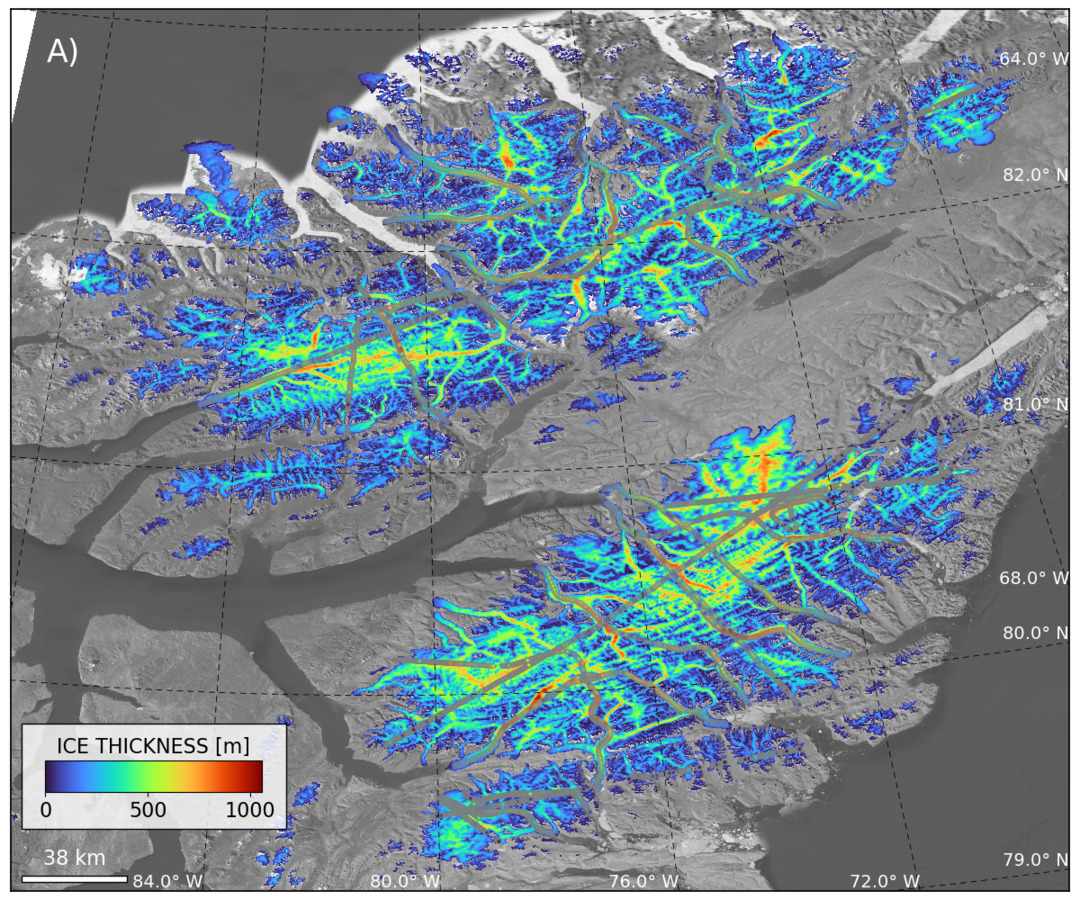}
    \includegraphics[width=.32\linewidth]{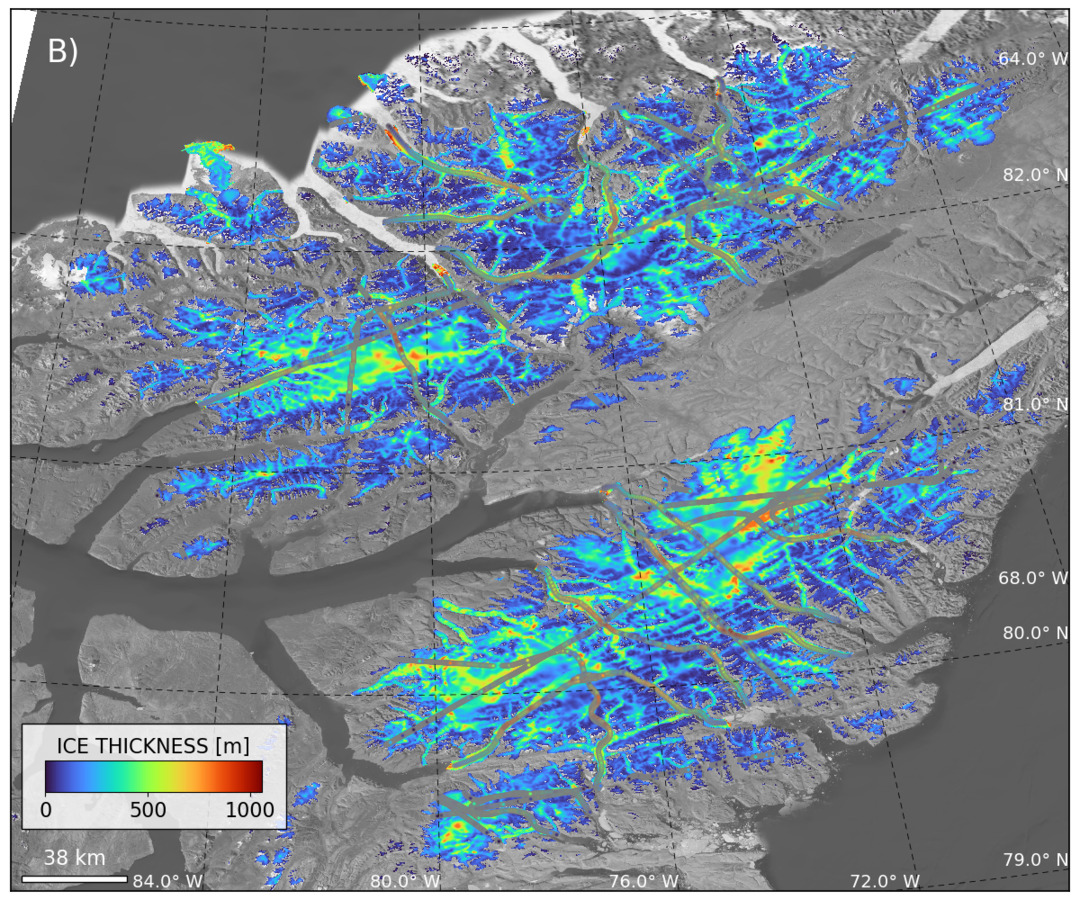}
    \includegraphics[width=.32\linewidth]{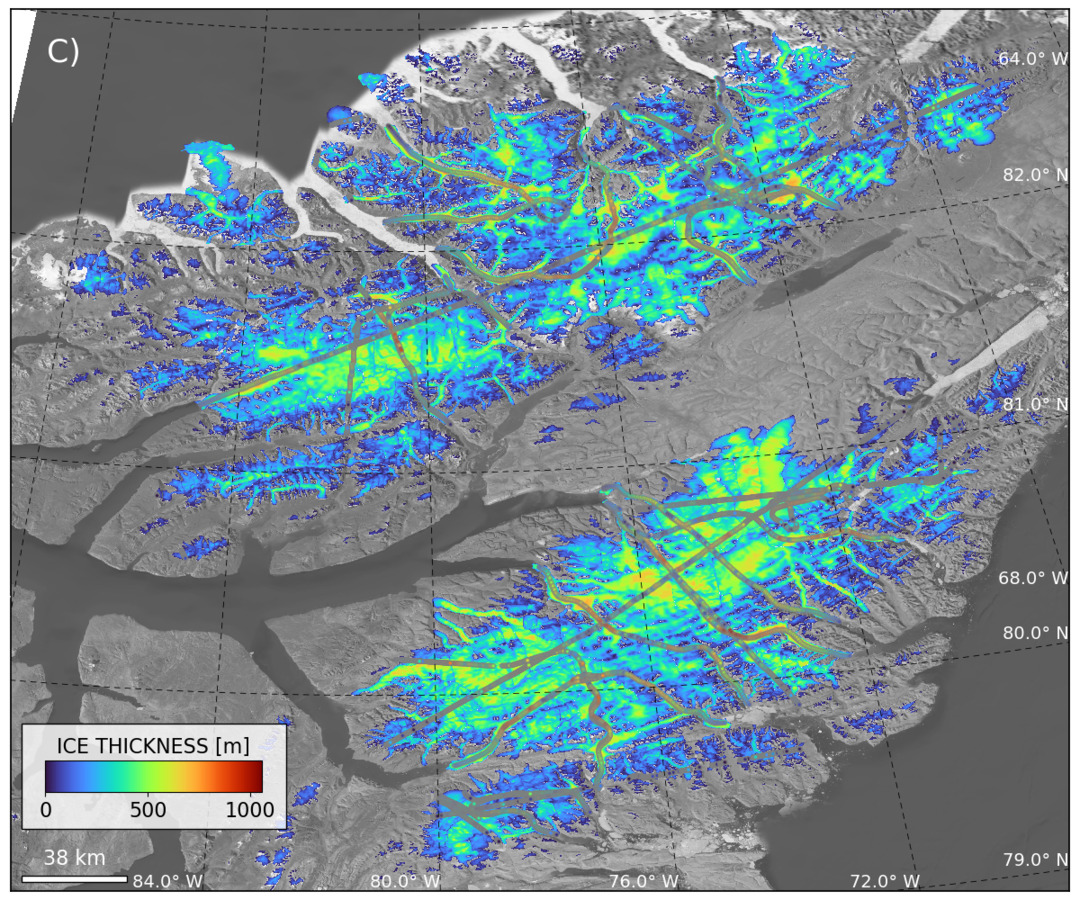}
    \label{fig:northern_ellesmere}\\
    \includegraphics[width=.32\linewidth]{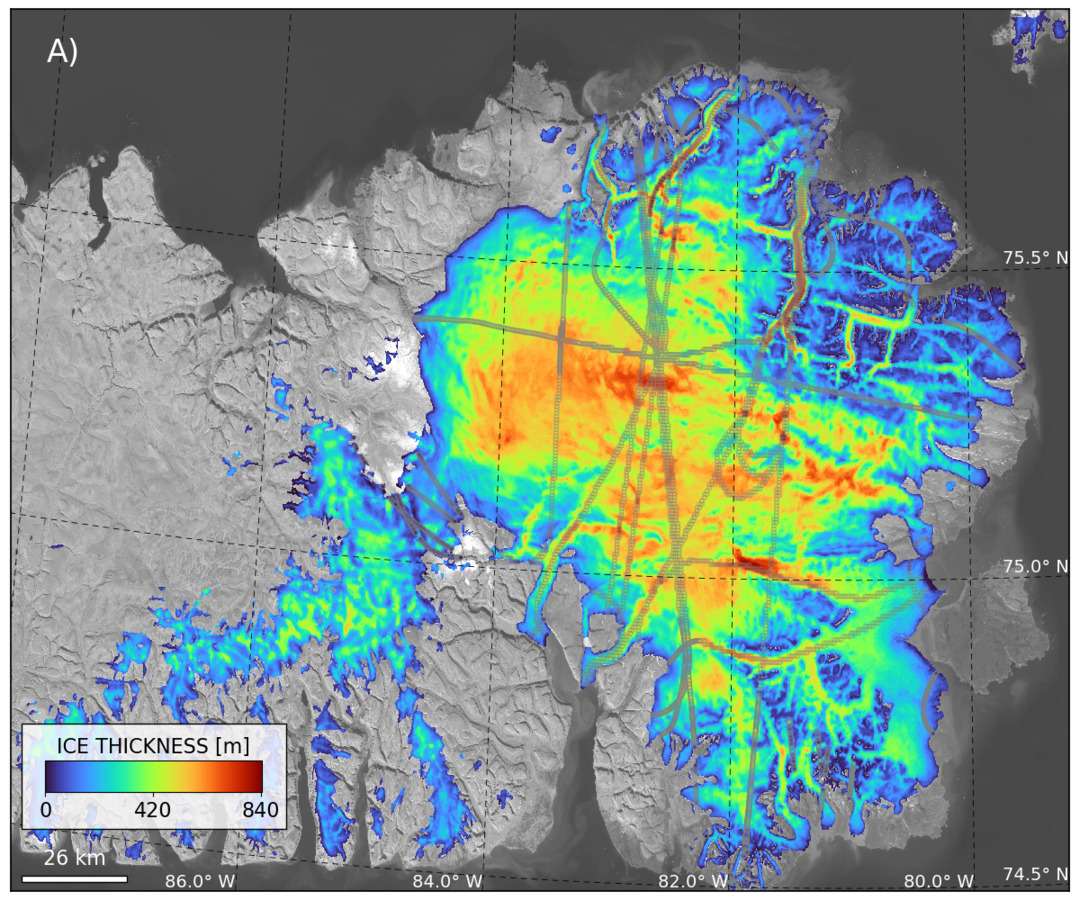}
    \includegraphics[width=.32\linewidth]{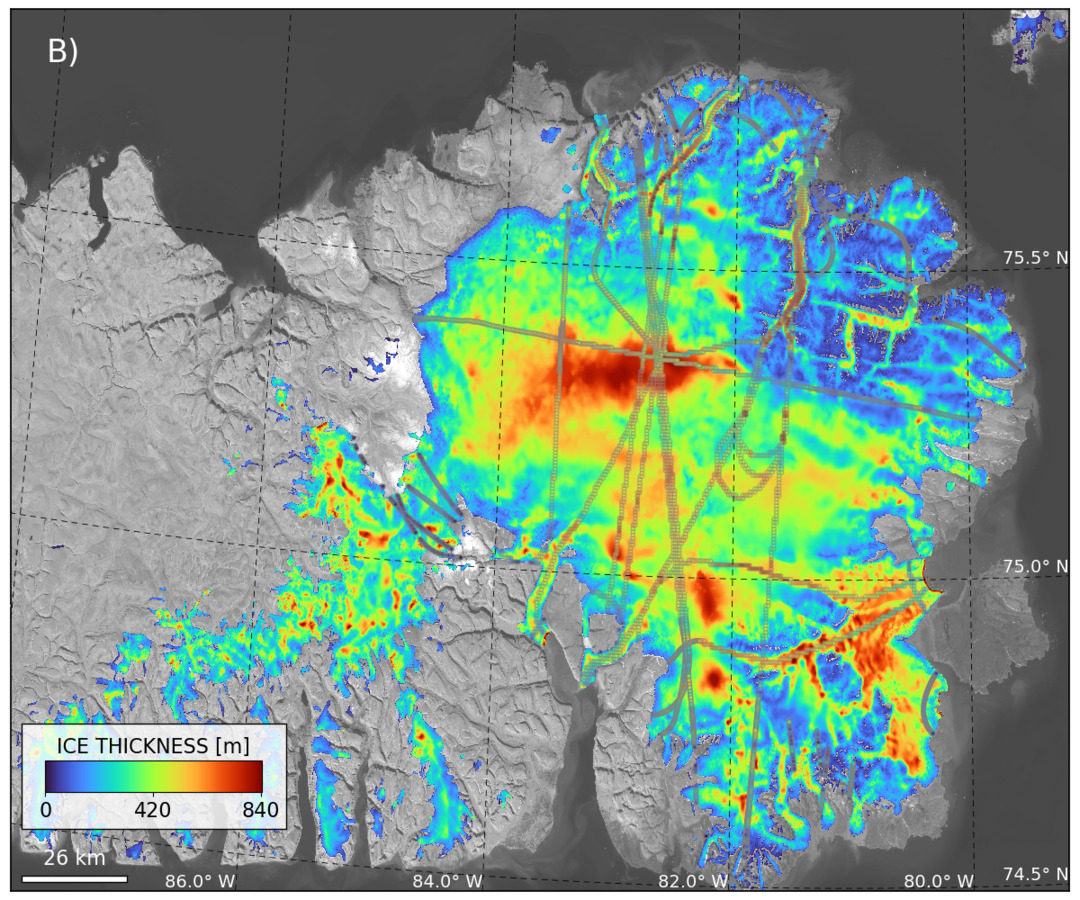}
    \includegraphics[width=.32\linewidth]{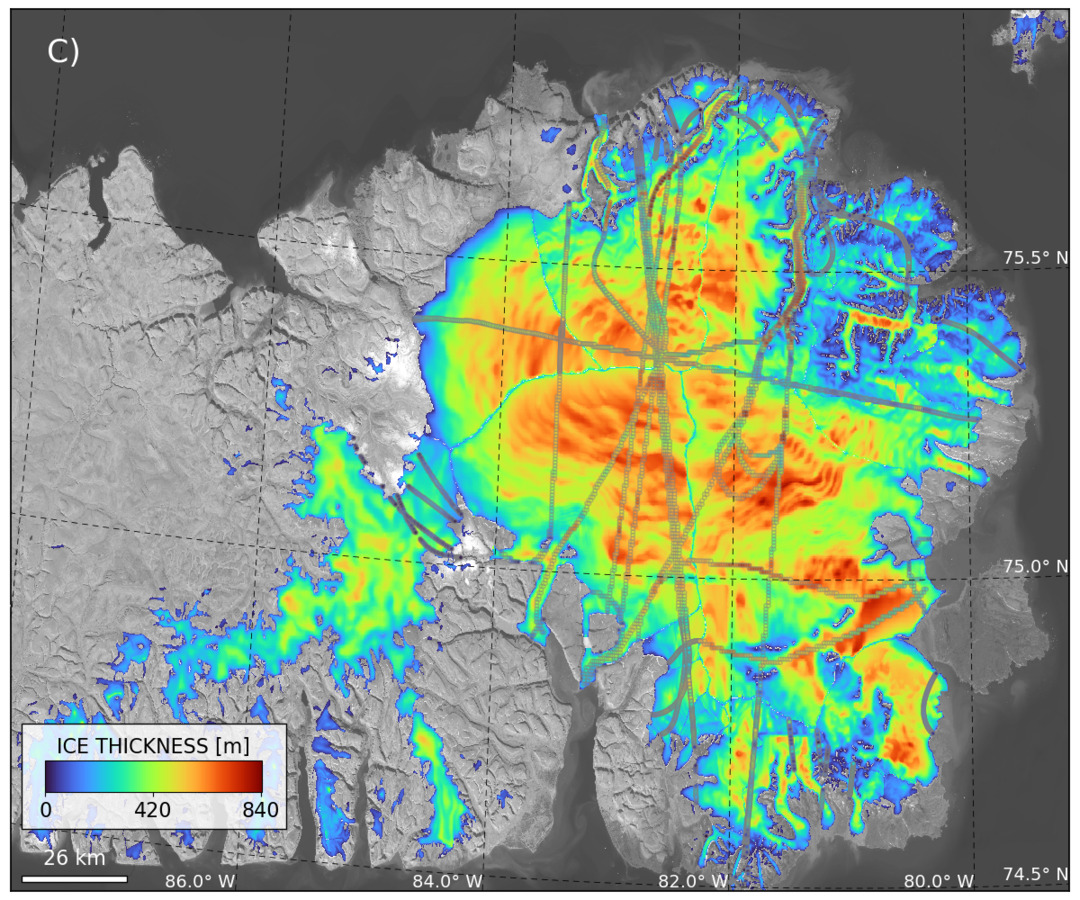}
    \label{fig:north_ellesmere_devon}
\end{figure}
\end{landscape}

% Canadian Arctic North: Northern Ellesmere Island
\noindent In the Northern Ellesmere Island (Fig. \ref{fig:north_ellesmere_devon}) deep ice channels are reproduced only by IceBoost. Millan's model is problematic at high speed-low slope termini. Farinotti’s model shows the least realistic thickness distribution.\\

% Canadian Arctic North: Devon ice cap
\noindent The Devon ice cap is modeled too thick compared to data by both Millan and Farinotti models, especially in the tidewater glacier terminations on the eastern side of the ice cap (Fig. \ref{fig:north_ellesmere_devon}). IceBoost models the southwest arm of the ice cap as thinner than the other reconstructions, by 100-200 meters on average. Very few measurements exist that can confirm this result, as the region remains unsurveyed.\\

\begin{landscape}
\begin{figure}[h!]
    \centering
    \caption{Canadian Arctic North. Top: Muller Icefield; Bottom: Prince of Wales, SydKap and Manson Icefields. A=IceBoost v2; B=Millan et al. 2022 \cite{millan2022}; C=Farinotti et al. \cite{farinotti2019}.}
    \includegraphics[width=.32\linewidth]{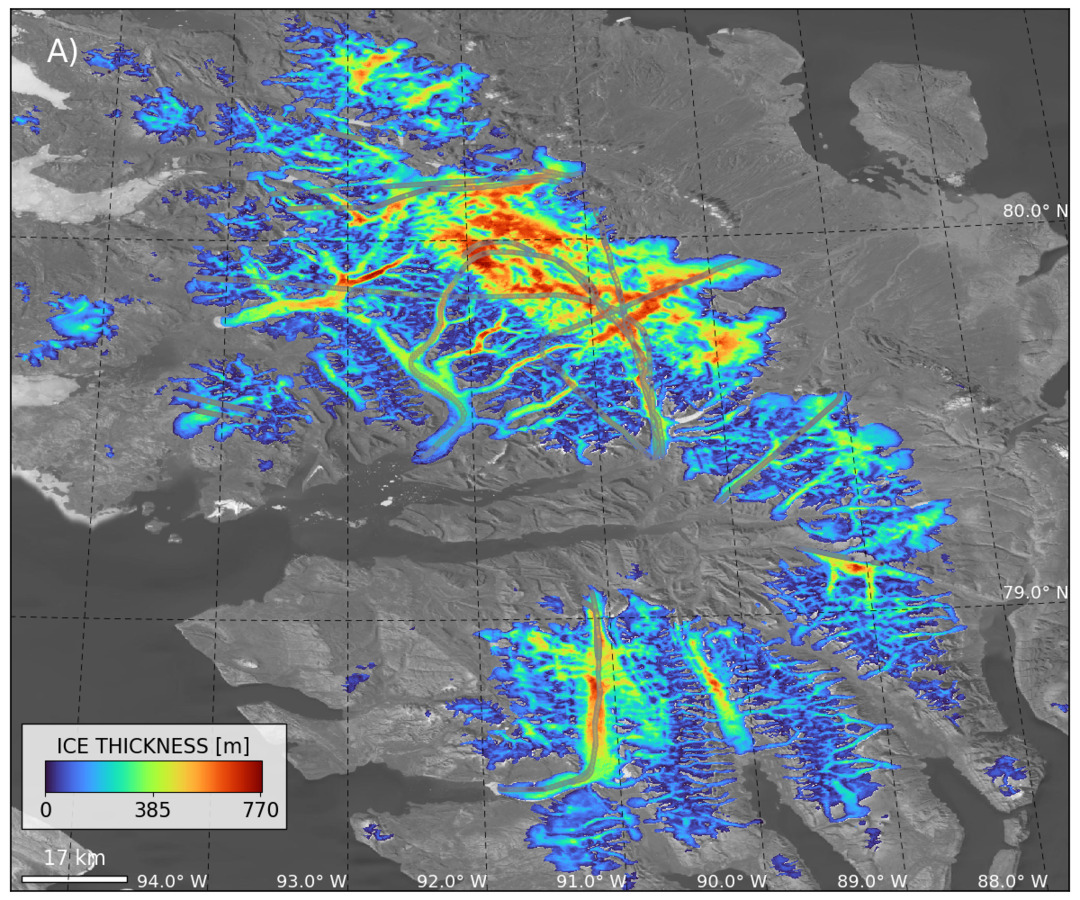}
    \includegraphics[width=.32\linewidth]{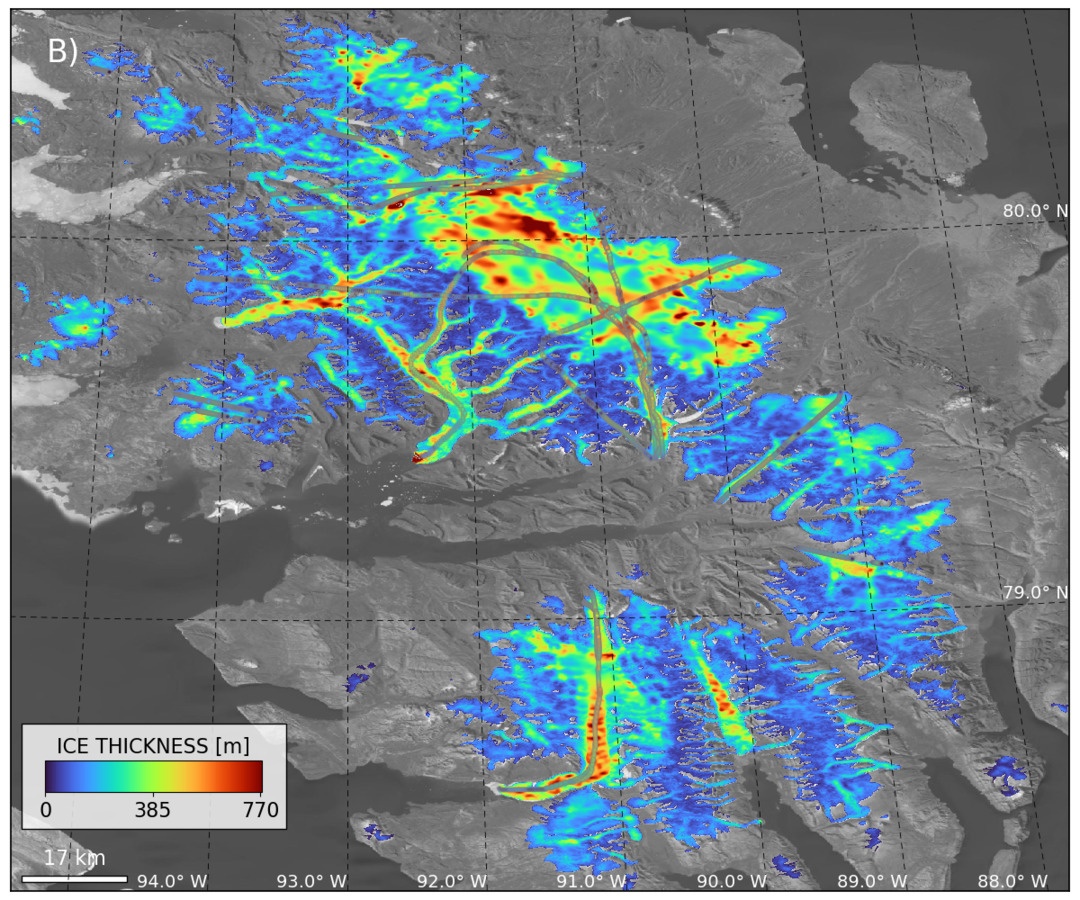}
    \includegraphics[width=.32\linewidth]{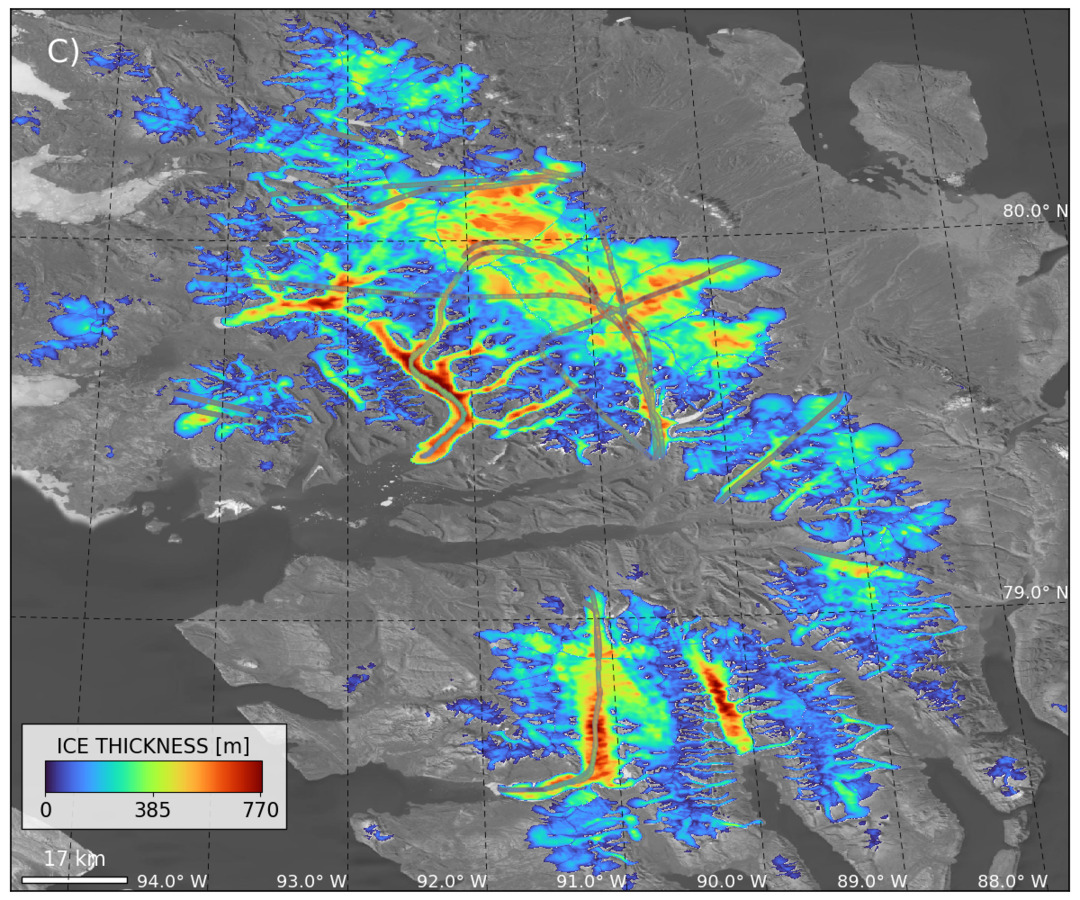}\\
    \includegraphics[width=.32\linewidth]{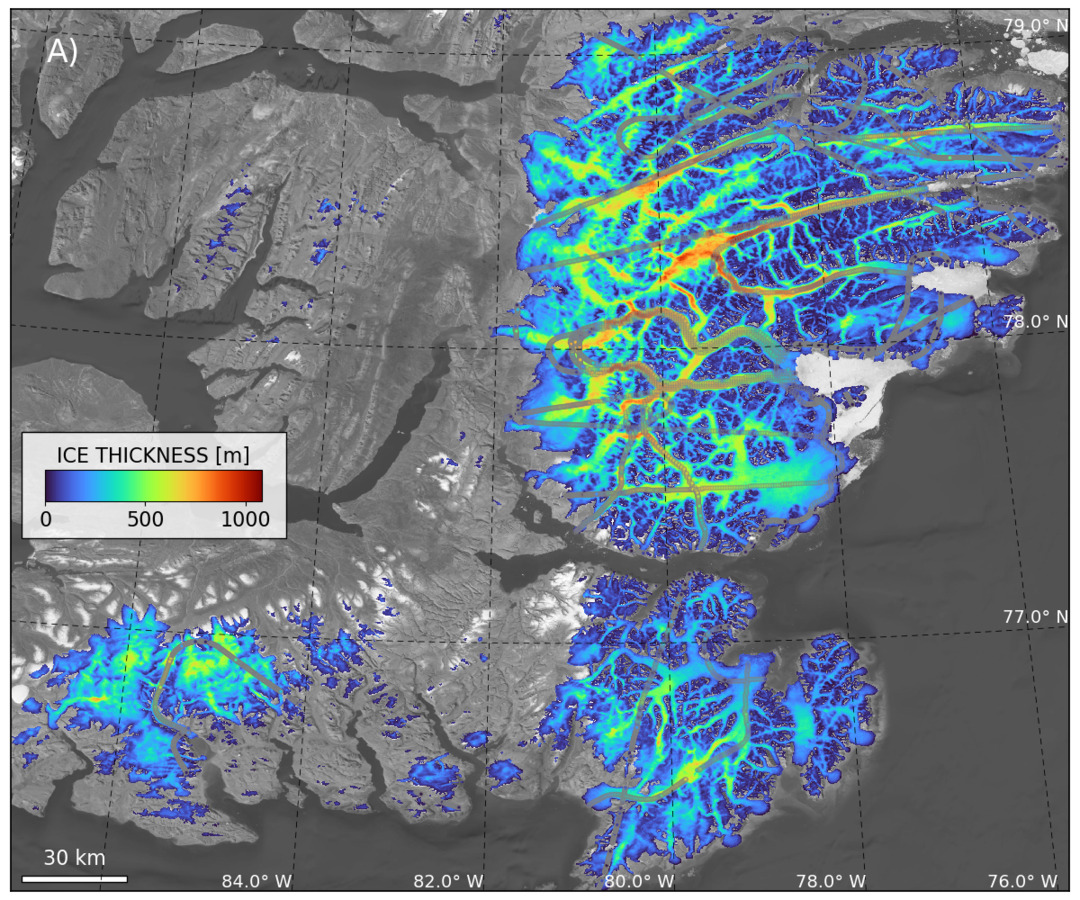}
    \includegraphics[width=.32\linewidth]{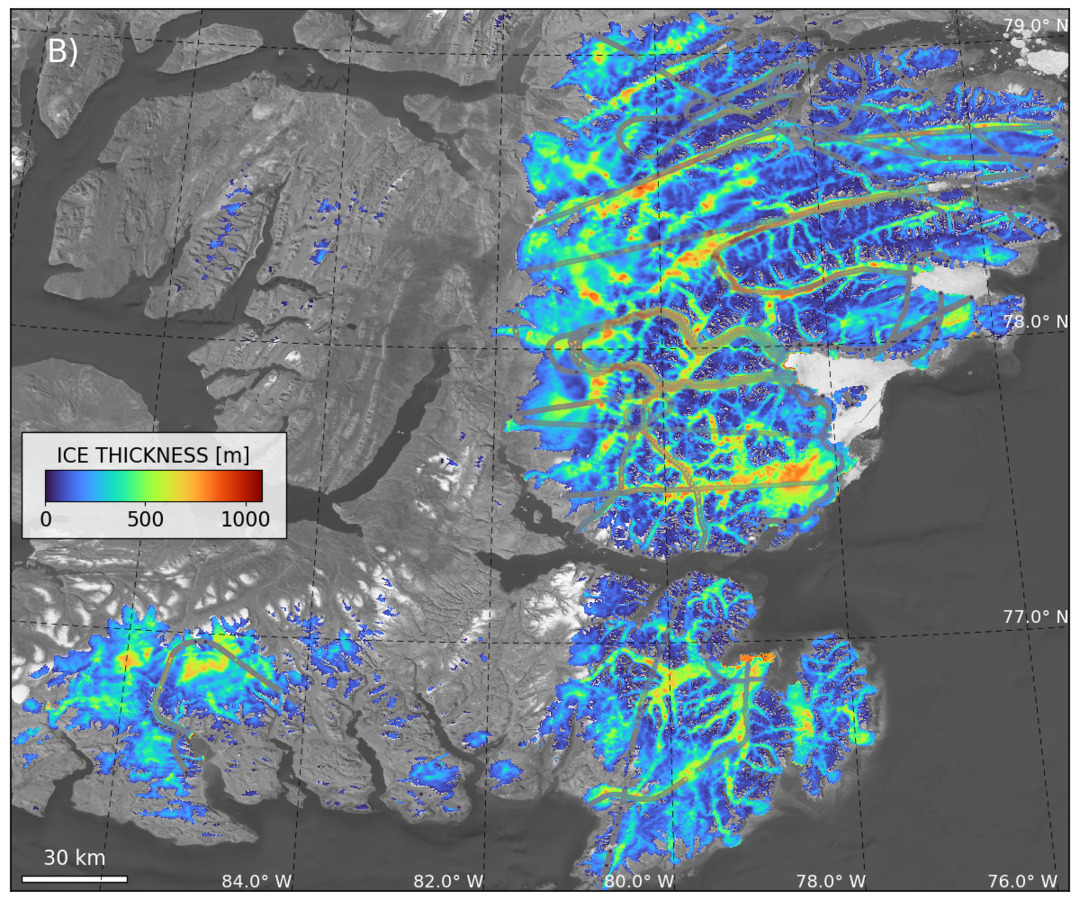}
    \includegraphics[width=.32\linewidth]{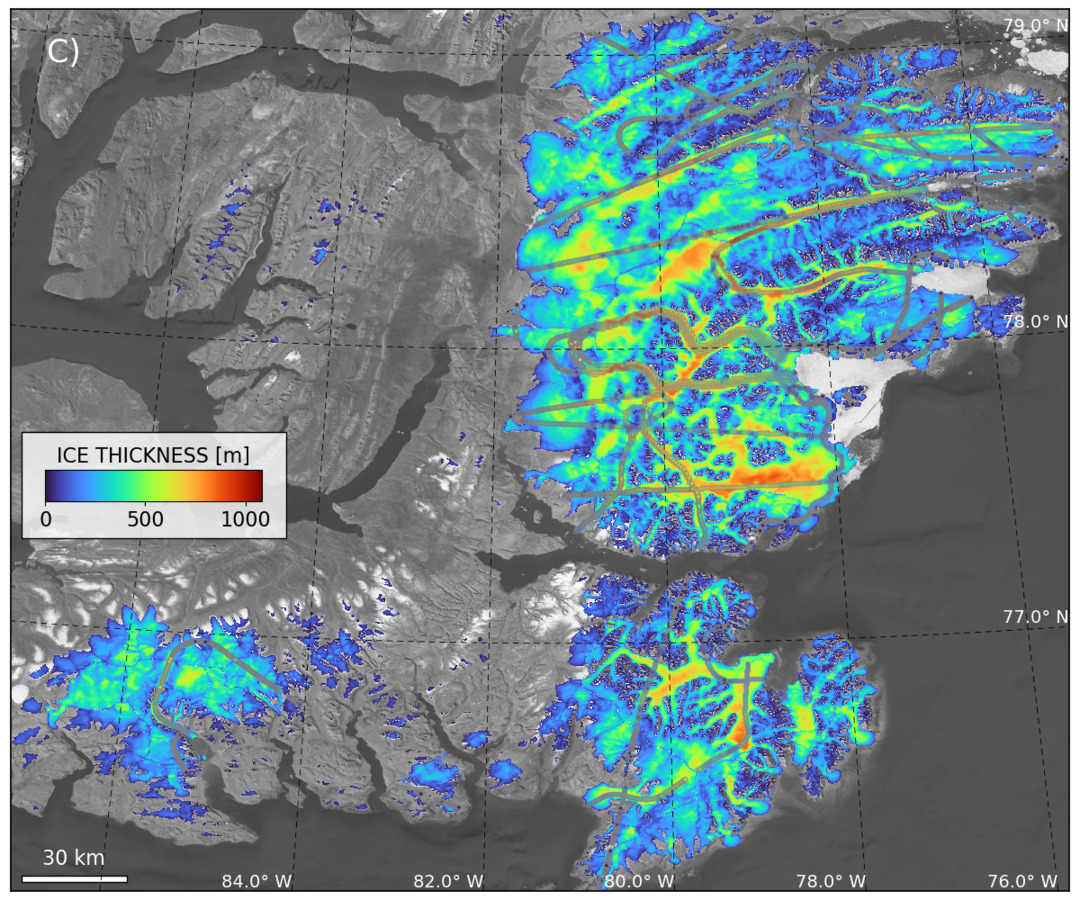}
    \label{fig:can_arctic_north_muller_prince_wales}
\end{figure}
\end{landscape}

% Muller Icefield 
% Prince of Wales, SydKap and Manson icefields
\noindent Over the Müller icefield (Fig. \ref{fig:can_arctic_north_muller_prince_wales}), IceBoost shows the best agreement with observations. Farinotti’s reconstruction is too thick. Millan's model performs better yet shows problems at outlet glaciers. Over the Prince of Wales, SydKap and Manson icefields, IceBoost and Millan's reconstructions works best (Fig. \ref{fig:can_arctic_north_muller_prince_wales}). IceBoost (and data) indicates that several glaciers terminating in the Baffin Bay -including Ekblaw, Cadogan, Trinity–Wykeham, and Mittie glaciers - are grounded below sea level for tens of kilometres inland, consistent with published literature \cite{dalton2022}. \\

\begin{figure}[h!]
    \caption{Baffin Island North (Can. Arctic South). A=IceBoost v2 ; B=Millan et al. 2022 \cite{millan2022}; C=Farinotti et al. \cite{farinotti2019}.}
    \includegraphics[width=.49\linewidth]{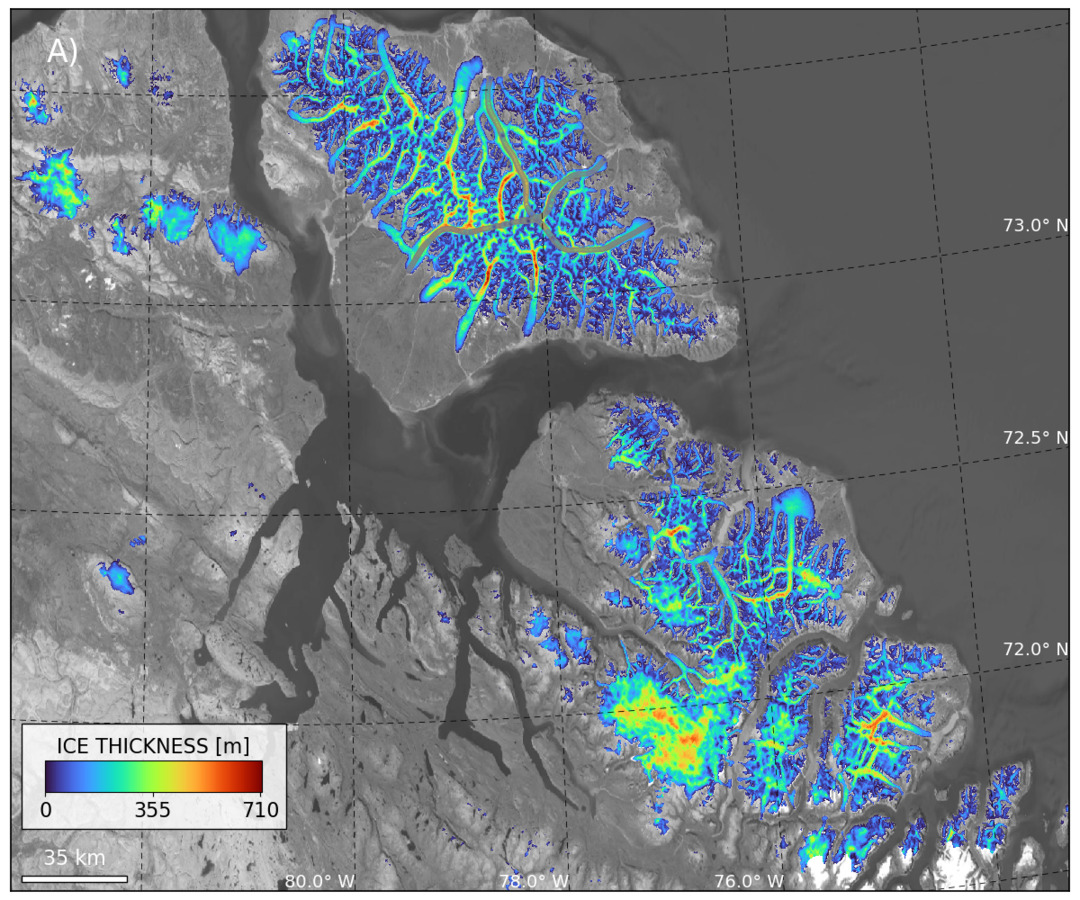}
    \includegraphics[width=.49\linewidth]{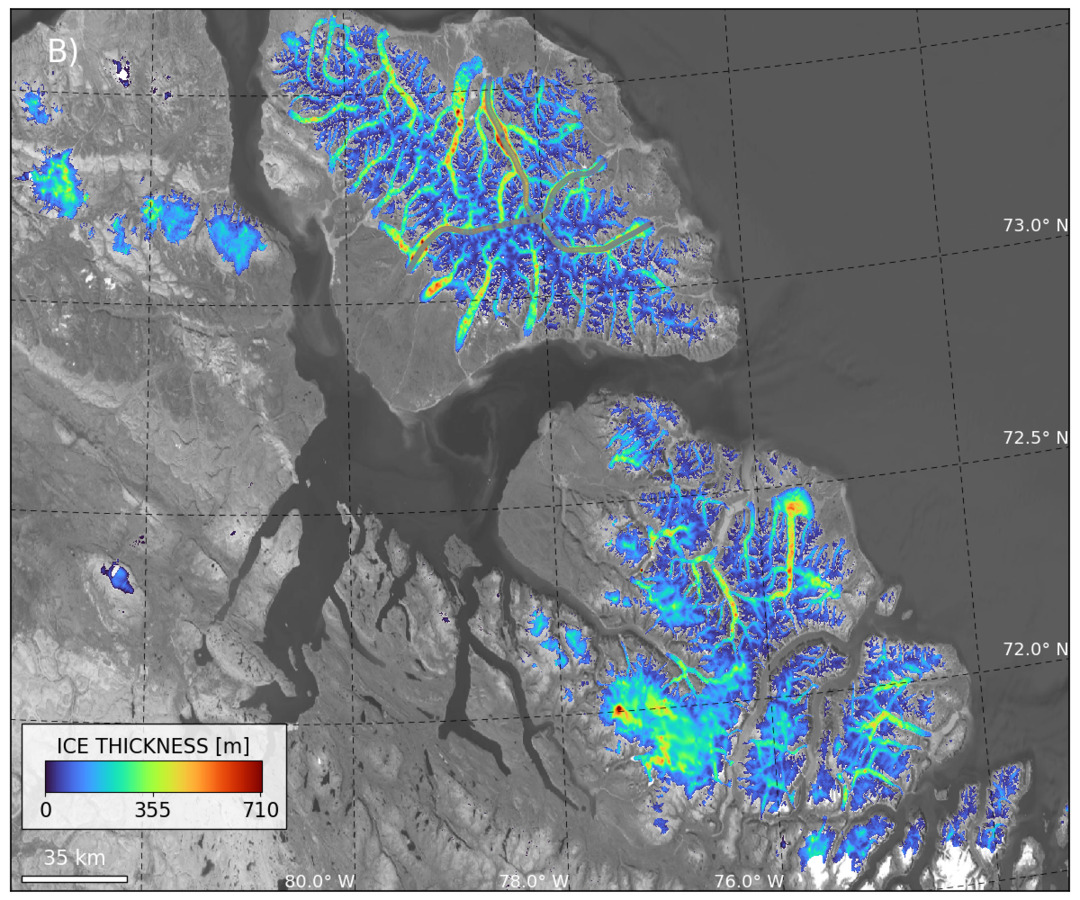}
    \includegraphics[width=.49\linewidth]{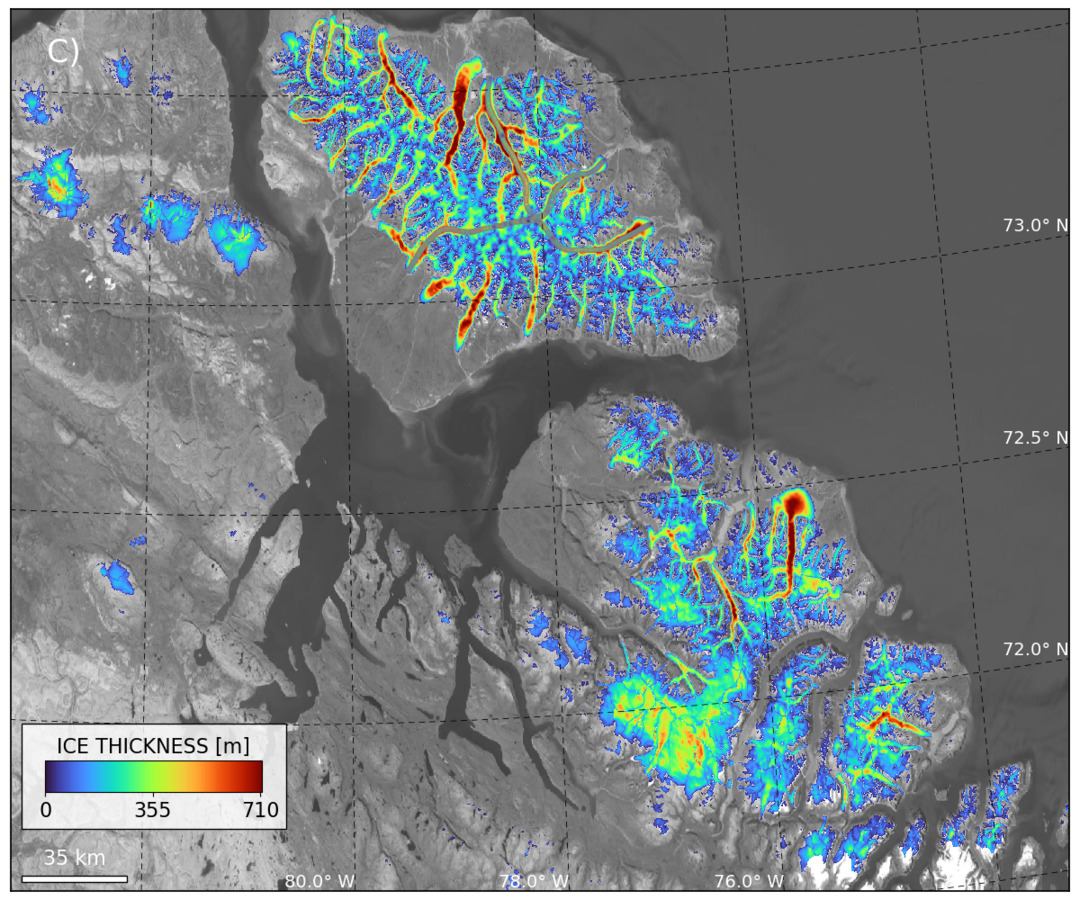}
    \label{fig:can_arctic_south_baffin_north}
\end{figure}

\begin{landscape}
\begin{figure}[h!]
    \centering
    \caption{Top: Baffin Island Central (Can. Arctic South). Bottom: Baffin Island South (Can. Arctic South). A=IceBoost v2; B=Millan et al. \cite{millan2022}; C=Farinotti et al. \cite{farinotti2019}.}
    \includegraphics[width=.32\linewidth]{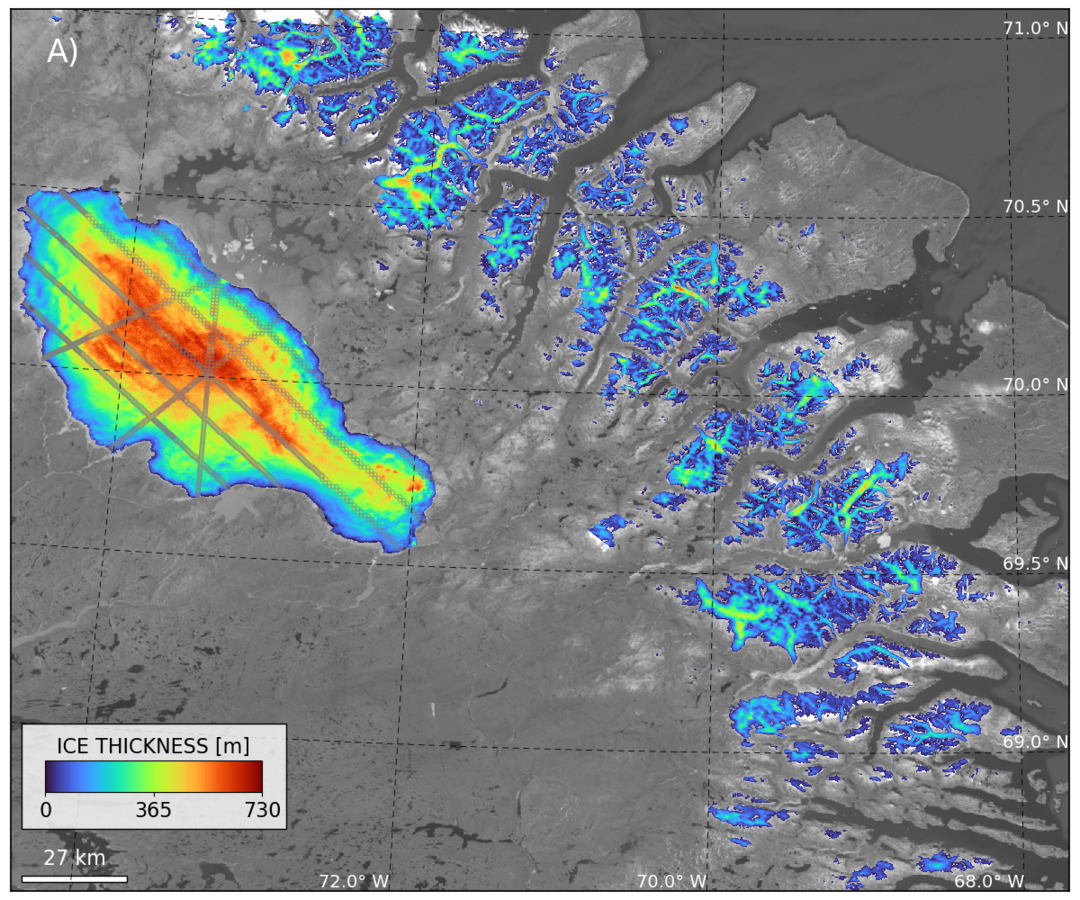}
    \includegraphics[width=.32\linewidth]{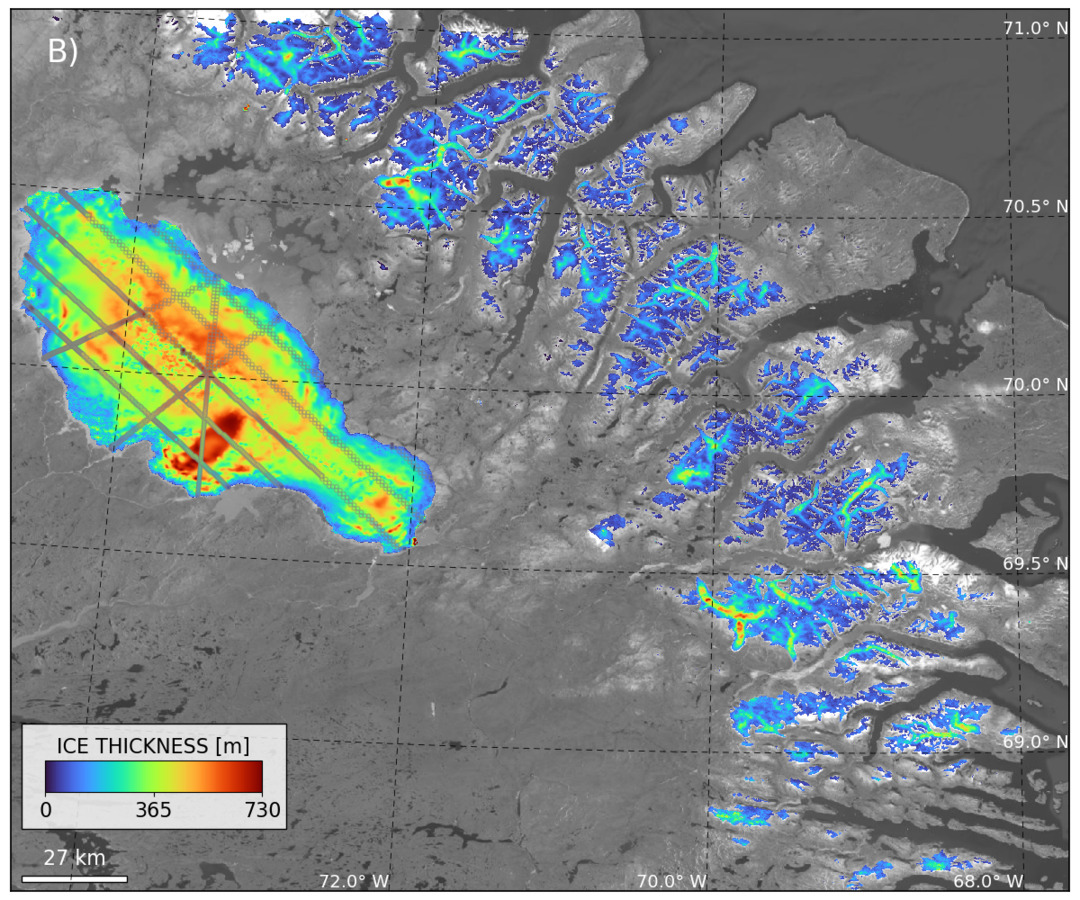}
    \includegraphics[width=.32\linewidth]{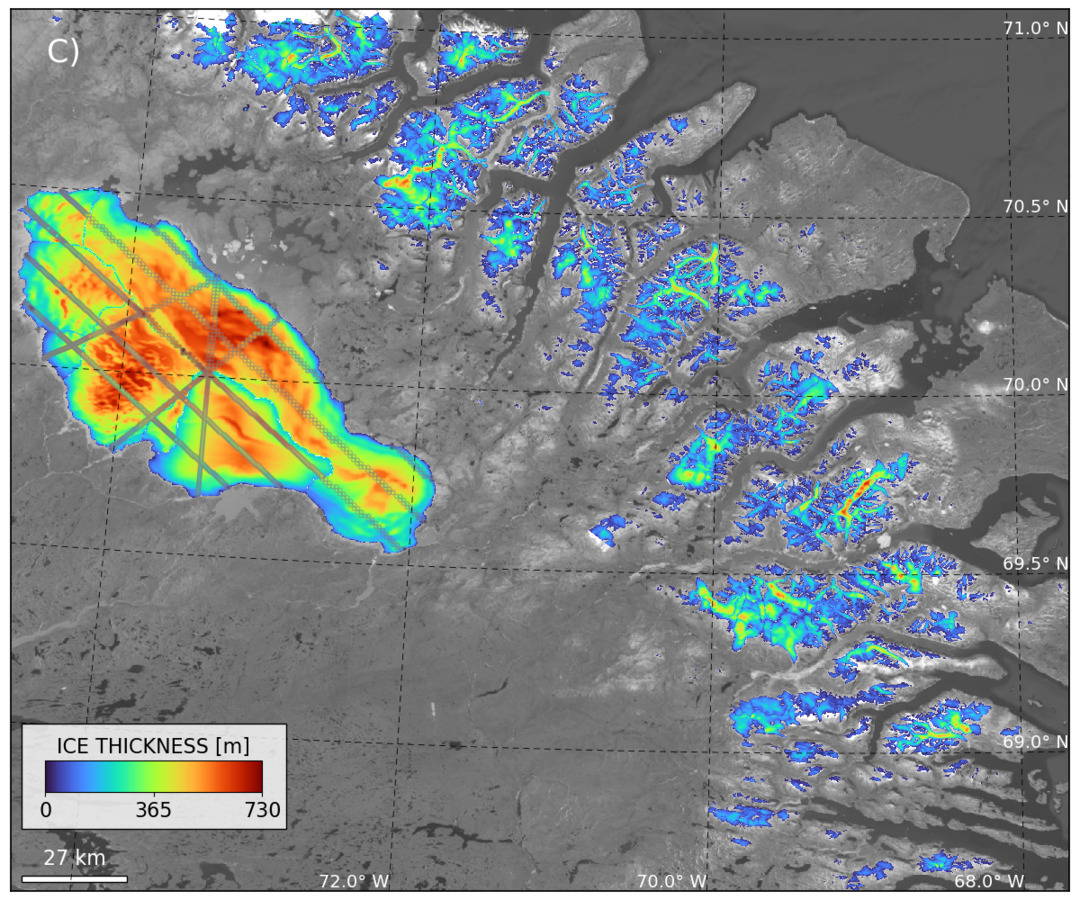}\\
    \includegraphics[width=.32\linewidth]{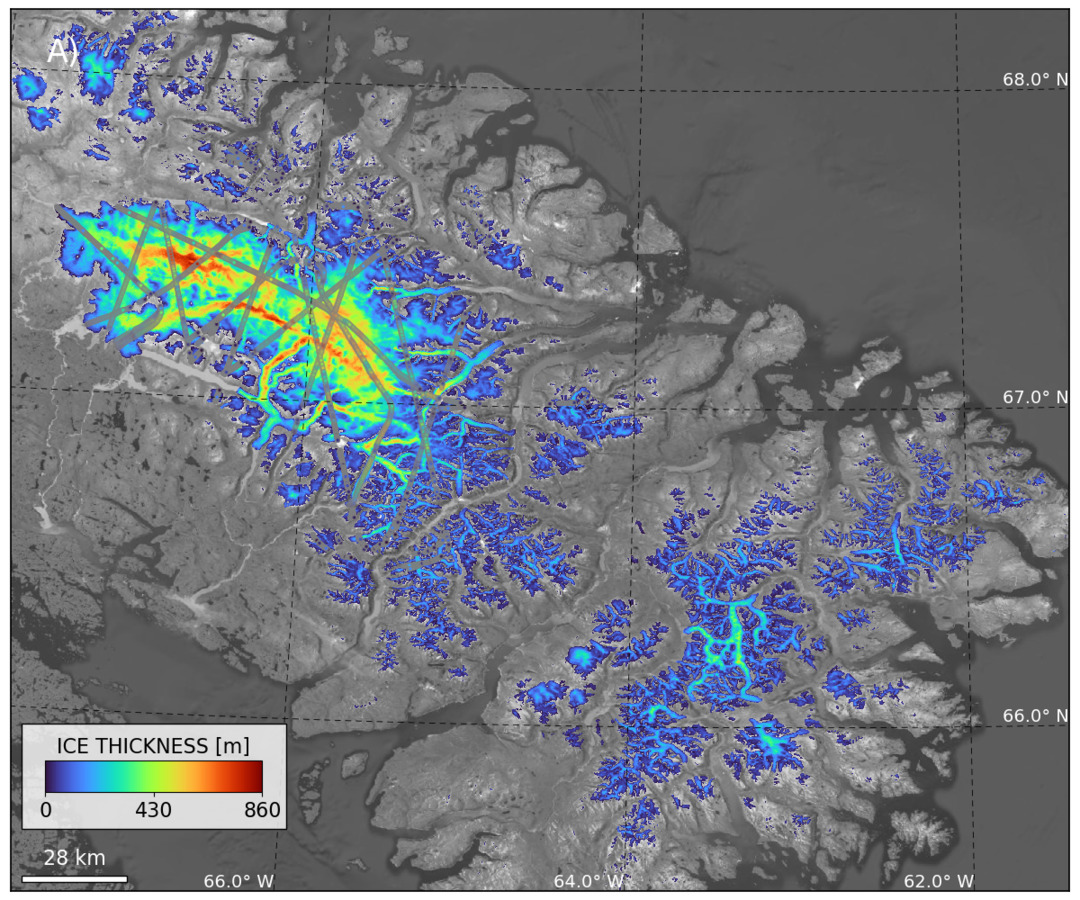}
    \includegraphics[width=.32\linewidth]{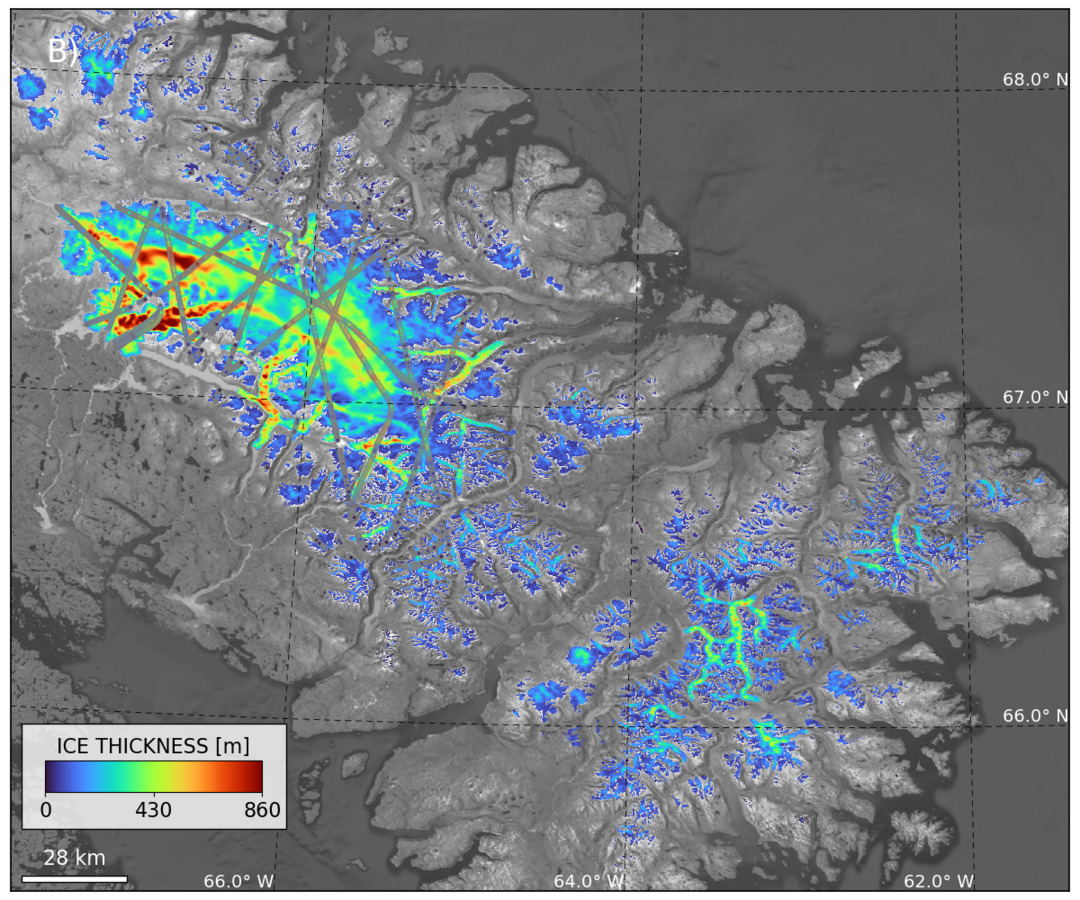}
    \includegraphics[width=.32\linewidth]{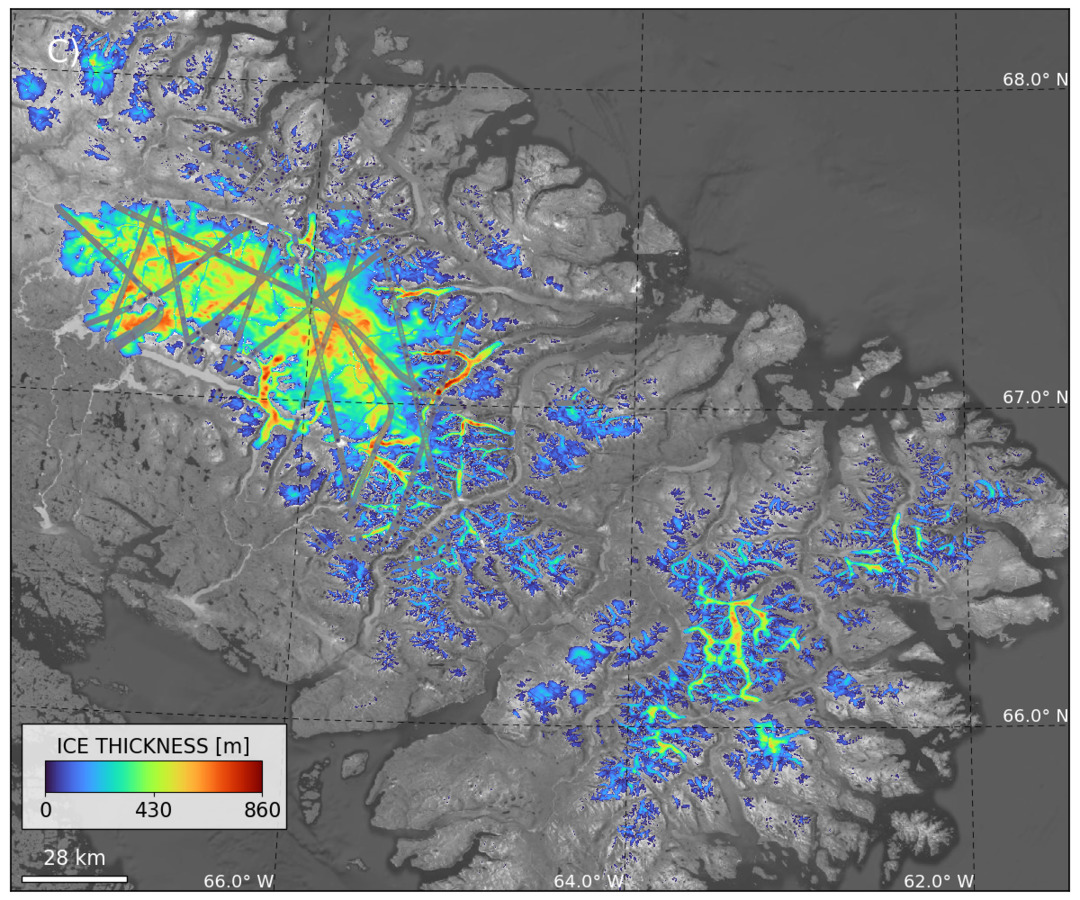}
    \label{fig:can_arctic_south_baffin_central_south}
\end{figure}
\end{landscape}

% Canadian Arctic South
\noindent In the Canadian Arctic South (Baffin Island, Figs. \ref{fig:can_arctic_south_baffin_north}-\ref{fig:can_arctic_south_baffin_central_south}) measurements acquired over glaciers in the Bylot Island, the Barnes and Penny ice caps indicate that IceBoost has the best agreement with data. IceBoost suggests that no glaciers in the Baffin Island are grounded below sea level.

\subsubsection{Russian Arctic (RGI 09)}
\label{sect:russian_arctic}

\begin{table}[h]
    \caption{Russian Arctic ice volumes estimated by different models. All units are \SI{e3}{\kilo\metre\tothe{3}}.}\label{table:russian_arctic}%
    \begin{tabular}{@{}lccc@{}}
    \toprule
    Russian Arctic (RGI 09) & IceBoost v2 & Millan et al. \cite{millan2022} & Farinotti et al. \cite{farinotti2019}\\
    \midrule
    Total & 12.8 ± 3.0 & 15.5 ± 3.9 & 14.6 ± 3.8  \\
    - Novaya Zemlya & 6.556 ± 1.307 & 7.595 & 7.080 \\
    - Severnaya Zemlya & 4.287 ± 1.023 & 4.929 & 4.976 \\
    - Franz Josef Land & 1.864 ± 0.6397 & 2.810 & 2.412 \\
    - Others & 0.093 ± 0.030 & 0.166 & 0.132 \\
    \bottomrule
    \end{tabular}
\end{table}

No measurements from the Russian Arctic are available in our training dataset to validate any of the models, so only qualitative comparisons are possible. Farinotti’s model exhibits strong discontinuities between neighboring basins. Millan’s inversion appears more realistic, but it predicts ice that is too thick at the high-velocity termini of the Academy of Sciences ice cap and near glacier margins. IceBoost produces the shallowest ice across all models, and its regionally integrated ice volume is lower than the other two estimates. Data is needed everywhere in the Russian Arctic.

\begin{figure}[h!]
    \caption{Franz Josef Land (Russian Arctic) modeled with IceBoost v2 (A), Millan et al. (B, \cite{millan2022}) and Farinotti et al. (C, \cite{farinotti2019}).}
    \includegraphics[width=.49\linewidth]{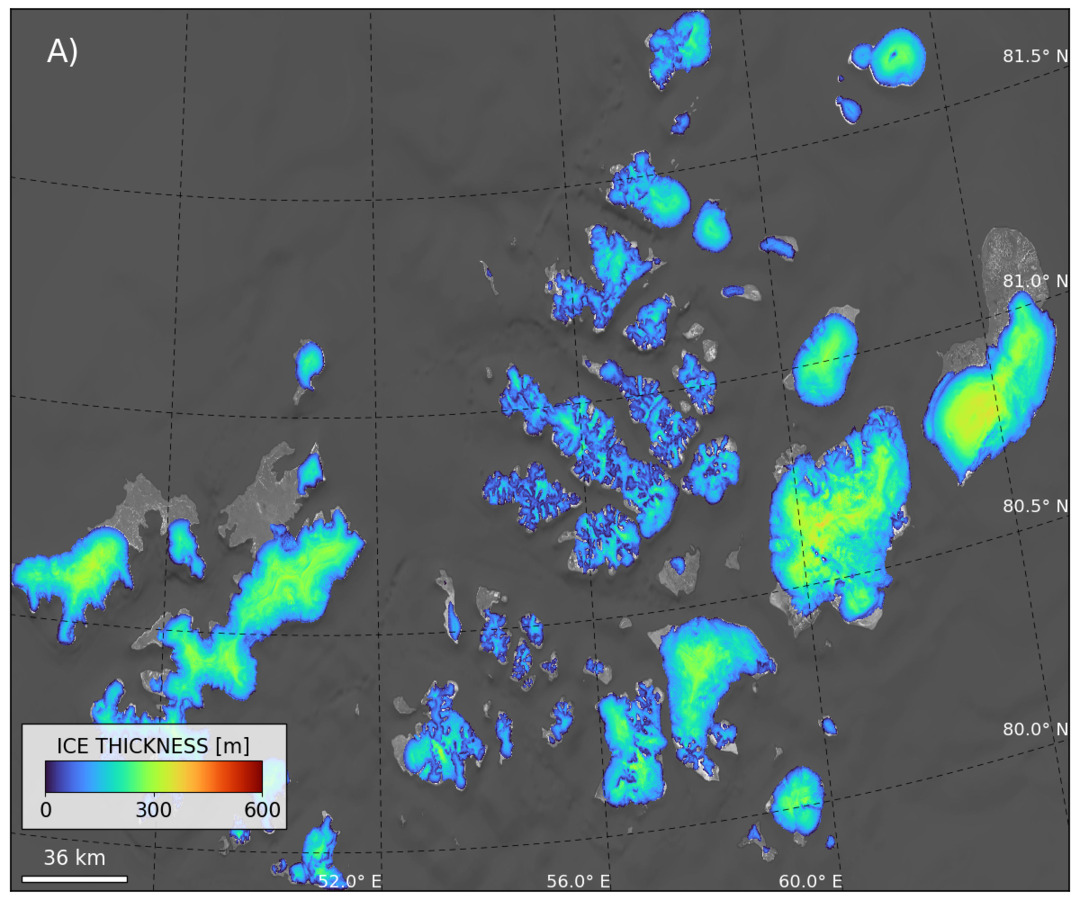}
    \includegraphics[width=.49\linewidth]{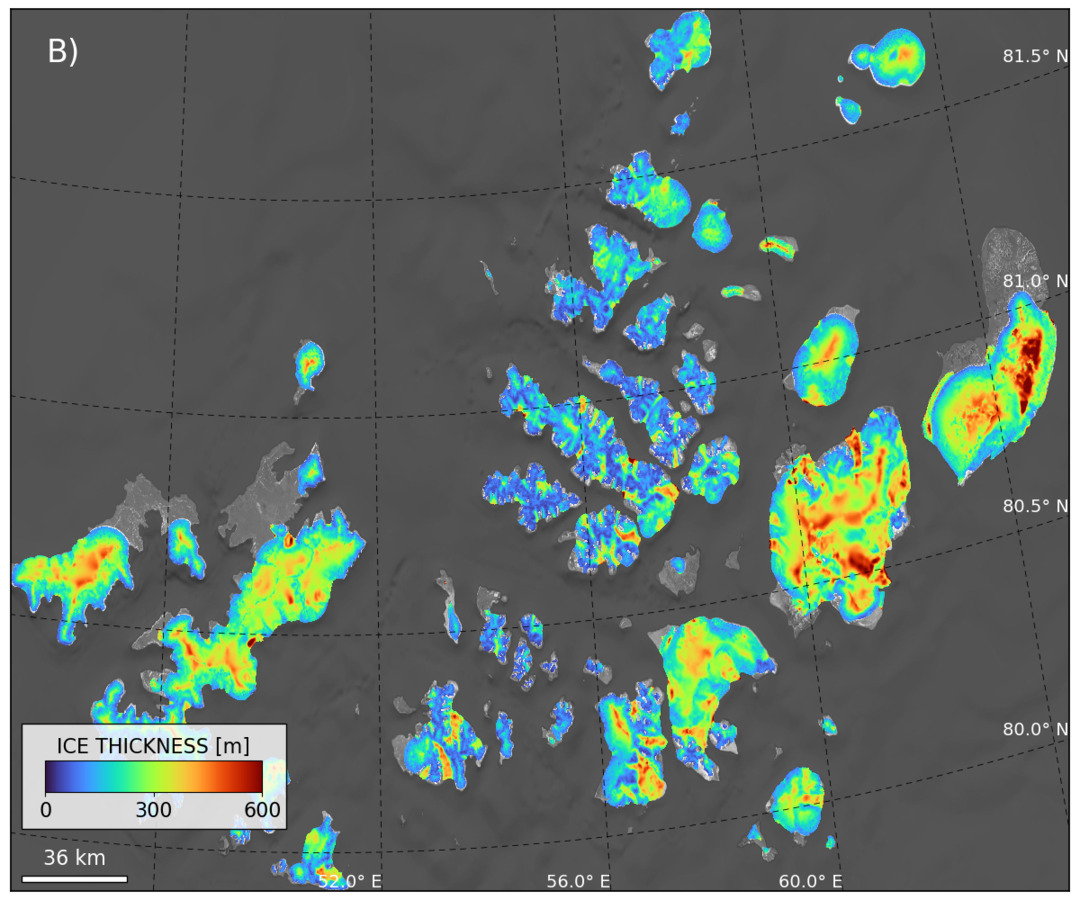}
    \includegraphics[width=.49\linewidth]{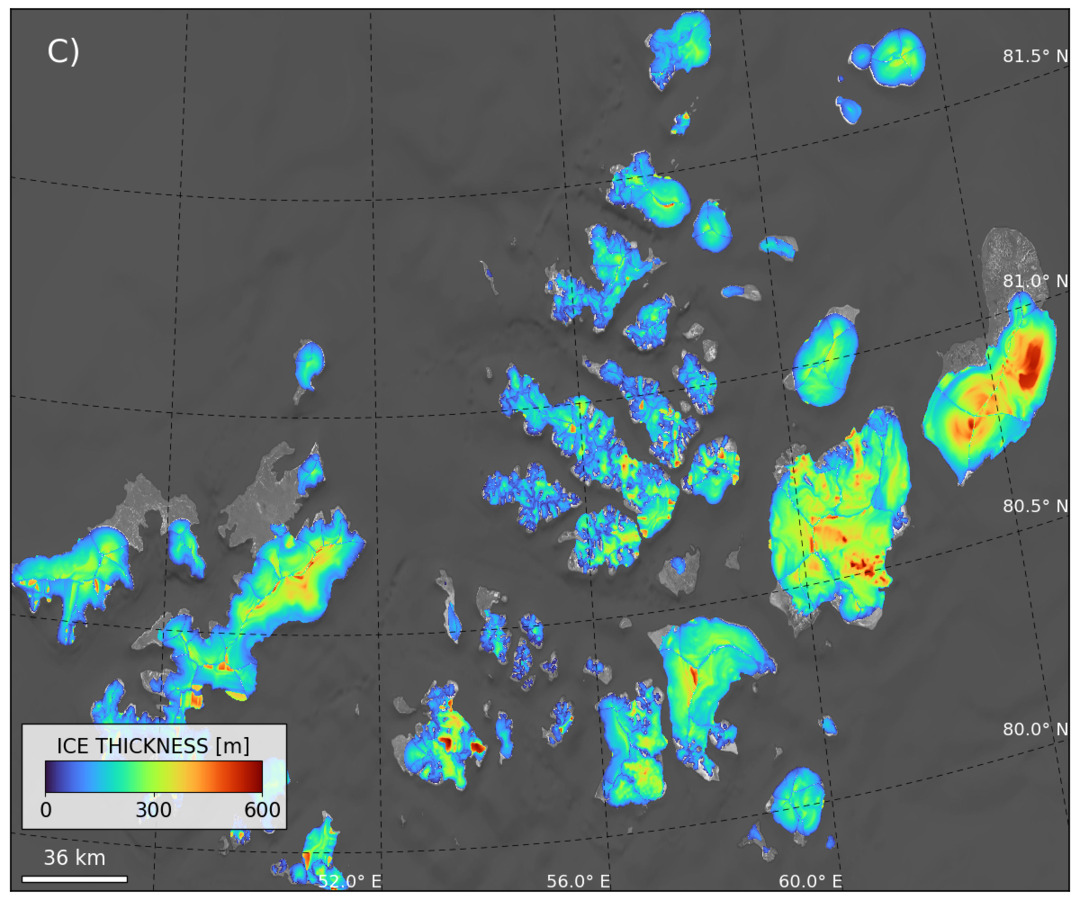}
    \label{fig:franz_joseph}
\end{figure}

\begin{landscape}
\begin{figure}[h!]
    \centering
    \caption{Russian Arctic. Top: Novaya Zemlya; Bottom: Severnaya Zemlya. A=IceBoost v2; B=Millan et al. \cite{millan2022}; C=Farinotti et al. \cite{farinotti2019}.}
    \includegraphics[width=.32\linewidth]{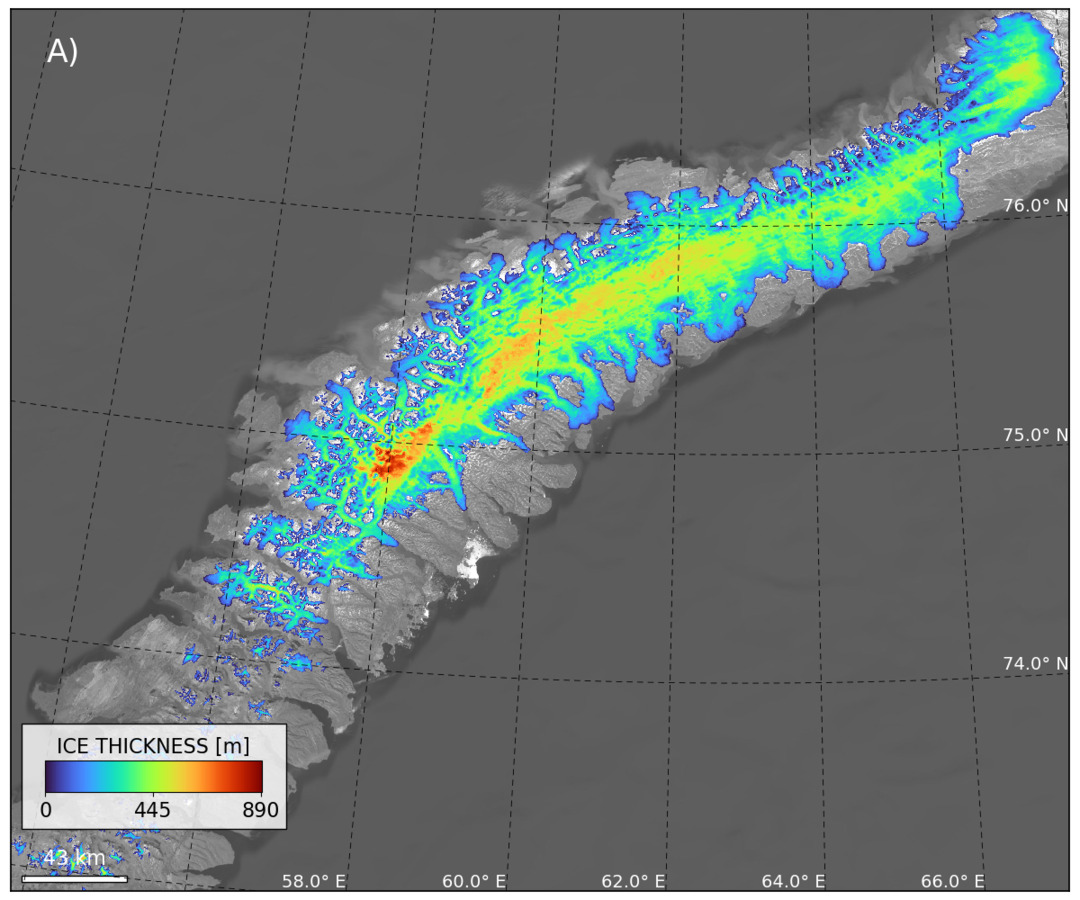}
    \includegraphics[width=.32\linewidth]{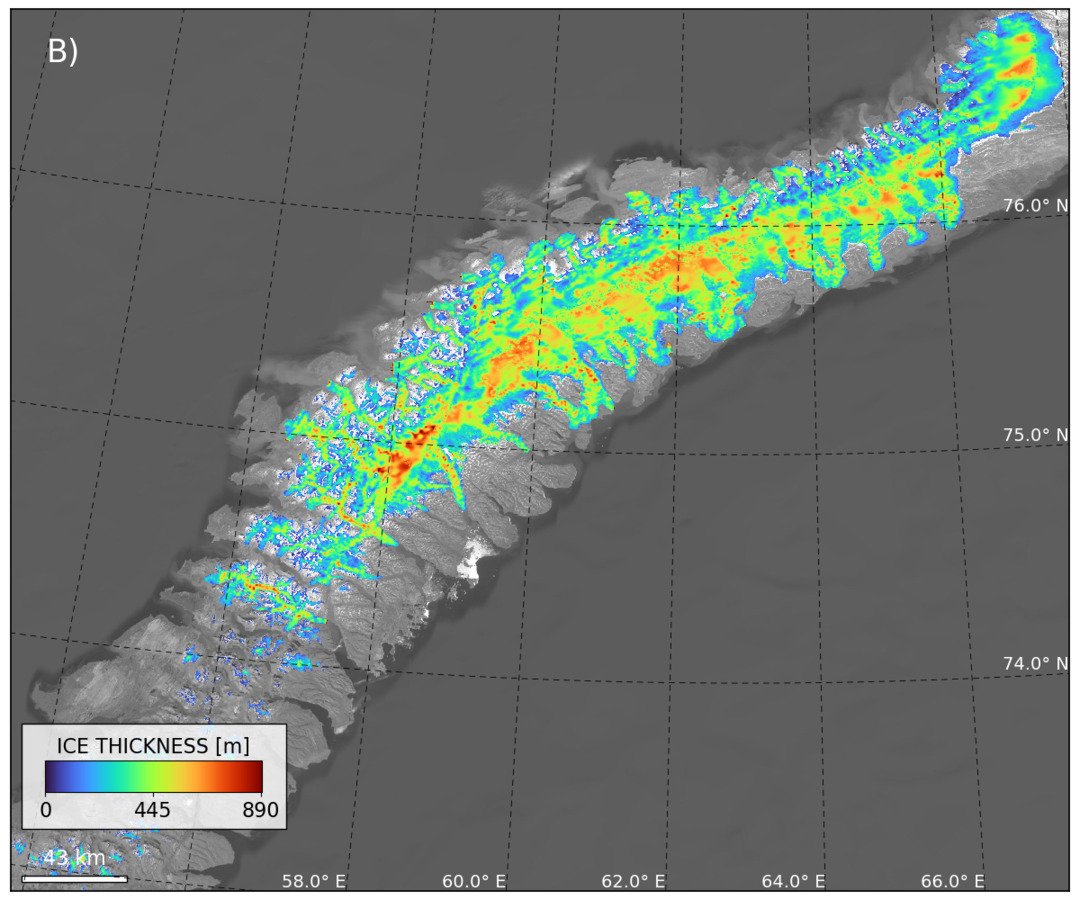}
    \includegraphics[width=.32\linewidth]{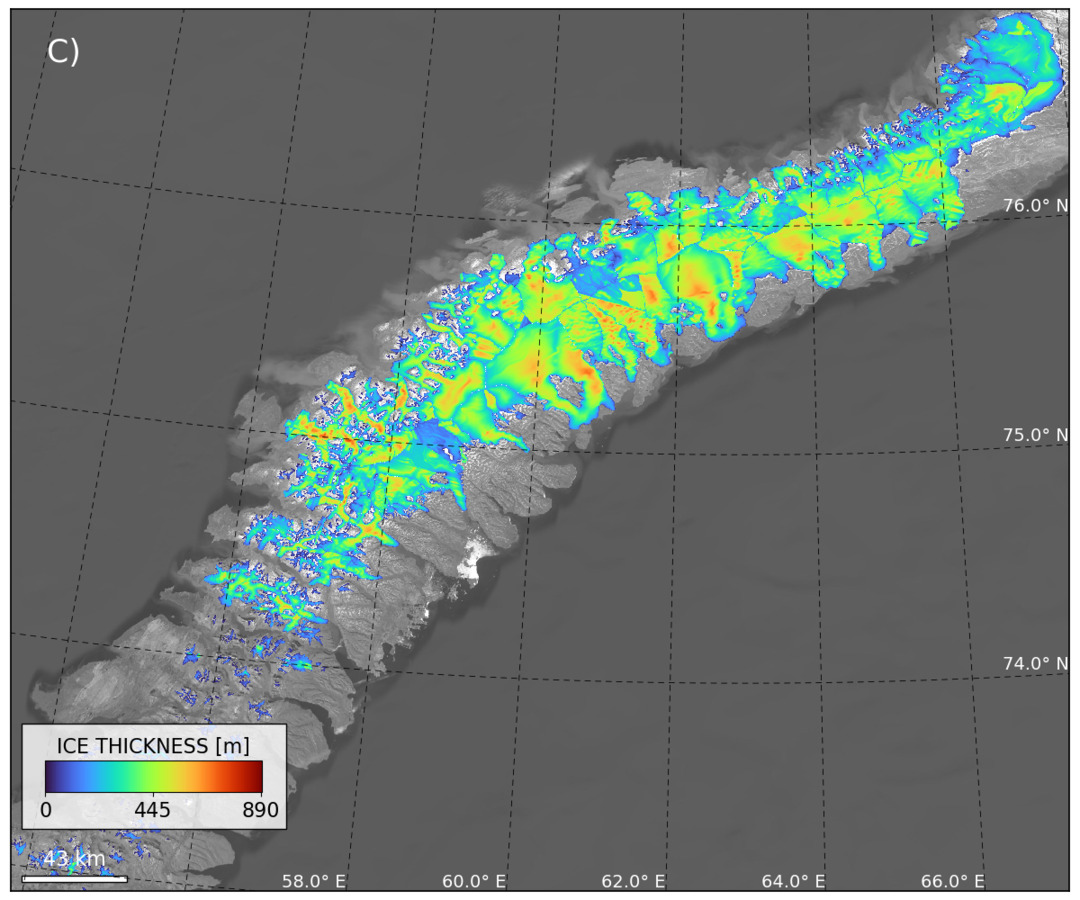}\\
    \includegraphics[width=.32\linewidth]{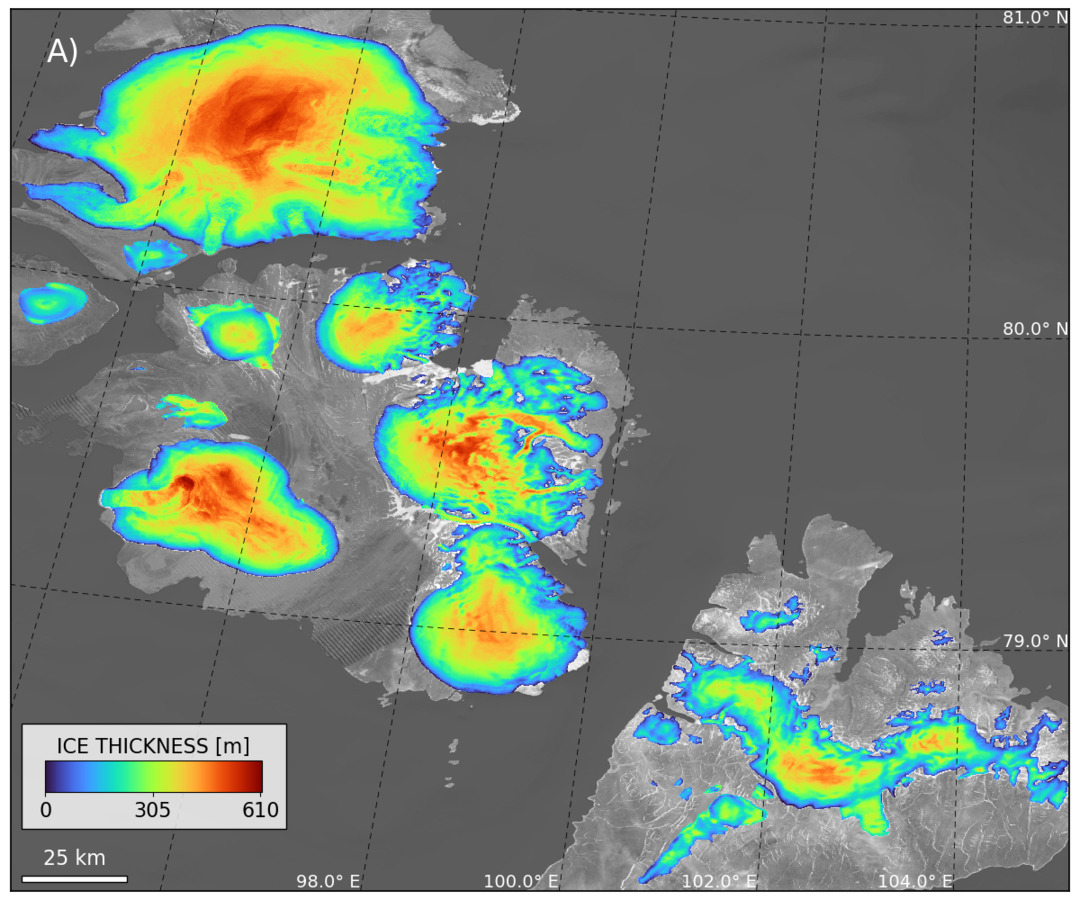}
    \includegraphics[width=.32\linewidth]{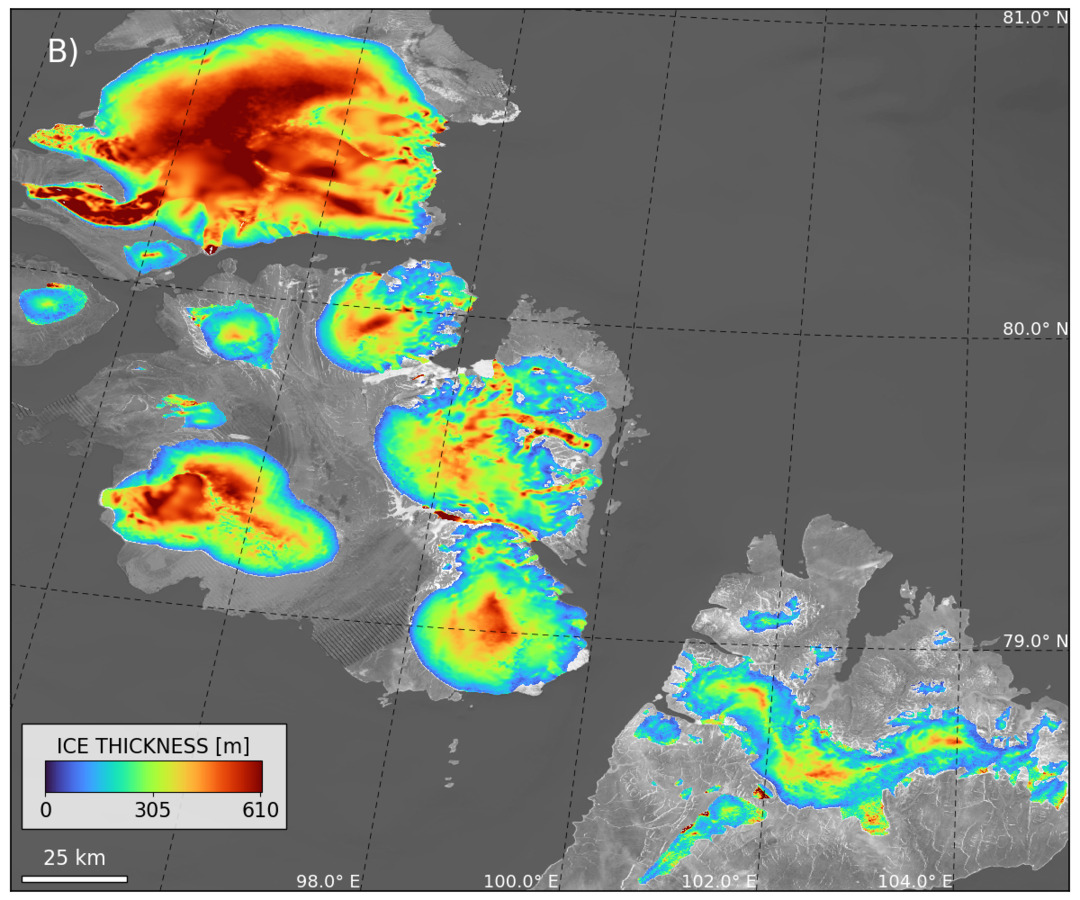}
    \includegraphics[width=.32\linewidth]{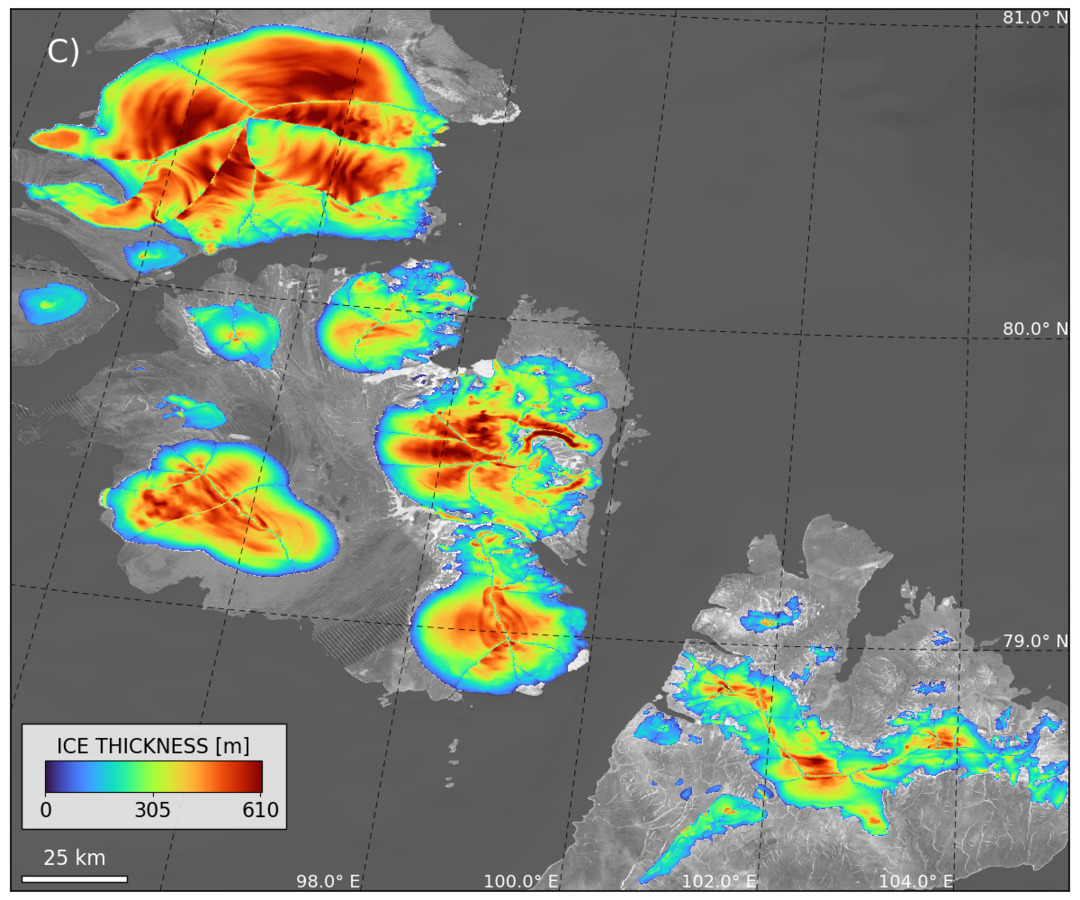}
    \label{fig:novaya_severnaya_zemlya}
\end{figure}
\end{landscape}

\subsubsection{Greenland periphery (RGI 05)}
\label{sect:greenland_periphery}

\begin{table}[h]
    \caption{Greenland periphery ice volumes estimated by different models. All units are \SI{e3}{\kilo\metre\tothe{3}}. \\ $^\star$ Central East Greenland includes 256 additional glaciers modeled with IceBoost, not included in \cite{millan2022, farinotti2019}.}\label{tab1}%
    \begin{tabular}{@{}lccc@{}}
    \toprule
    Greenland periphery (RGI 05) & IceBoost v2 & Millan et al. \cite{millan2022} & Farinotti et al. \cite{farinotti2019}\\
    \midrule
    Total & 13.2 ± 4.6 & 11.8 ± 3.7  & 15.7 ± 4.1 \\
    - North Greenland and Hans Tausen Ice Cap & 2.461 ± 1.009 & 1.734 & 2.556 \\
    - Flade Isblink & 3.450 ± 1.014 & 3.689 & 4.844 \\
    - Sukkertoppen Ice Cap & 1.17920 ± 0.3496& 0.852 & 1.33788 \\
    - Central East Greenland$^\star$ & 2.611 ± 0.777$^\star$ & 1.882 & 2.182 \\
    - Others & 3.4988 ± 1.45 & 3.643 & 4.78012 \\
    \midrule
     Glaciers with connection to the ice sheet & IceBoost v2 & BedMachine v5 \cite{morlighem2022_bedmac_Gr} & \\
    - Geikie plateau & 9.57061 ± 1. 953& 5.100 ± 1.516 & \\
    \bottomrule
    \end{tabular}
\end{table}

% north Greenland
Ice thickness estimates differ depending on the modeling approach in the Northern Greenland periphery (Fig. \ref{fig:hans_tausen_central_east_greenland}). IceBoost and Millan’s solutions diverge markedly over the Hans Tausen Ice Cap and the Freuchen Land peninsula: IceBoost predicts thicker ice and $\sim$40\% more volume. No measurements exist to validate any of the models in this region.\\

% Central East Greenland
\noindent In Central East Greenland (Scoresby Land), differences are also present (Fig. \ref{fig:hans_tausen_central_east_greenland}), though they are largely confined to the deepest portions of fjord glaciers and are on the order of 100–300 m. IceBoost shows the best agreement with radar profiles over the Renland Ice Cap.\\

% Flade Isblink
\noindent Significant discrepancies also appear across Flade Isblink, Greenland’s largest ice cap (Fig. \ref{fig:flade_isblink_sukkertoppen}). IceBoost agrees most closely with existing data, indicating that the ice cap is everywhere no more than $\sim$150 m above sea level. Millan and Farinotti both predict ice that is too thick. Smaller glaciers on the Kronprins Christian Land peninsula are consistent across models.\\

% Sukkertoppen
\noindent In western Greenland, over the Sukkertoppen Ice Cap, IceBoost estimates ice thicknesses of up to $\sim$500 meters in the northern sector and up to $\sim$1000 meters in the southern basin.\\

\begin{figure}[h!]
    \centering
    \includegraphics[width=.49\linewidth]{figs/fig2/fig_geikie_ice.jpg}
    \includegraphics[width=.49\linewidth]{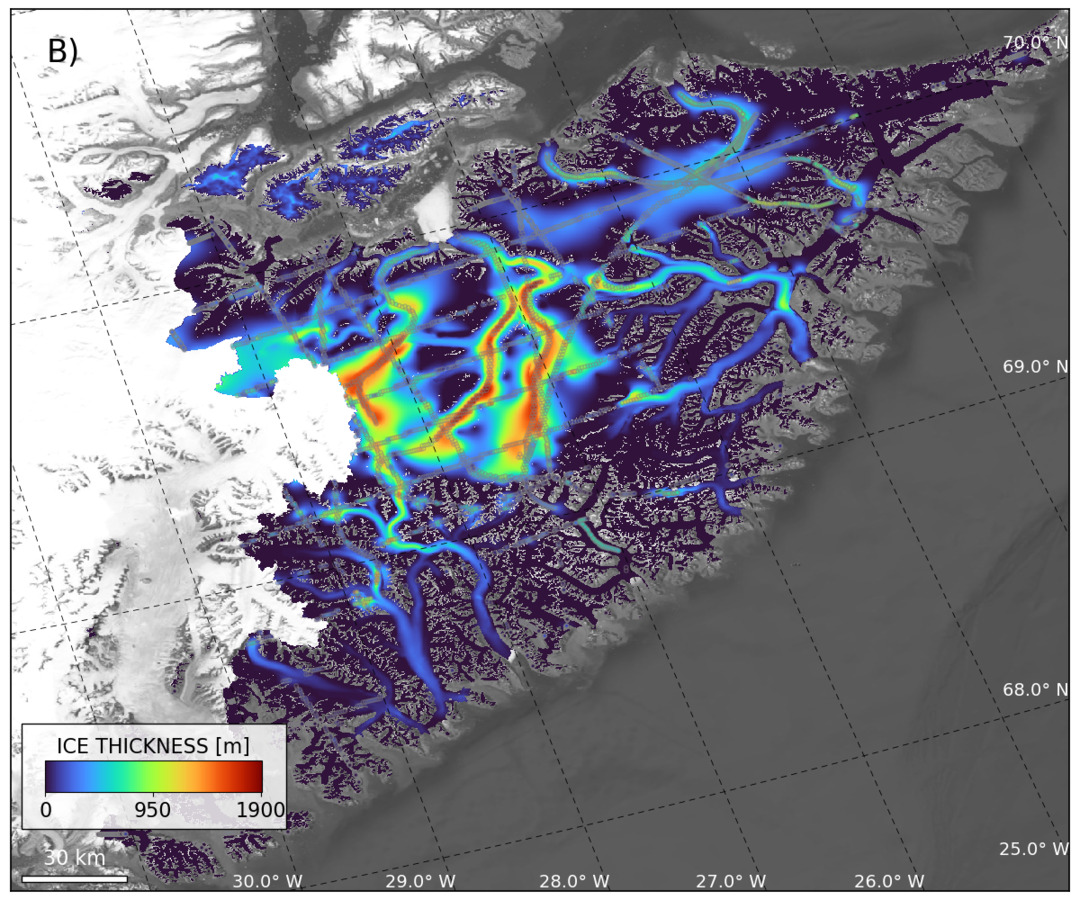}
    \caption{Geikie Plateau (East Greenland). Comparison between two models. A=IceBoost v2; B=BedMachine v5 \cite{morlighem2022_bedmac_Gr}.}
    \label{fig:geikie_plateau}
\end{figure}

\begin{landscape}
\begin{figure}[h!]
    \centering
    \caption{Greenland Periphery. Top: Hans Tausen ice cap; Bottom: Central East Greenland. A=IceBoost v2; B=Millan et al. \cite{millan2022}; C=Farinotti et al. \cite{farinotti2019}. Central East Greenland includes 256 additional glacier polygons modeled with IceBoost.}
    \includegraphics[width=.32\linewidth]{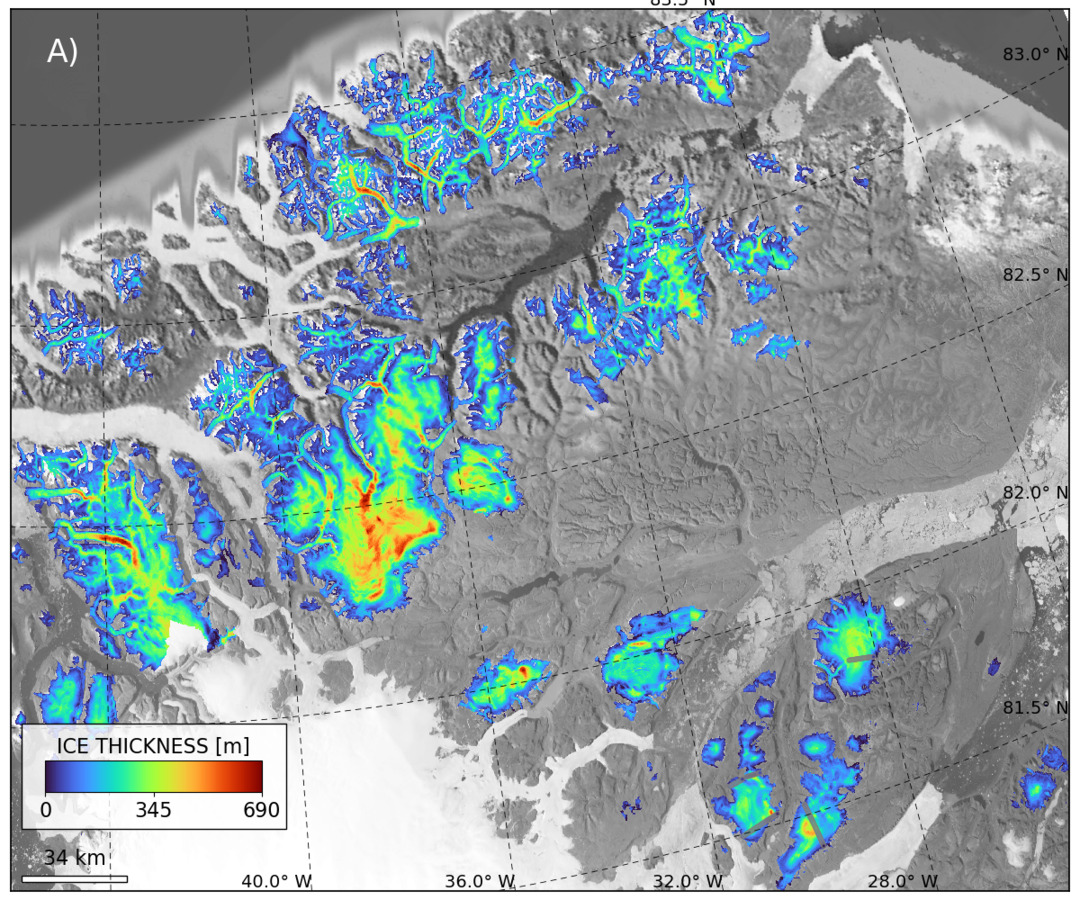}
    \includegraphics[width=.32\linewidth]{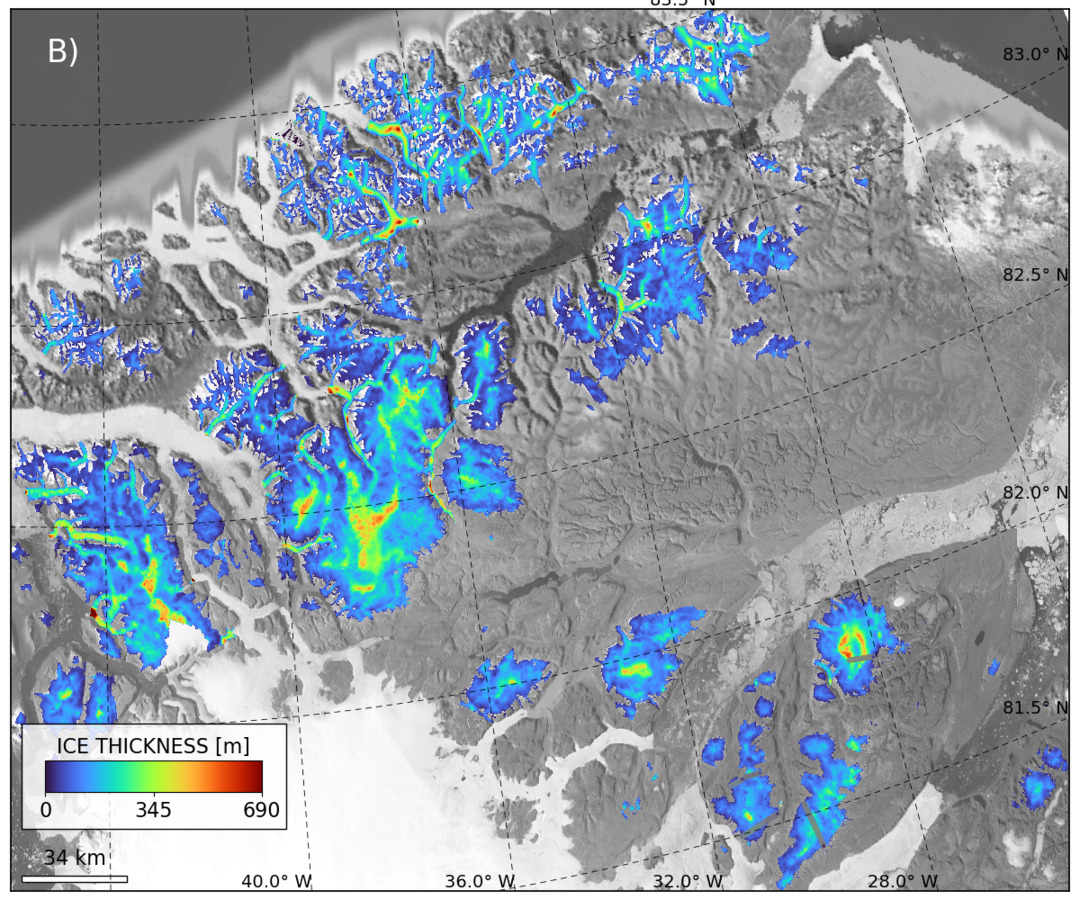}
    \includegraphics[width=.32\linewidth]{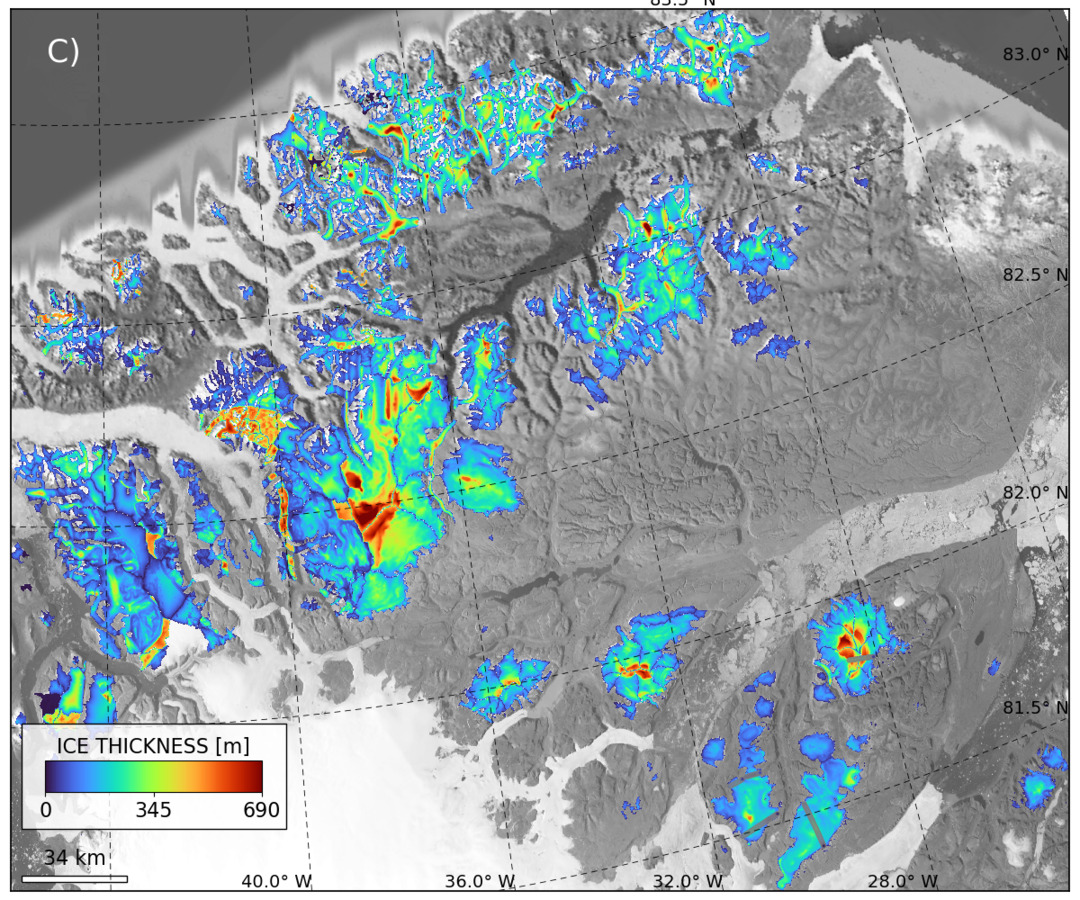}\\
    \includegraphics[width=.32\linewidth]{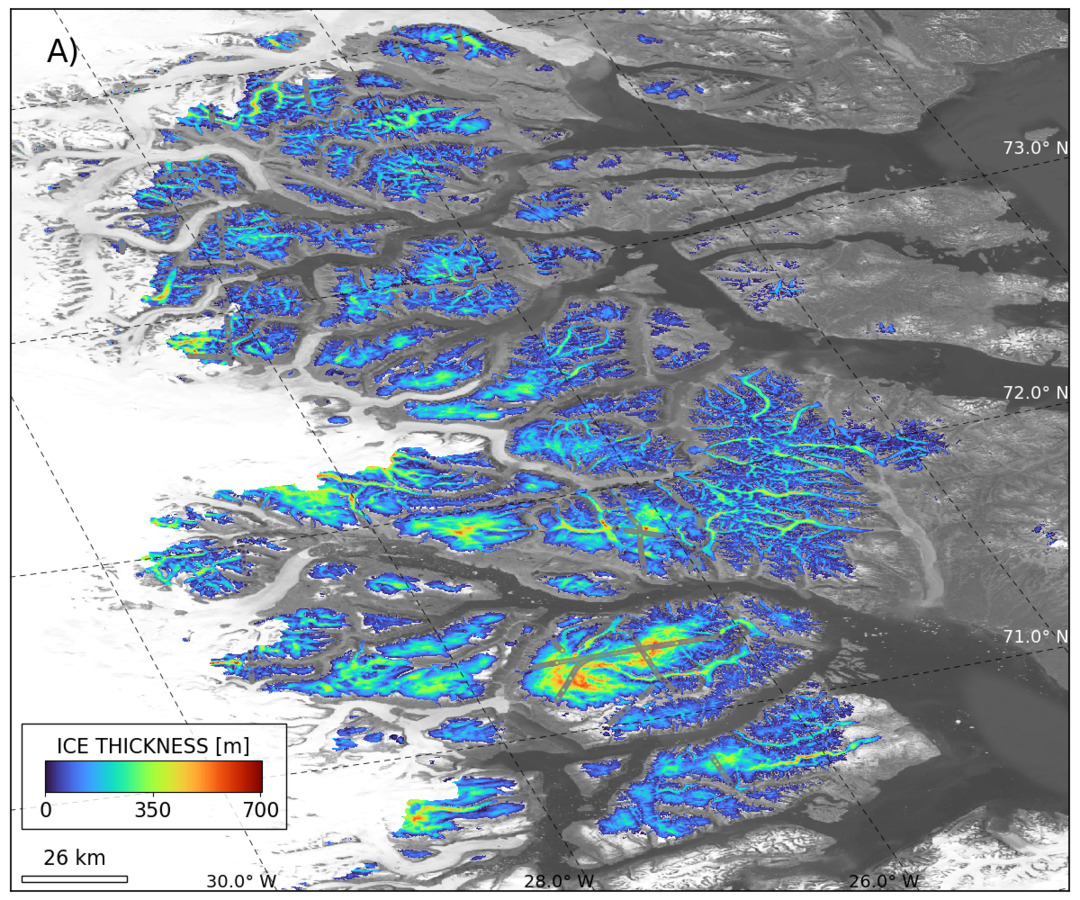}
    \includegraphics[width=.32\linewidth]{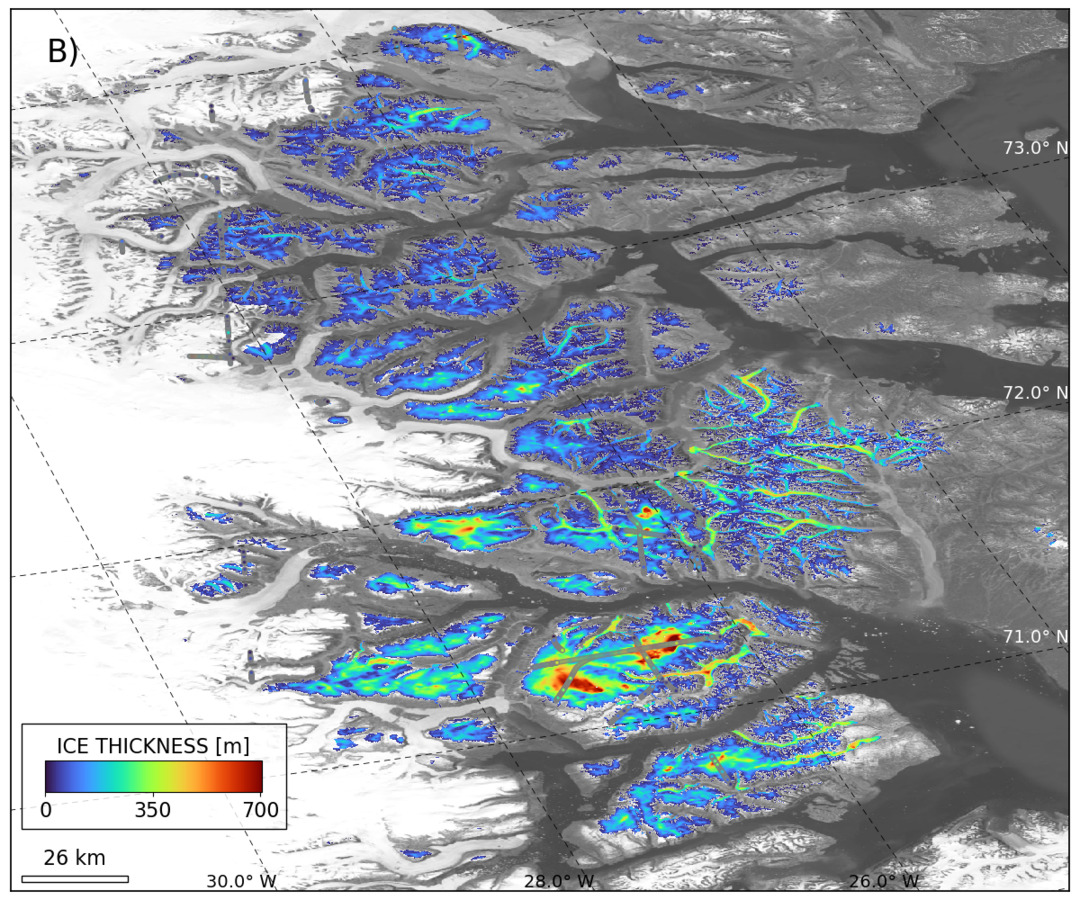}
    \includegraphics[width=.32\linewidth]{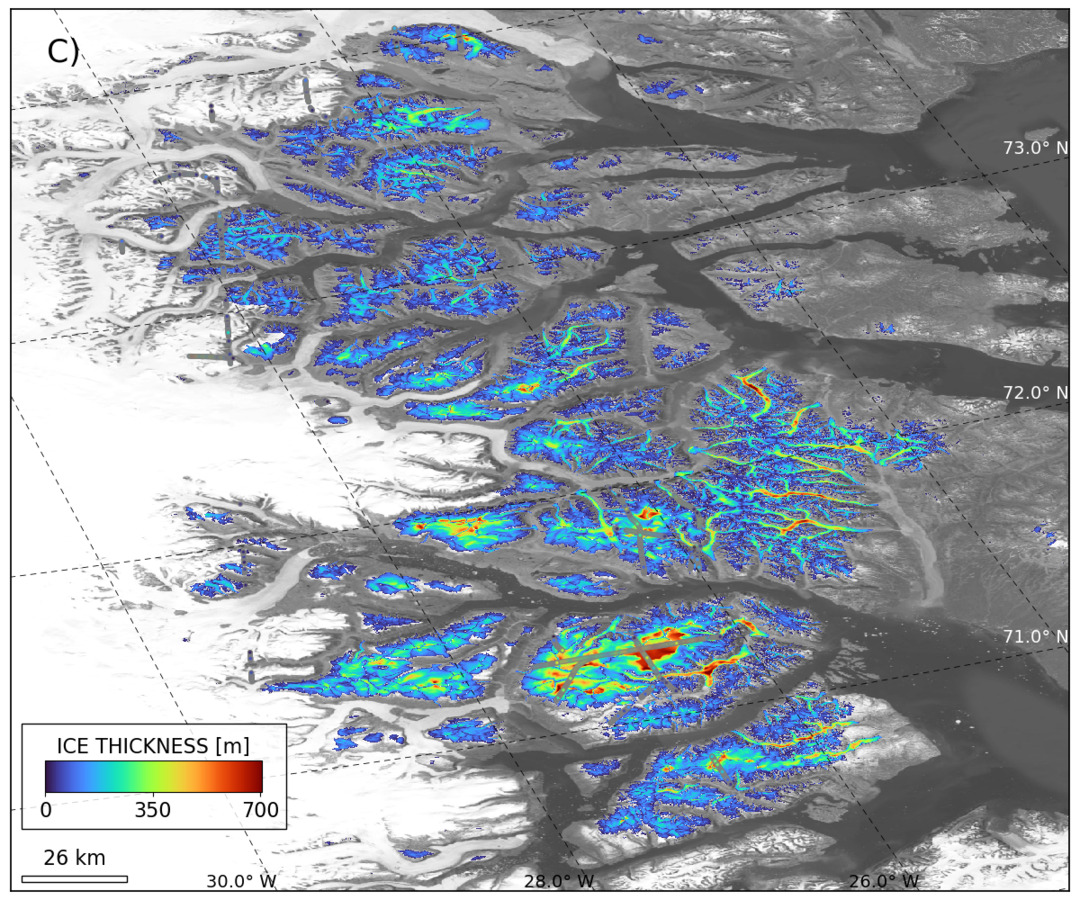}
    \label{fig:hans_tausen_central_east_greenland}
\end{figure}
\end{landscape}

\begin{landscape}
\begin{figure}[h!]
    \centering
    \caption{Greenland Periphery. Top: Flade Isblink; Bottom: Sukkertoppen ice cap. A=IceBoost v2; B=Millan et al. \cite{millan2022}; C=Farinotti et al. \cite{farinotti2019}.}
    \includegraphics[width=.32\linewidth]{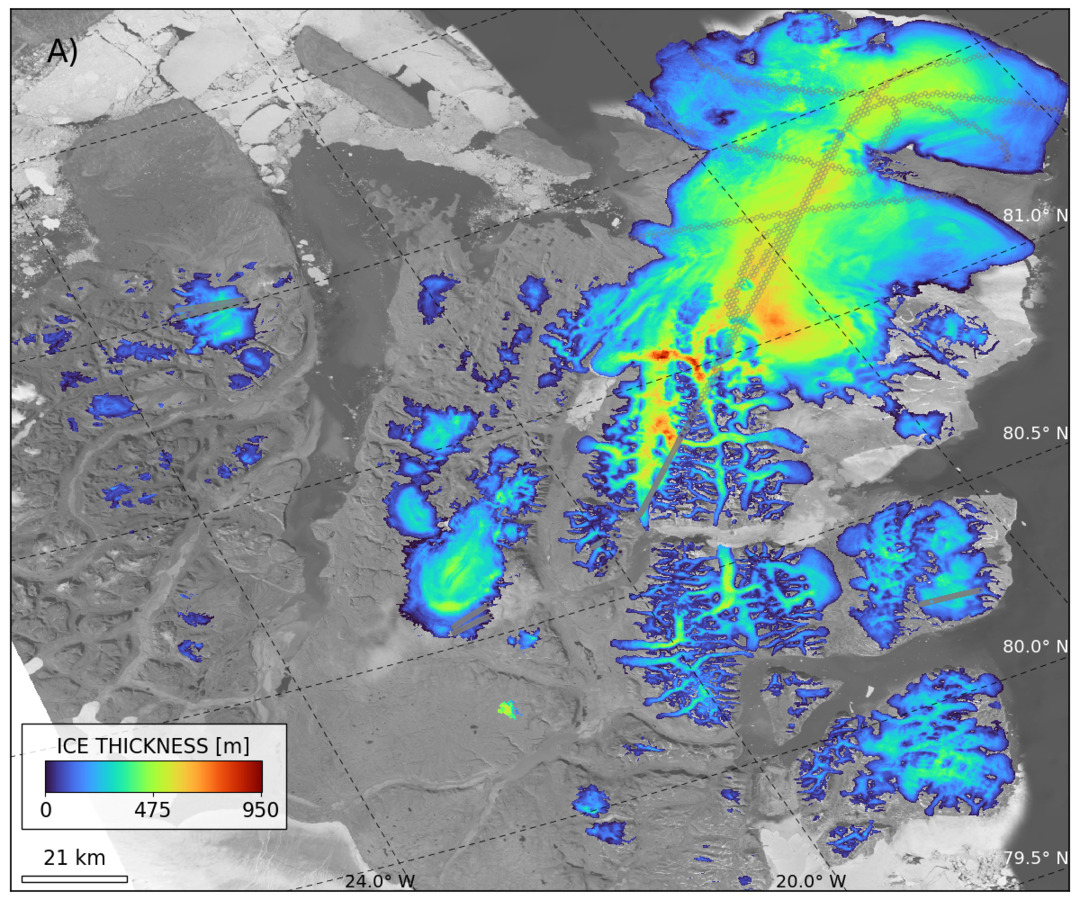}
    \includegraphics[width=.32\linewidth]{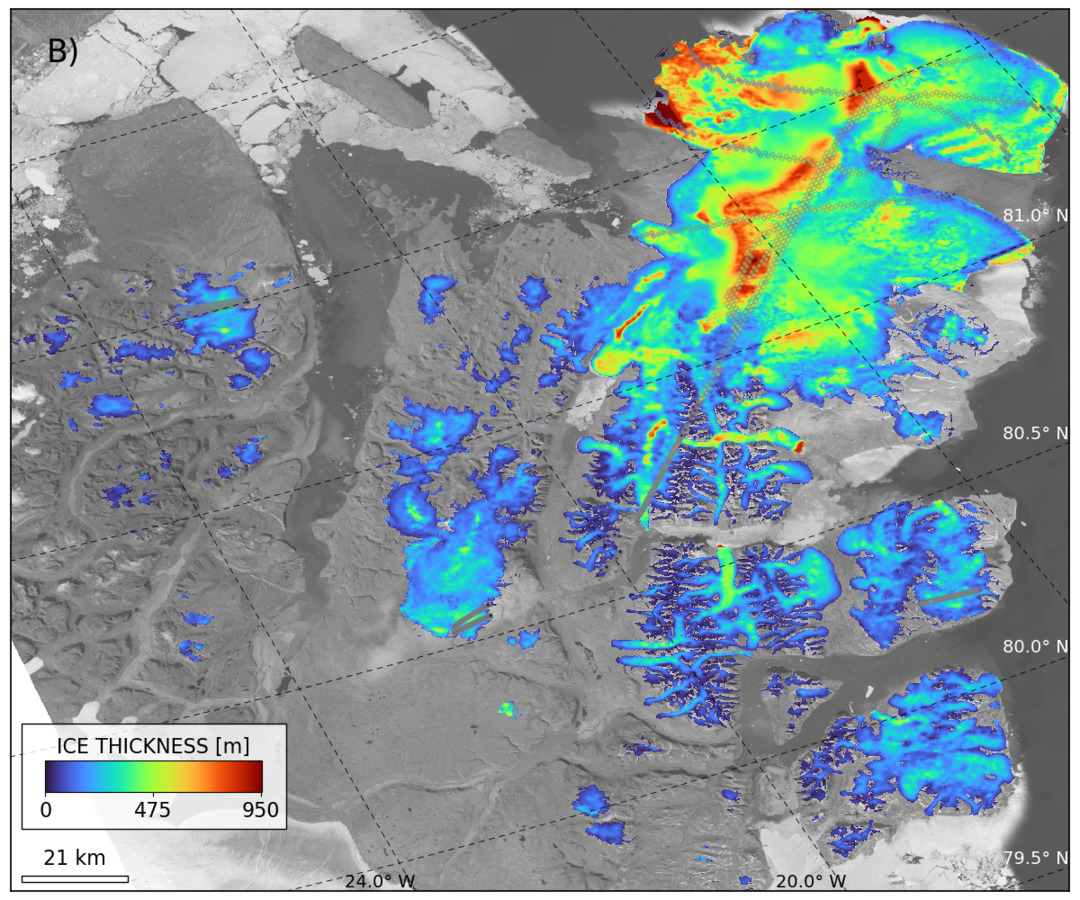}
    \includegraphics[width=.32\linewidth]{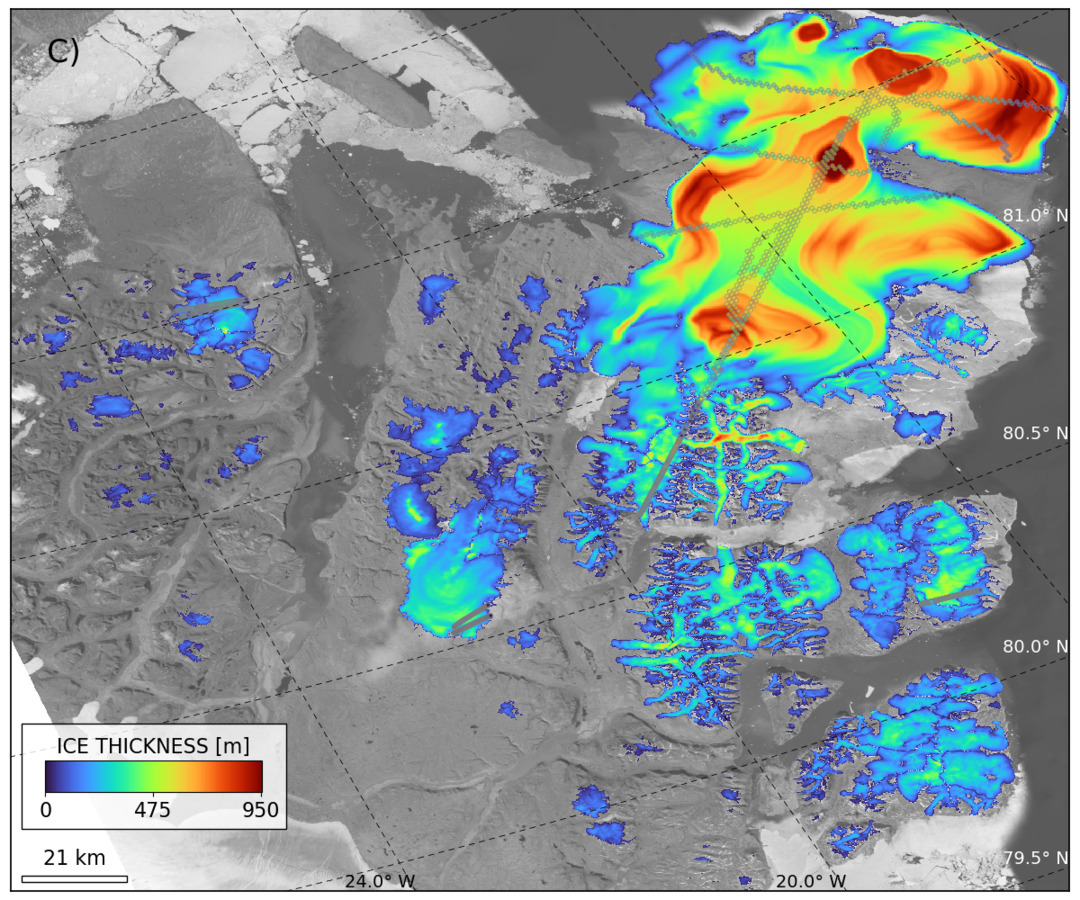}\\
    \includegraphics[width=.32\linewidth]{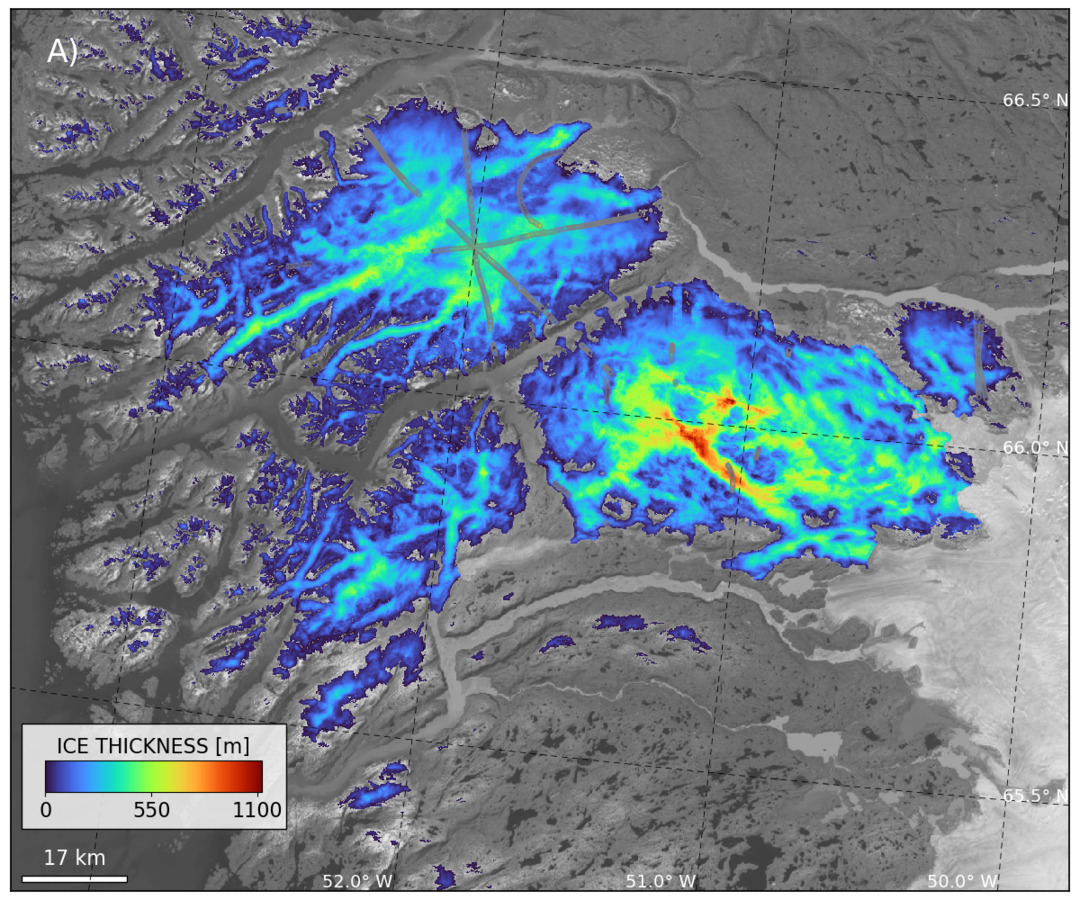}
    \includegraphics[width=.32\linewidth]{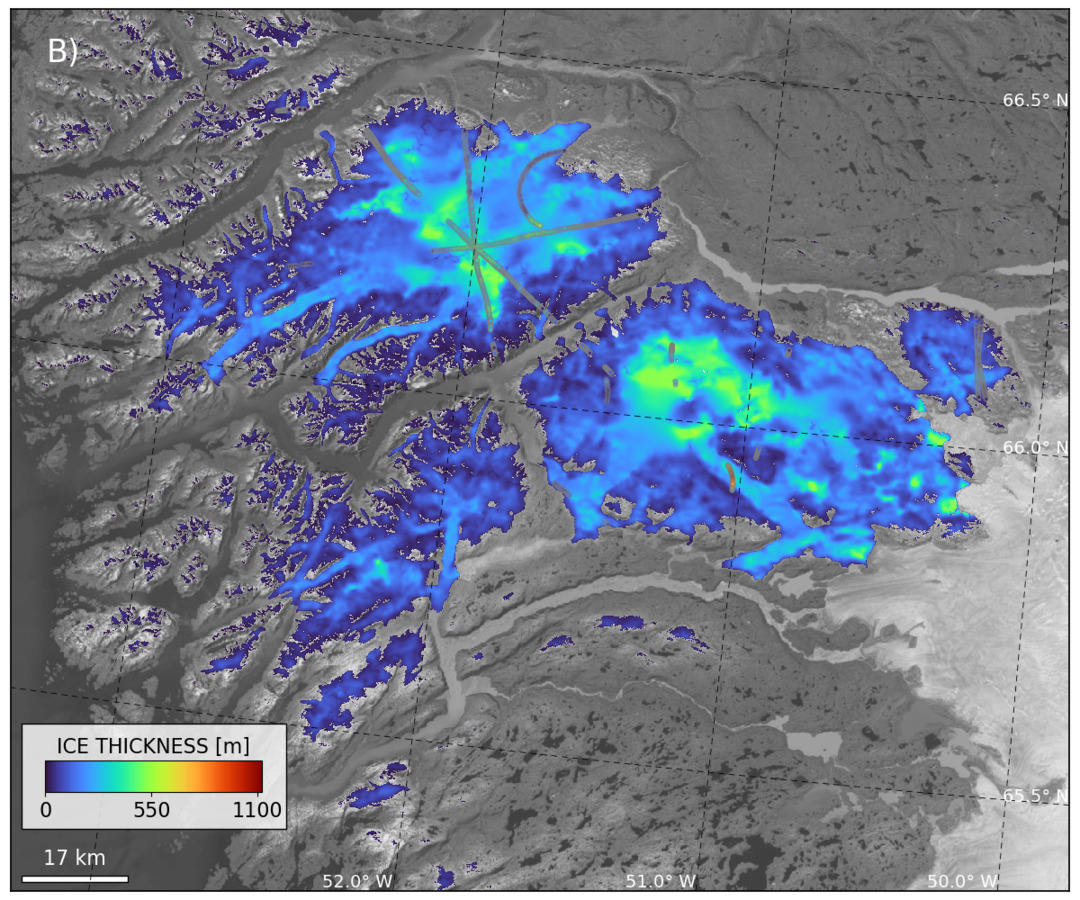}
    \includegraphics[width=.32\linewidth]{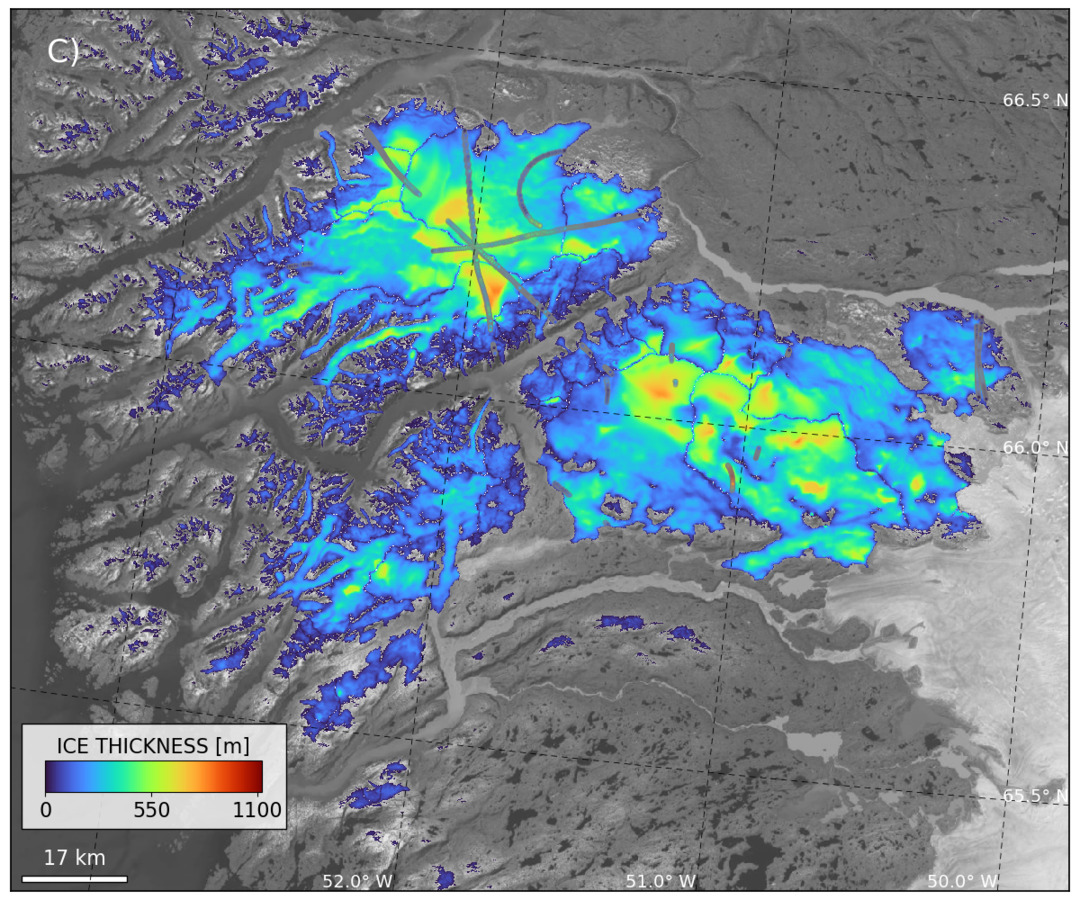}
    \label{fig:flade_isblink_sukkertoppen}
\end{figure}
\end{landscape}

% geikie plateau
\noindent The comparison between IceBoost and BedMachine v5 \cite{morlighem2022_bedmac_Gr} over the Geikie Plateau (Fig. \ref{fig:geikie_plateau}) suggests that the kriging and streamline-diffusion techniques used in BedMachine are unable to resolve many local features and ice streams in this region. IceBoost predicts nearly twice the total ice volume. We note that the forthcoming BedMachine v6 release will provide an updated thickness map for this area.

% conclusions on greenland ?

\subsubsection{Asia (RGI 13-14-15)}
\label{sect:asia}

\begin{table}[h]
    \caption{Asia ice volumes estimated by different models. All units are \SI{e3}{\kilo\metre\tothe{3}}.}\label{tab1}%
    \begin{tabular}{@{}lccc@{}}
    \toprule
    Region & IceBoost v2 & Millan et al. \cite{millan2022} & Farinotti et al. \cite{farinotti2019}\\
    \midrule
    Total Central Asia (RGI 13) & 3.8 ± 2.2 & 4.4 ± 2.7  & 3.27 ± 0.85\\
    - Pamir & 0.589 ± 0.275 & 0.671 & 0.497\\
    - Tian Shan & 0.463 ± 0.197 & 0.552 & 0.405 \\
    - Western Kulun Mountains & 0.587 ± 0.267 & 0.624 & 0.388\\
    - Others & 2.161 ± 1.461 & 2.553 & 1.98 \\
    \midrule
    Total South Asia West (RGI 14) & 3.8 ± 1.5 & 3.8 ± 2.4 & 2.87 ± 0.74 \\
    - Karakoram range & 2.991 ± 1.164 & 2.710 & 2.062 \\
    - Others & 0.809 ± 0.336 & 1.09 & 0.808  \\
    \midrule
    Total South Asia East (RGI 15) & 1.0 ± 0.5 & 1.2 ± 0.8  & 0.88 ± 0.23 \\
    - Eastern Himalayas & 0.233 ± 0.085 & 0.328 & 0.182 \\
    - Others & 0.767 ± 0.415 & 0.872 & 0.698 \\
    \bottomrule
    \end{tabular}
\end{table}

\begin{figure}[h!]
    \caption{Tian Shan (Central Asia, RGI 13). A=IceBoost v2; B=Millan et al. \cite{millan2022}; C=Farinotti et al. \cite{farinotti2019}.}
    \includegraphics[width=.49\linewidth]{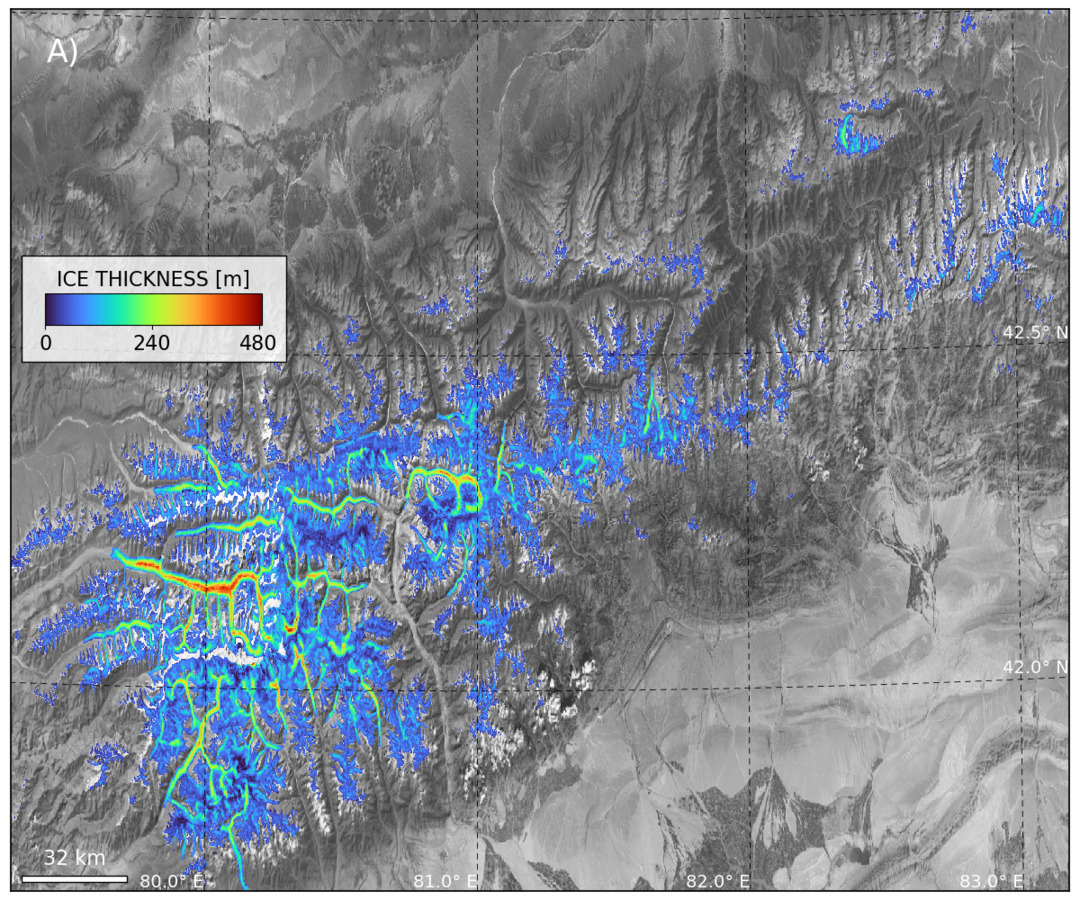}
    \includegraphics[width=.49\linewidth]{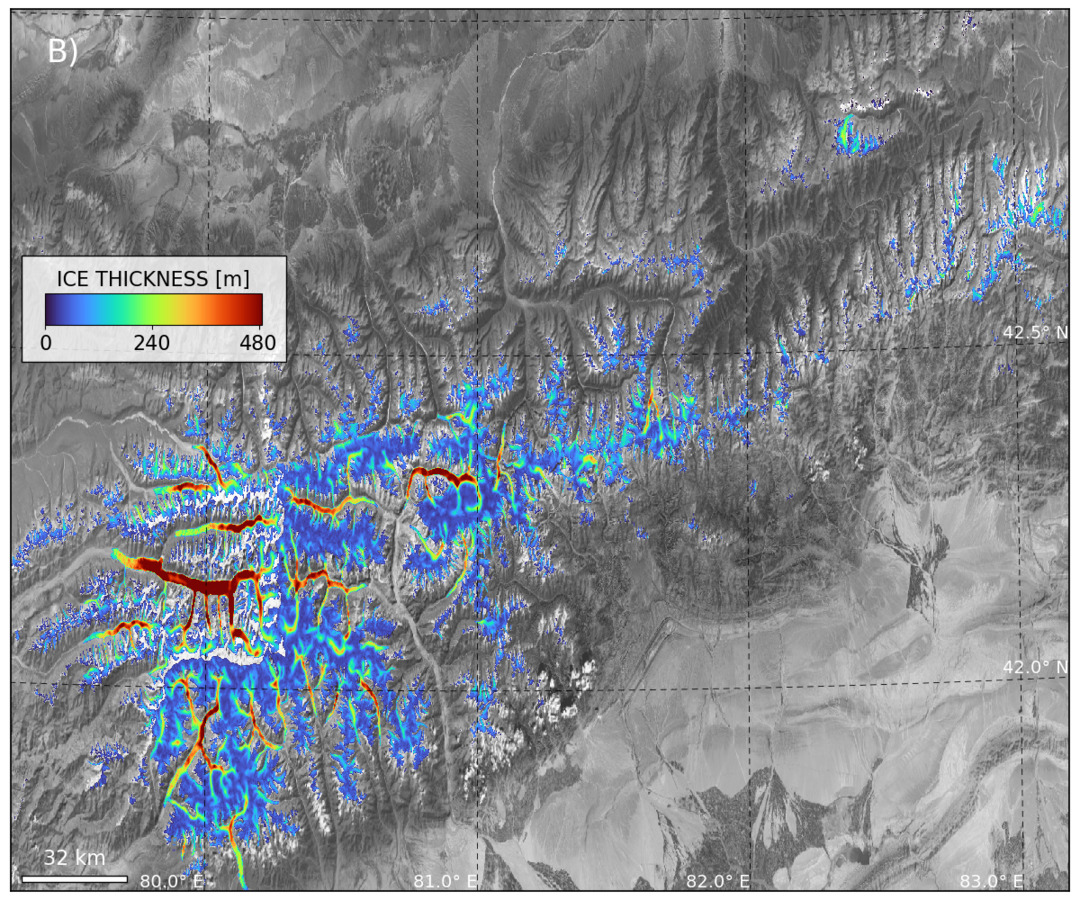}
    \includegraphics[width=.49\linewidth]{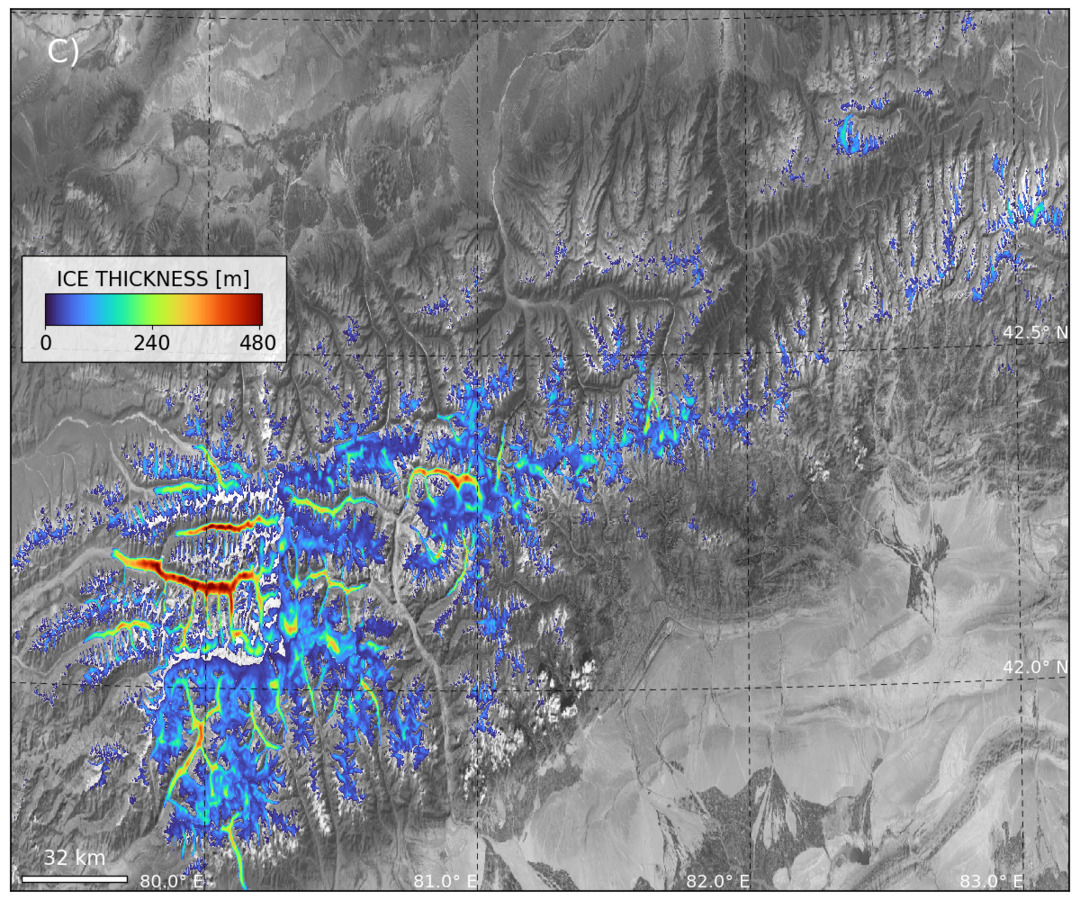}
    \label{fig:tian_shan}
\end{figure}

\begin{landscape}
\begin{figure}[h!]
    \centering
    \caption{Central Asia. Top: Pamir range (Fedchenko glacier in the center); Bottom: West Kulun mountains. A=IceBoost v2; B=Millan et al. \cite{millan2022}; C=Farinotti et al. \cite{farinotti2019}.}
    \includegraphics[width=.32\linewidth]{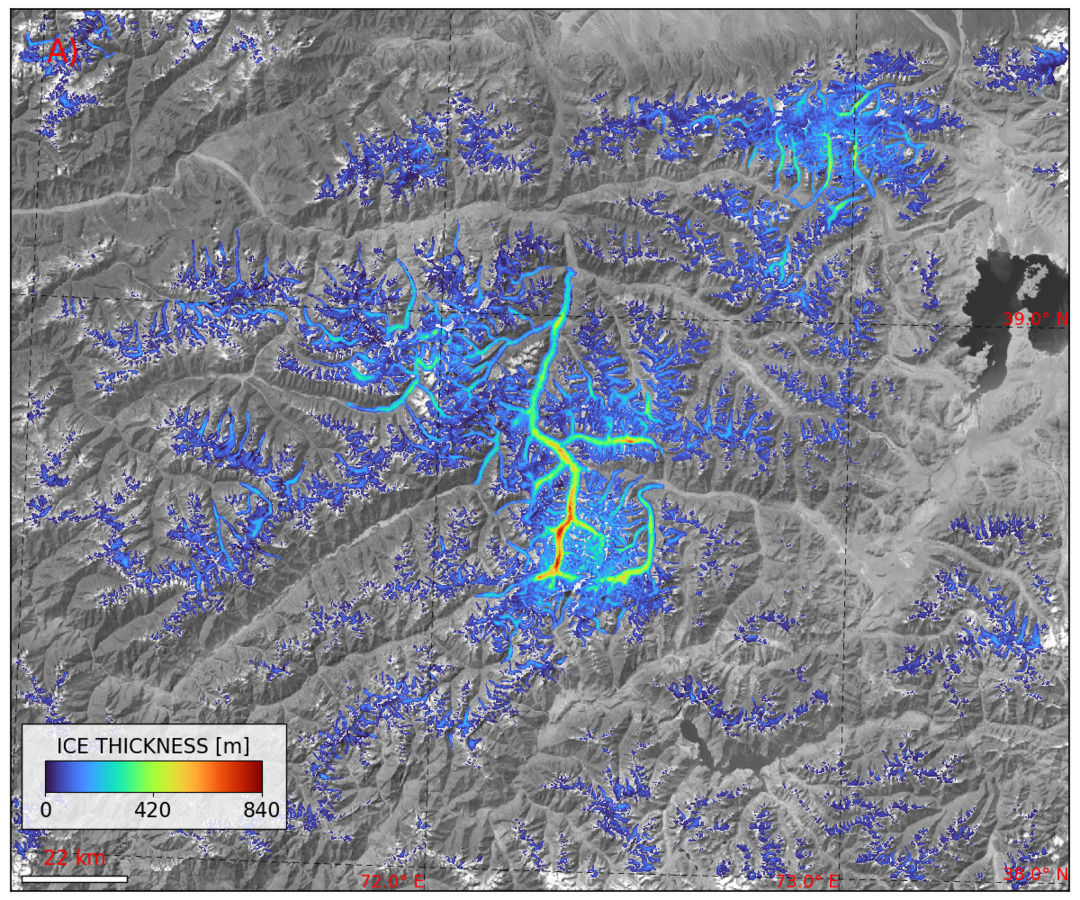}
    \includegraphics[width=.32\linewidth]{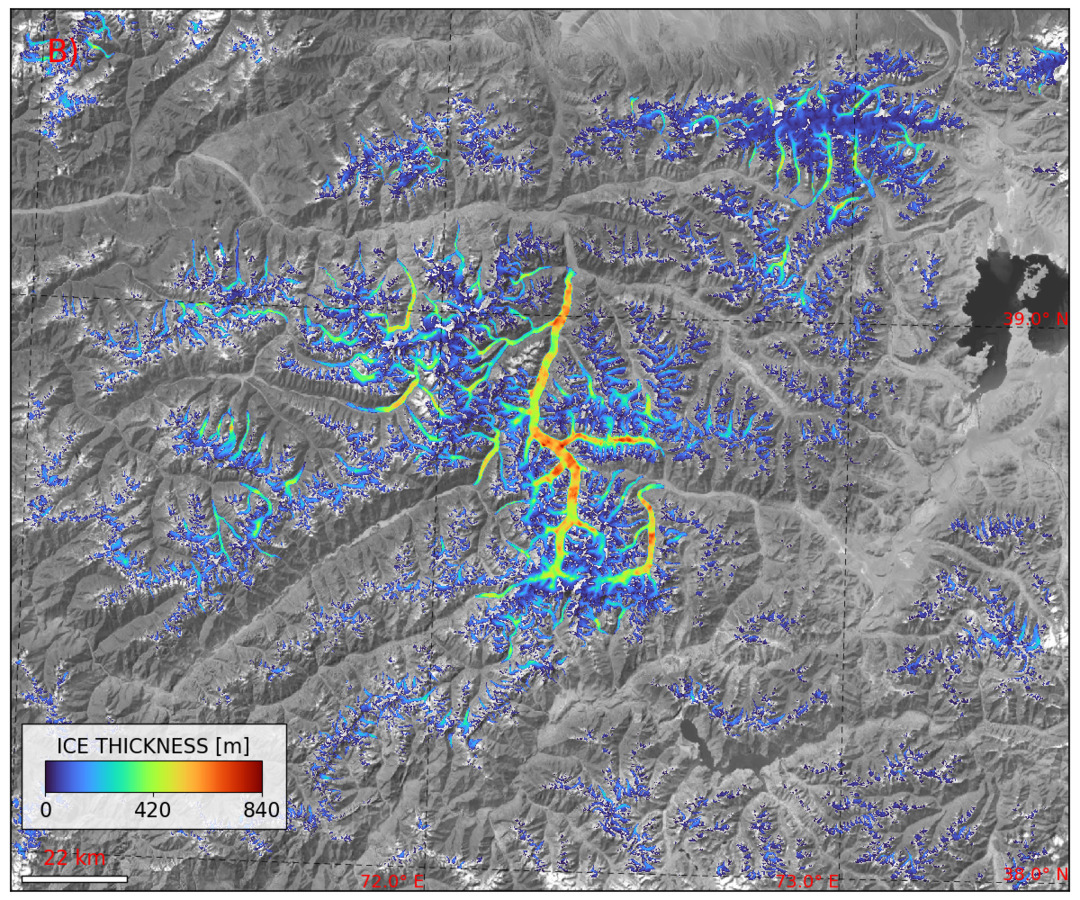}
    \includegraphics[width=.32\linewidth]{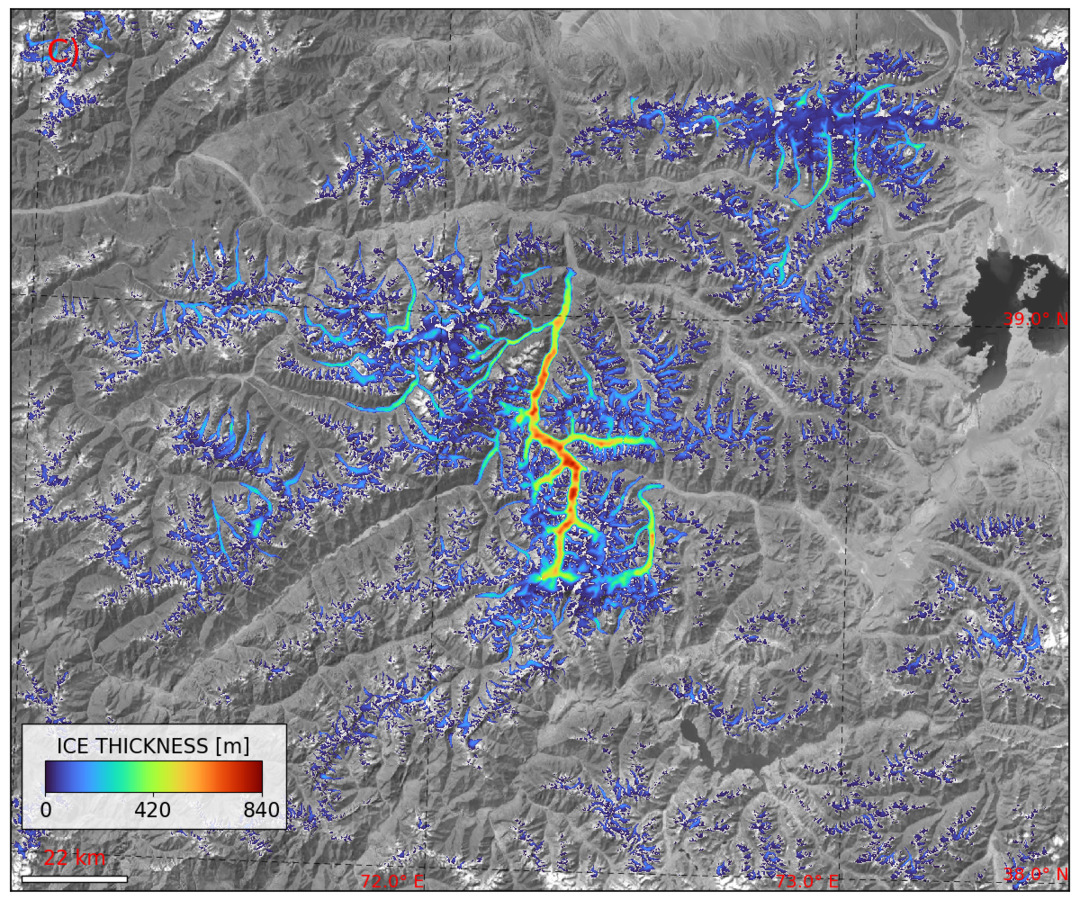}\\
    \includegraphics[width=.32\linewidth]{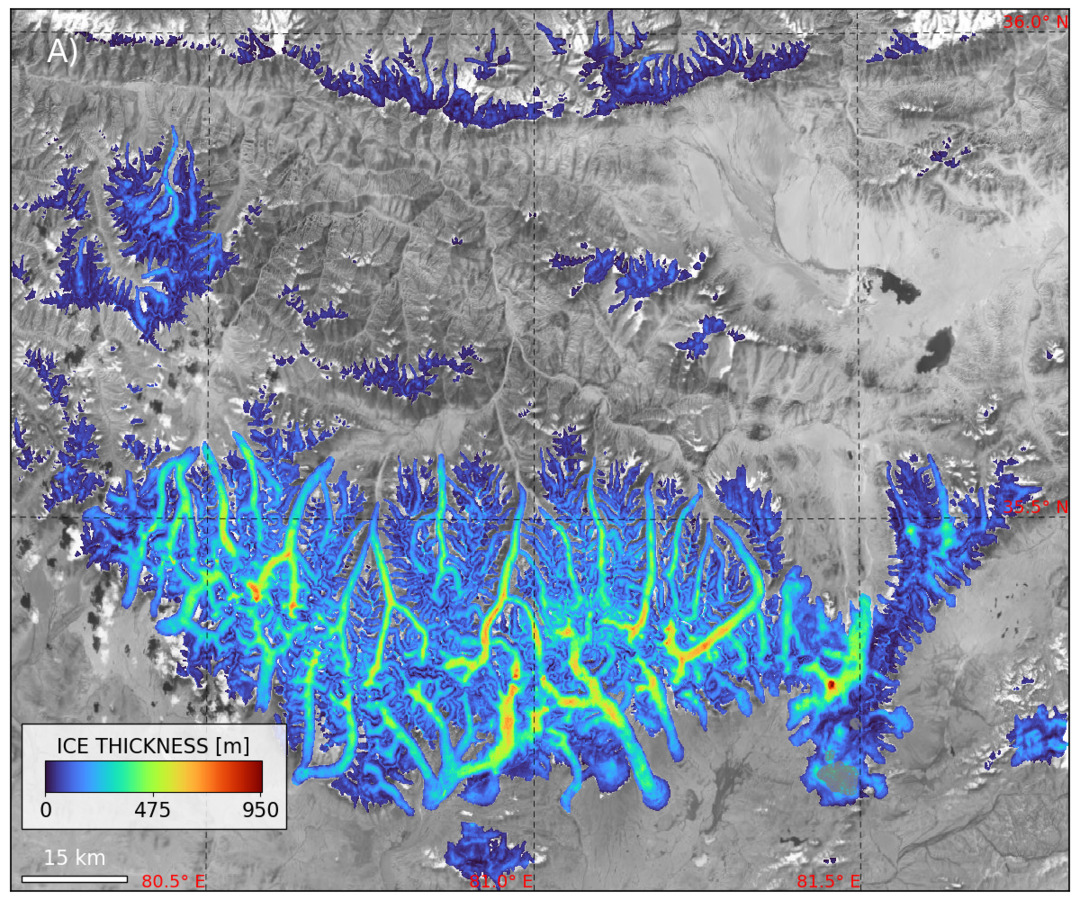}
    \includegraphics[width=.32\linewidth]{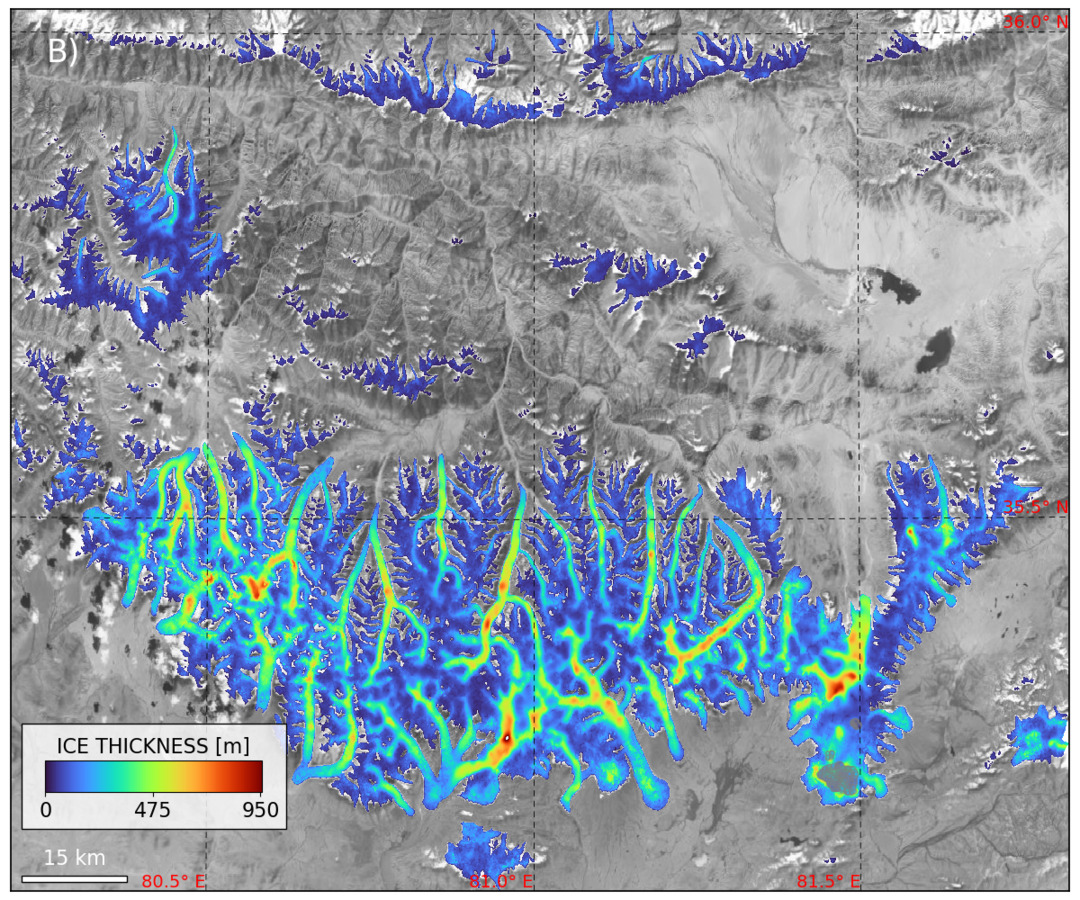}
    \includegraphics[width=.32\linewidth]{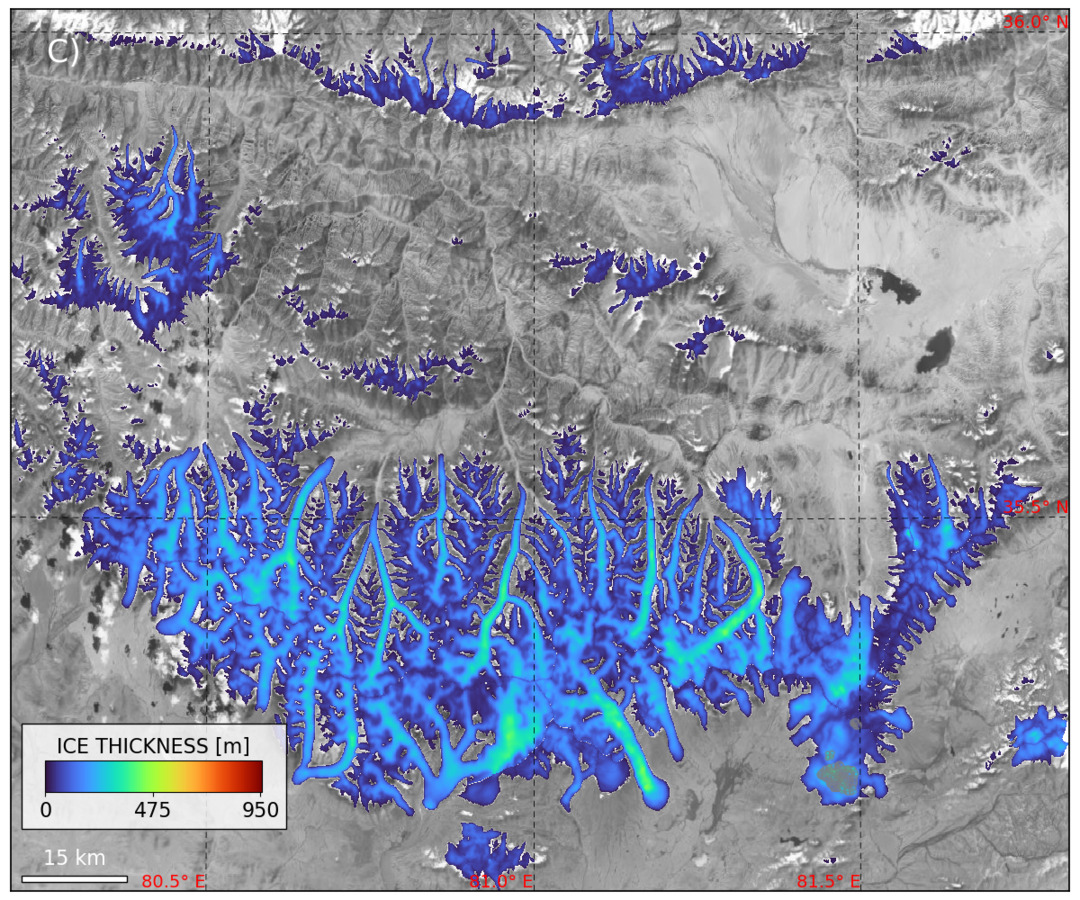}
    \label{fig:pamir_west_kulun}
\end{figure}
\end{landscape}

\begin{landscape}
\begin{figure}[h!]
    \centering
    \caption{Top: Karakoram range (RGI 14); Bottom: Eastern Himalayas (RGI 15). A=IceBoost v2; B=Millan et al. \cite{millan2022}; C=Farinotti et al. \cite{farinotti2019}.}
    \includegraphics[width=.32\linewidth]{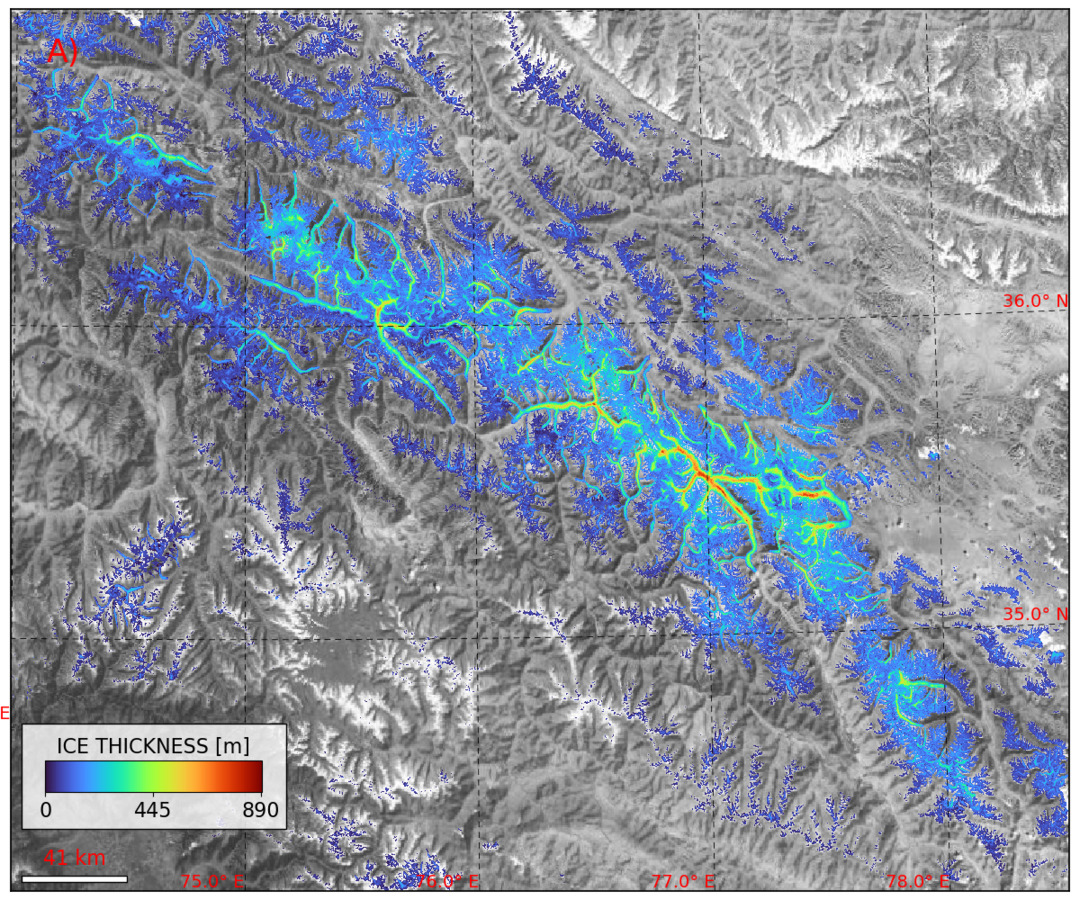}
    \includegraphics[width=.32\linewidth]{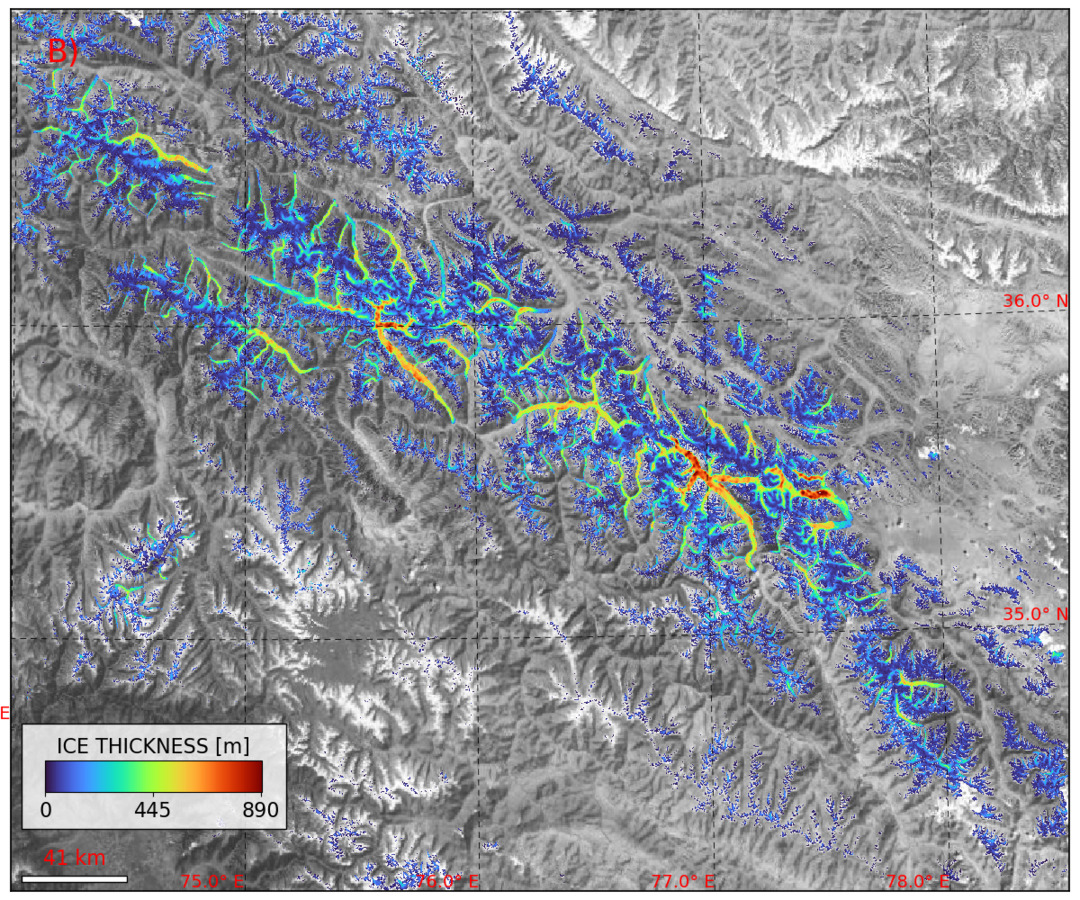}
    \includegraphics[width=.32\linewidth]{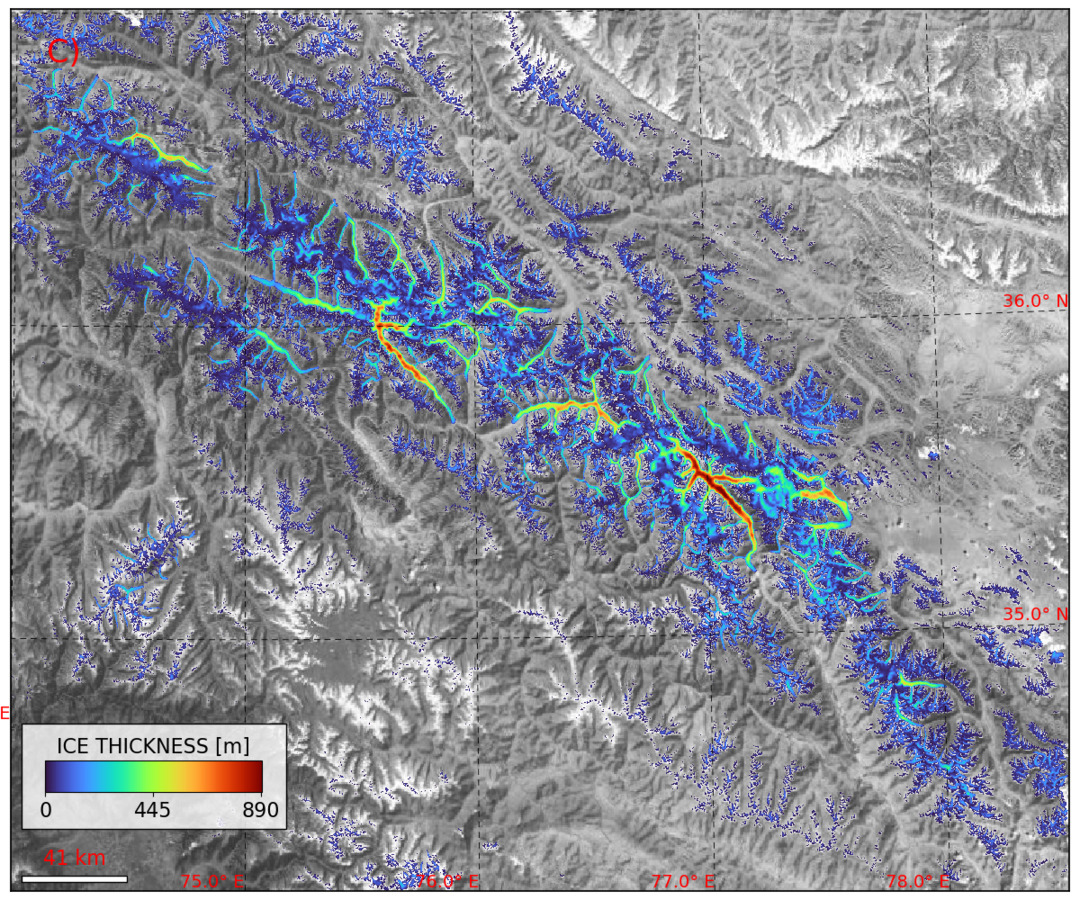}\\
    \includegraphics[width=.32\linewidth]{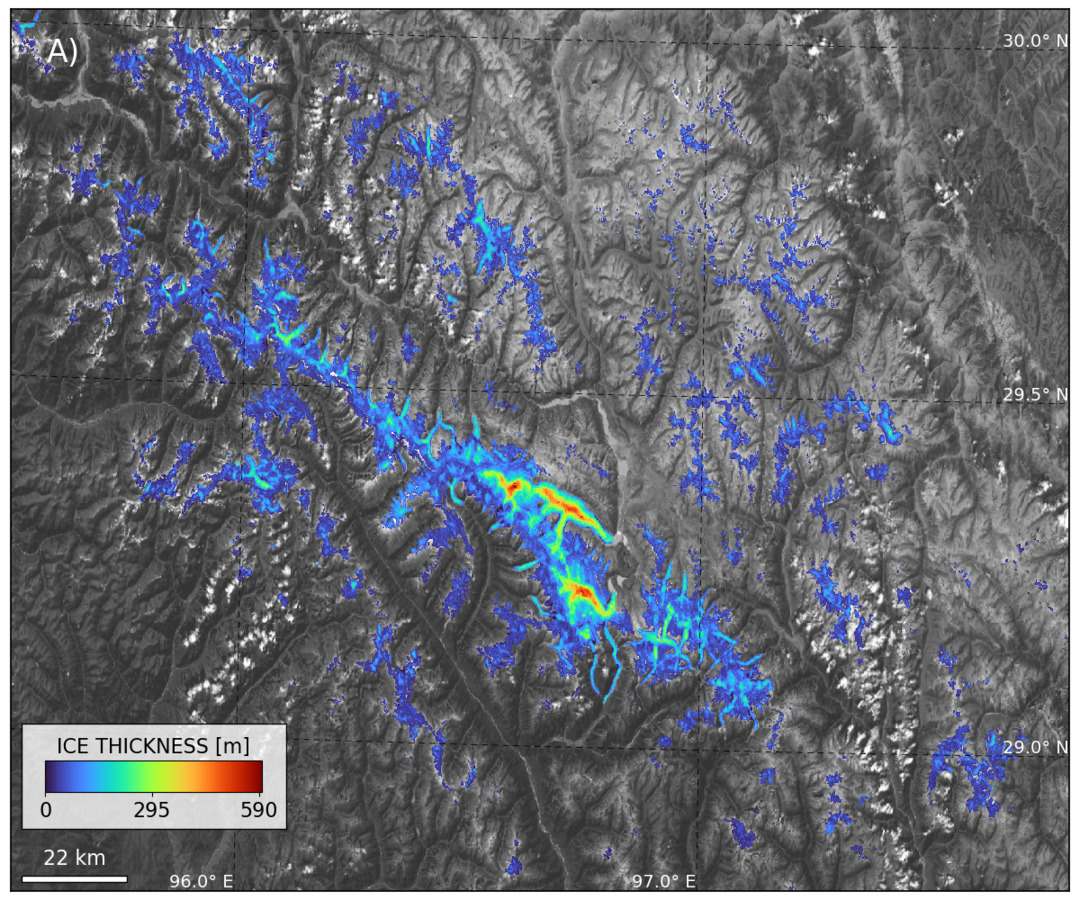}
    \includegraphics[width=.32\linewidth]{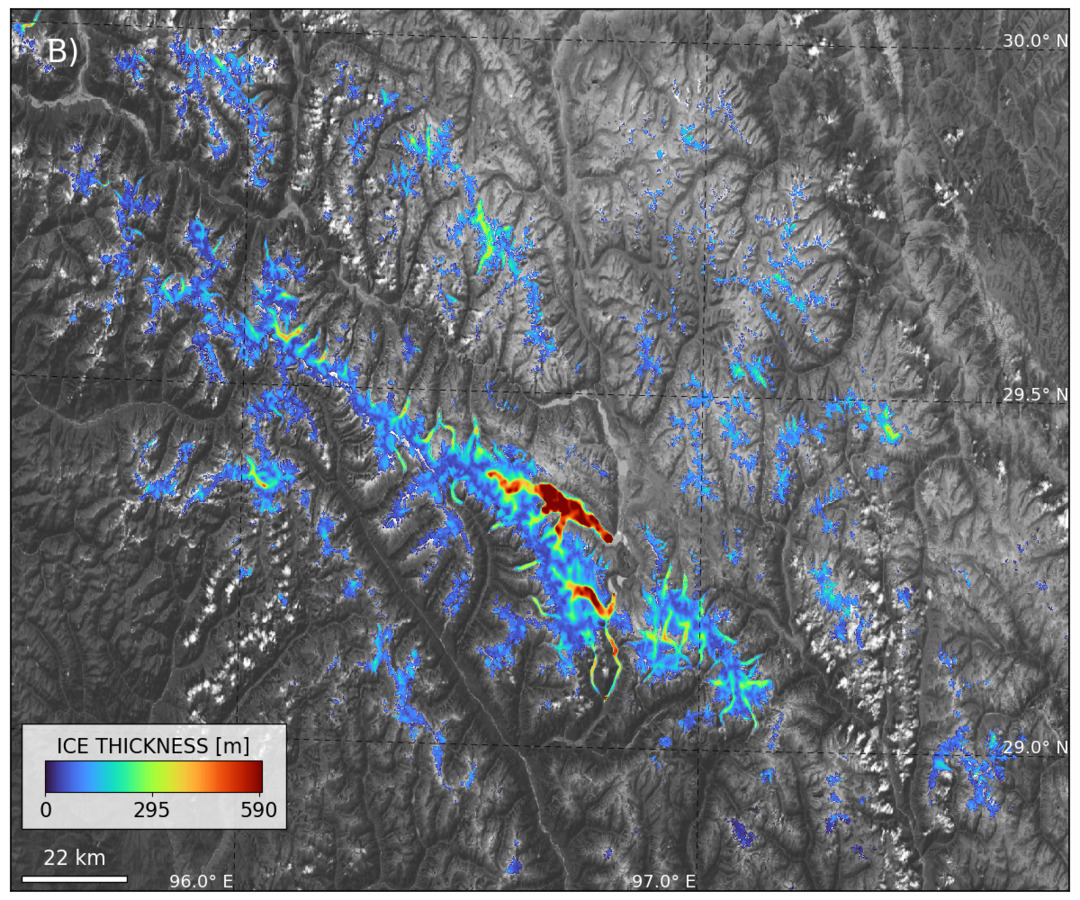}
    \includegraphics[width=.32\linewidth]{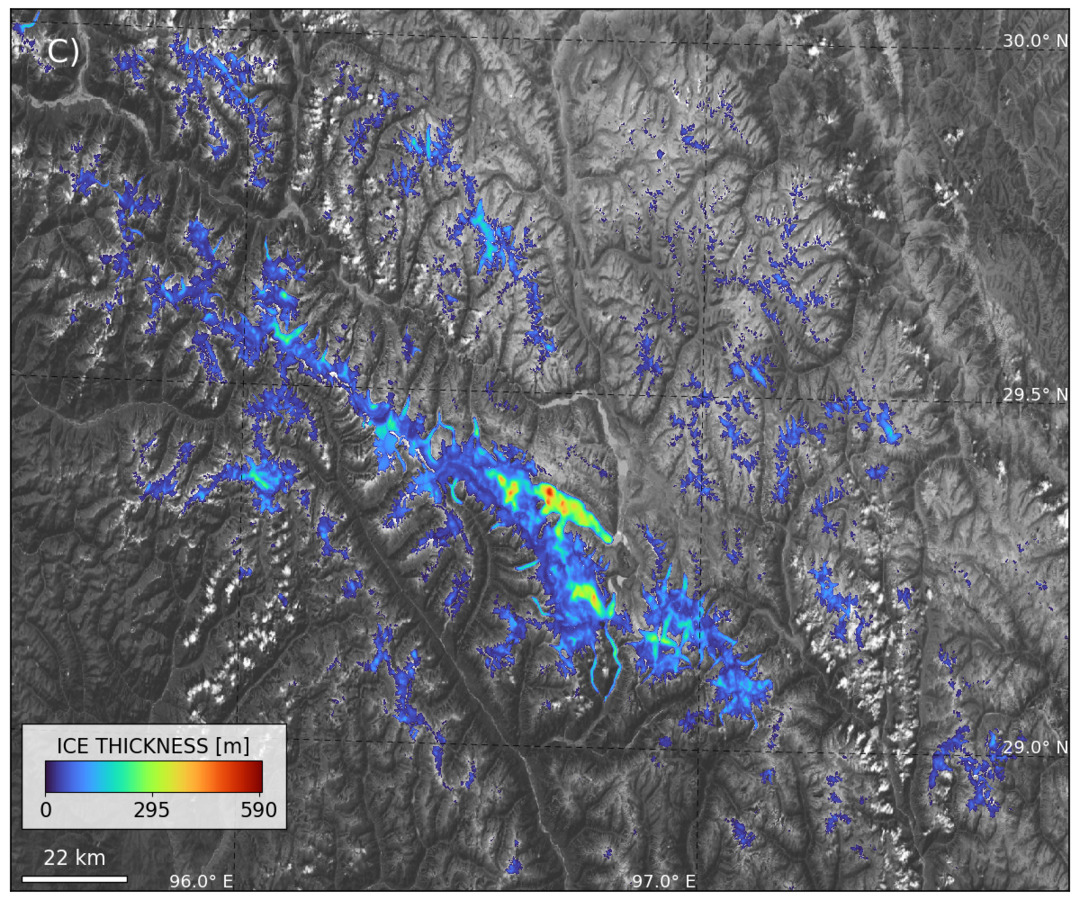}
    \label{fig:karakoram_eastern_himalayas}
\end{figure}
\end{landscape}

Few-to-no measurements have been collected in Asia. Unlike other data-sparse Arctic regions (such as the Russian Arctic), the feature space in this region is likely unrepresented or outside the domain covered by the training data. As a result, the model operates in a highly generative regime. Elevations can be extreme, slopes very steep, velocity fields contain large gaps and significant outliers, and the mass-balance input may be a crude approximation. Yet, the IceBoost inversion appears realistic (Figs. \ref{fig:tian_shan}, \ref{fig:pamir_west_kulun}, \ref{fig:karakoram_eastern_himalayas}). The Karakoram stands out: IceBoost produces shallower thick-ice and thicker shallow-ice regions compared to both Millan and Farinotti (Fig. \ref{fig:karakoram_eastern_himalayas}). However, given the absence of ground-truth data and the strongly out-of-distribution feature space, we cannot claim that the machine-learning approach yields an improvement over existing methods in this region.

\subsubsection{Southern Andes (RGI 17)}
\label{sect:southern_andes}

\begin{table}[h]
    \caption{Southern Andes ice volumes estimated by different models. All units are \SI{e3}{\kilo\metre\tothe{3}}.}\label{tab1}%
    \begin{tabular}{@{}lccccc@{}}
    \toprule
    \makecell[l]{Southern Andes \\ (RGI 17)} & IceBoost v2 & \makecell[c]{Millan 2022 \\ \cite{millan2022}} & \makecell[c]{Farinotti 2019 \\ \cite{farinotti2019}} & \makecell[c]{Fürst 2024 \\ \cite{furst2024}} & \makecell[c]{Millan 2019 \\ \cite{millan2019_patagonia}} \\
    \midrule
    Total & 6.8 ± 1.3 & 5.9 ± 1.6  & 5.3 ± 1.4 & - & - \\
    \makecell[l]{- Northern Patagonian \\ Icefield} & 1.627 ± 0.272 & 1.224 & 1.105 & 1.150 & 1.147 \\
    \makecell[l]{- Southern Patagonian \\ Icefield} & 4.287 ± 0.701 & 3.928 & 3.507 & 4.182 & 3.826 \\
    - Cordillera Darwin & 0.357 ± 0.097 & 0.250 & 0.307 & - & - \\
    - Others & 0.529±0.23 & 0.498 & 0.381 & - & - \\
    \bottomrule
    \end{tabular}
\end{table}

Over the Patagonian Icefields, we add to the set of model comparisons the inversions by Fürst et al. 2024 \cite{furst2024} and by Millan et al. 2019 \cite{millan2019_patagonia}. The former uses a mass-conservation approach refined by a shallow ice approximation; the second uses a gravity-based inversion.\\

% NPI
\noindent In the interior of the Northern Patagonian Icefield (NPI), IceBoost and the models by Fürst et al. (2024) \cite{furst2024} and Millan et al. (2019) \cite{millan2019_patagonia} reproduce the high-thickness measurements. However, IceBoost predicts thicker ice than the other models along the steep, mountainous terrain near ice-free nunataks. In the eastern sector of the icefield, none of the models can reproduce the thick-ice measurements included in the training dataset. These values may be unrealistically high and could bias IceBoost toward predicting excessively thick ice.\\
\noindent Overall, both the data and three of the five models indicate that the central portion of the icefield is likely about 1000–1200 m thick. The shallower, more rugged areas are more uncertain and show larger discrepancies among solutions. IceBoost may be positively biased, and its bootstrap performance analysis (Supp. Info. Fig. S1) supports this possibility. Its error in the Southern Andes is the highest among all global regions, suggesting that some training data there may be unreliable.\\
\noindent In the Northern Patagonian Icefield, we are not more confident in IceBoost than in the other models. Suspect thick-ice measurements in the eastern NPI make the reliability of IceBoost in this region uncertain, although it may still perform well over the thick central region and outlet glaciers. The San Quintín Glacier and Steffen Glacier are modeled to be grounded below sea level. The model is supported by data.\\

% SPI
\noindent Reconstructions of the Southern Patagonian Icefield (SPI) using IceBoost, the method by Fürst et al. (2024) and the shallow-ice approximation \cite{millan2022} are broadly similar. In contrast, Farinotti’s ensemble appears generally too shallow. The gravity inversion by Millan et al. (2019) \cite{millan2019_patagonia} shows very shallow ice over steep mountainous terrain and very thick ice elsewhere. We find that the termini of the Pío XI Glacier and Occidental Glacier are grounded below sea level, by up to roughly 500 m. This result cannot be confirmed by measurements, as these areas remain unsurveyed. Both IceBoost and observations indicate that the termini of the George Montt Glacier, Bernardo Glacier, Upsala Glacier and Tyndall Glacier are also grounded below sea level.

% concluding statement about Patagonian Icefields
\noindent Across both Patagonian Icefields, available reconstructions still diverge substantially. Disagreement occurs in both thick- and thin-ice regions. We argue that measurements would be beneficial over thin ice over the NPI, and everywhere over the SPI.

\begin{figure}[h!]
    \caption{Northern Patagonian Icefield (RGI 17). A=IceBoost v2; B=Millan et al. 2022 \cite{millan2022}; C=Farinotti et al. \cite{farinotti2019}; D=Fürst et al. \cite{furst2024}; E=Millan et al. 2019 \cite{millan2019_patagonia}.}
    \includegraphics[width=.49\linewidth]{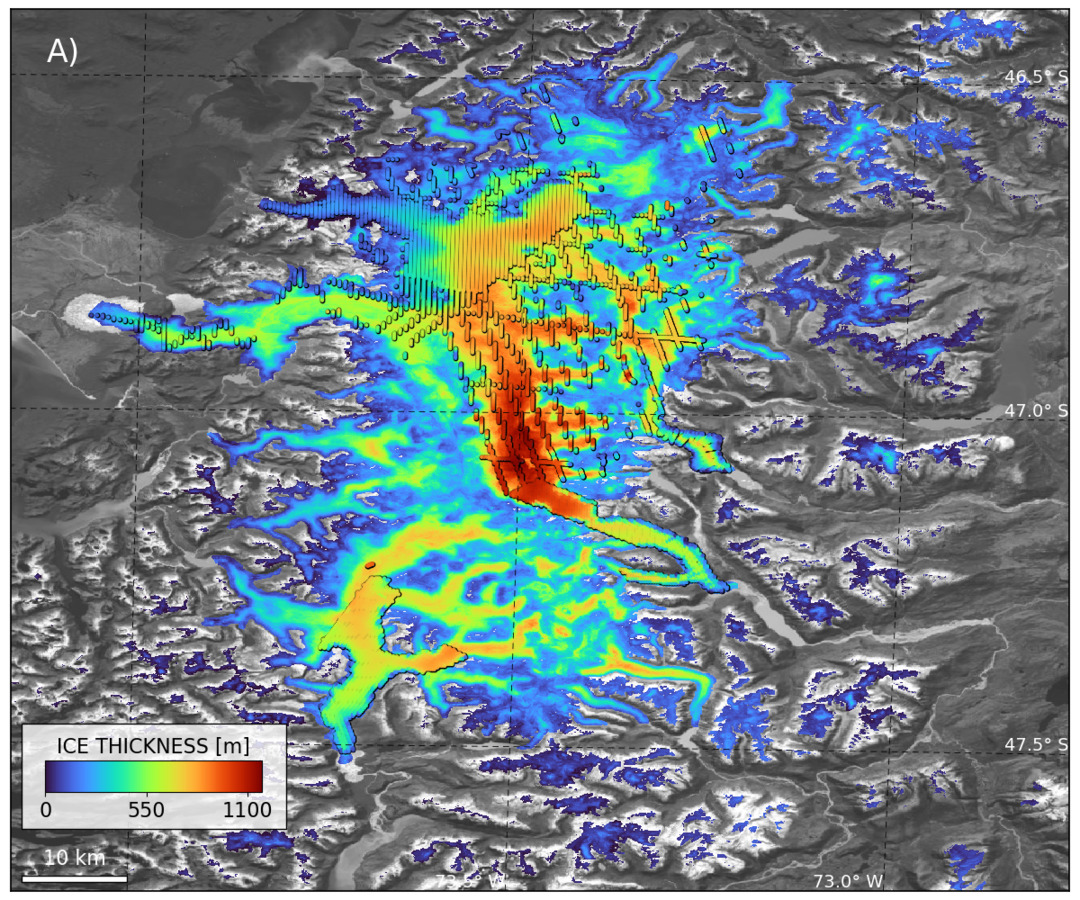}
    \includegraphics[width=.49\linewidth]{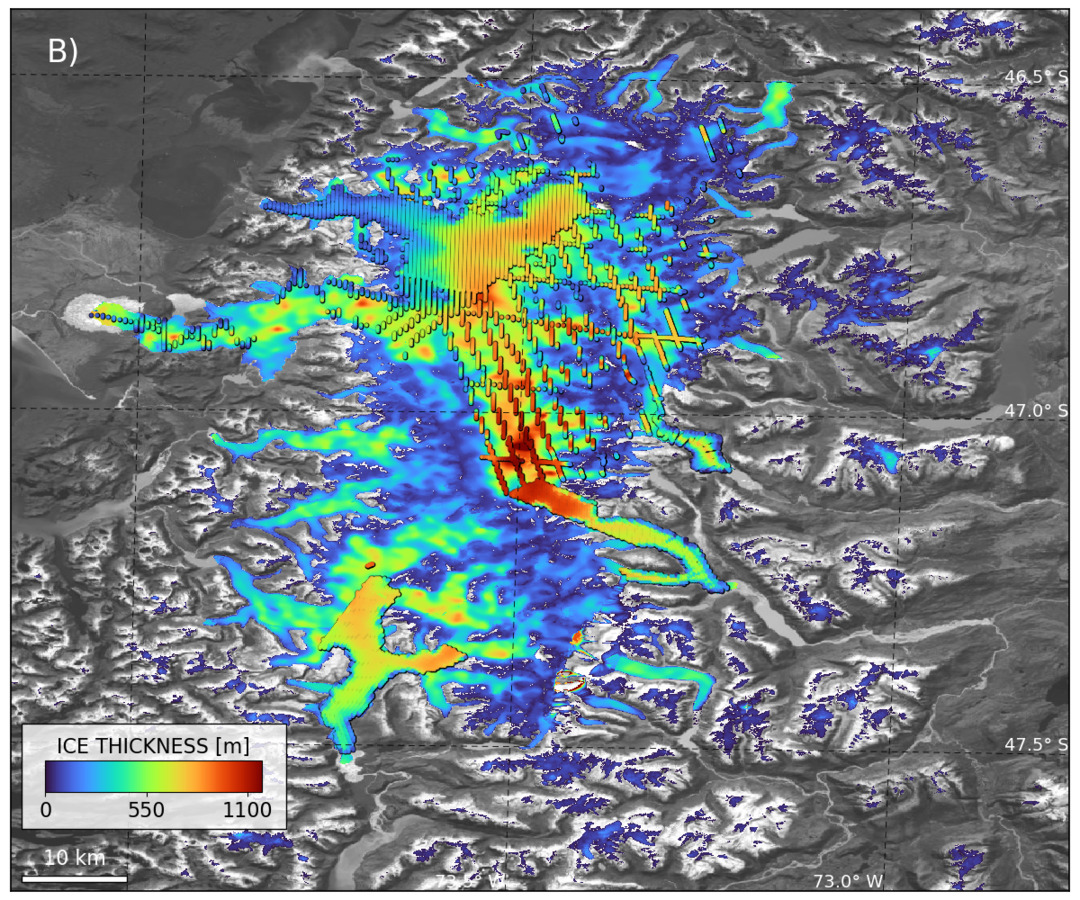}\\
    \includegraphics[width=.49\linewidth]{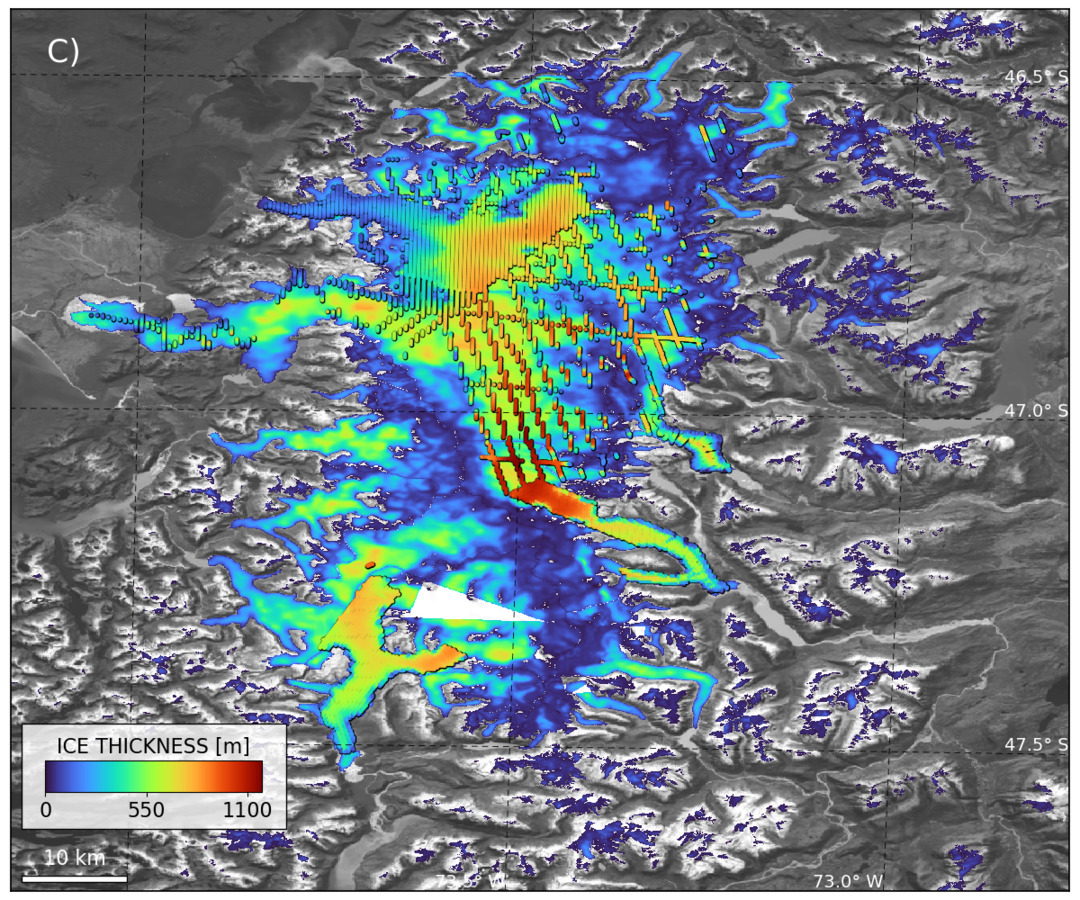}
    \includegraphics[width=.49\linewidth]{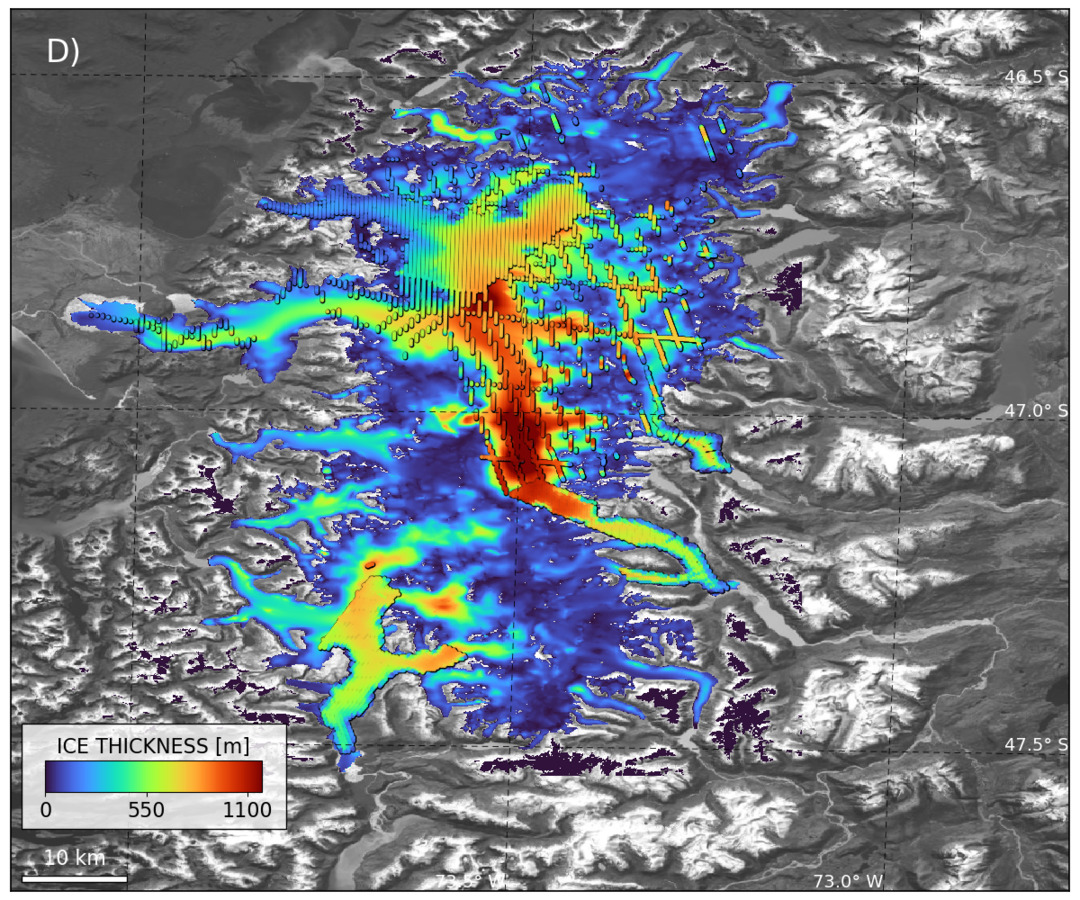}\\
    \includegraphics[width=.49\linewidth]{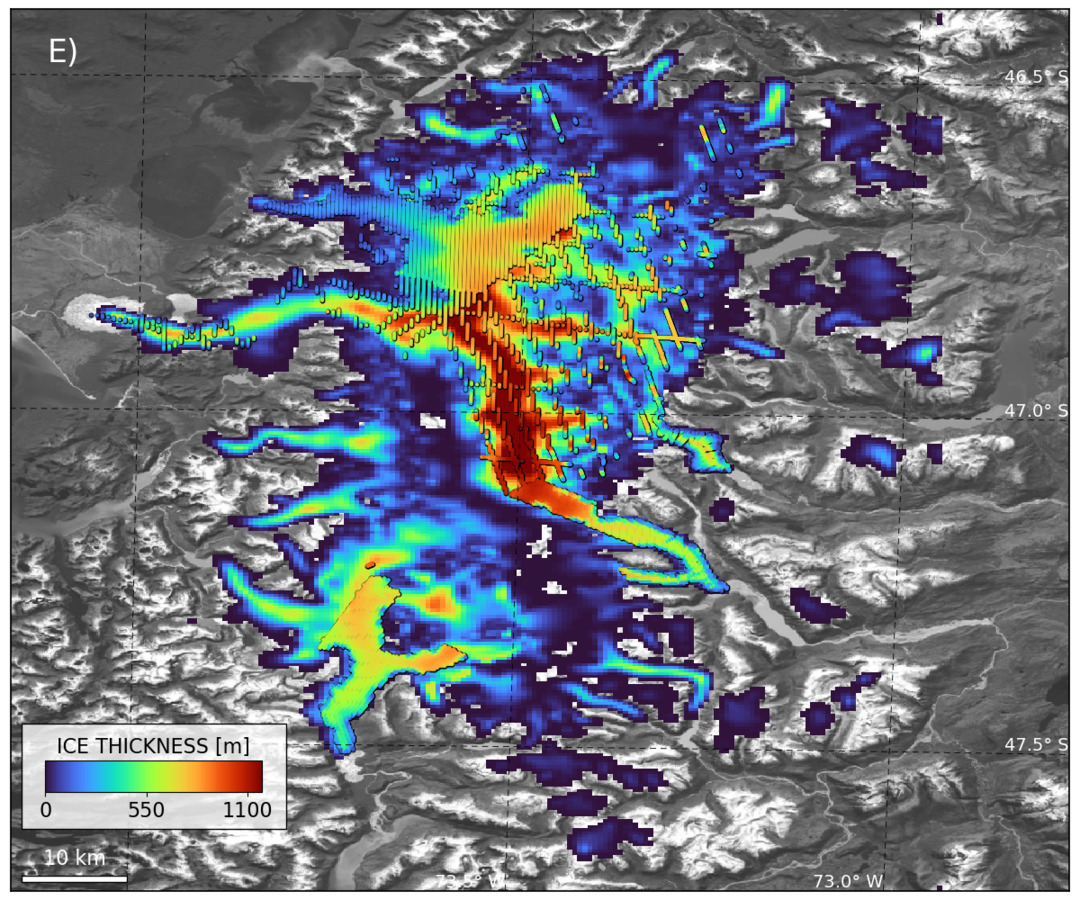}
    \label{fig:southern_andes_npi}
\end{figure}

\begin{figure}[h!]
    \caption{Southern Patagonian Icefield (RGI 17). A=IceBoost v2; B=Millan et al. 2022 \cite{millan2022}; C=Farinotti et al. \cite{farinotti2019}; D=Fürst et al. \cite{furst2024}; E=Millan et al. 2019 \cite{millan2019_patagonia}.}
    \includegraphics[width=.49\linewidth]{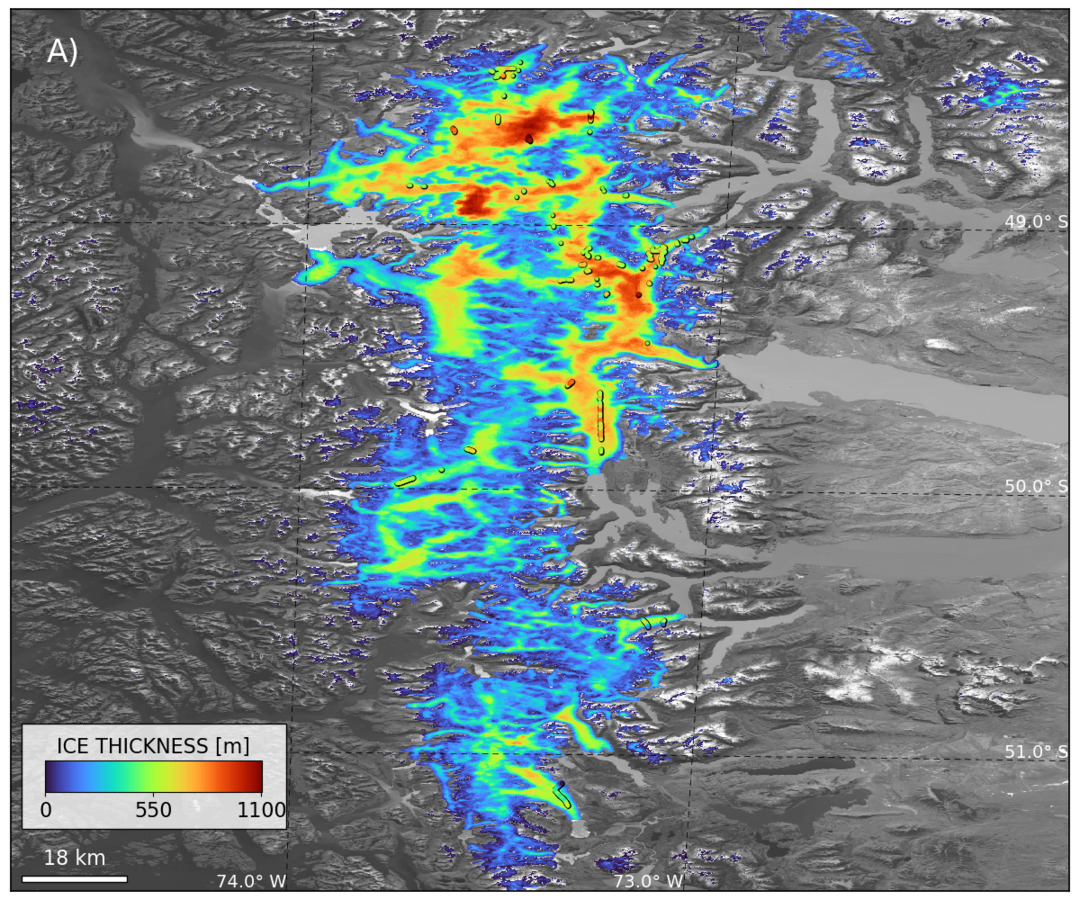}
    \includegraphics[width=.49\linewidth]{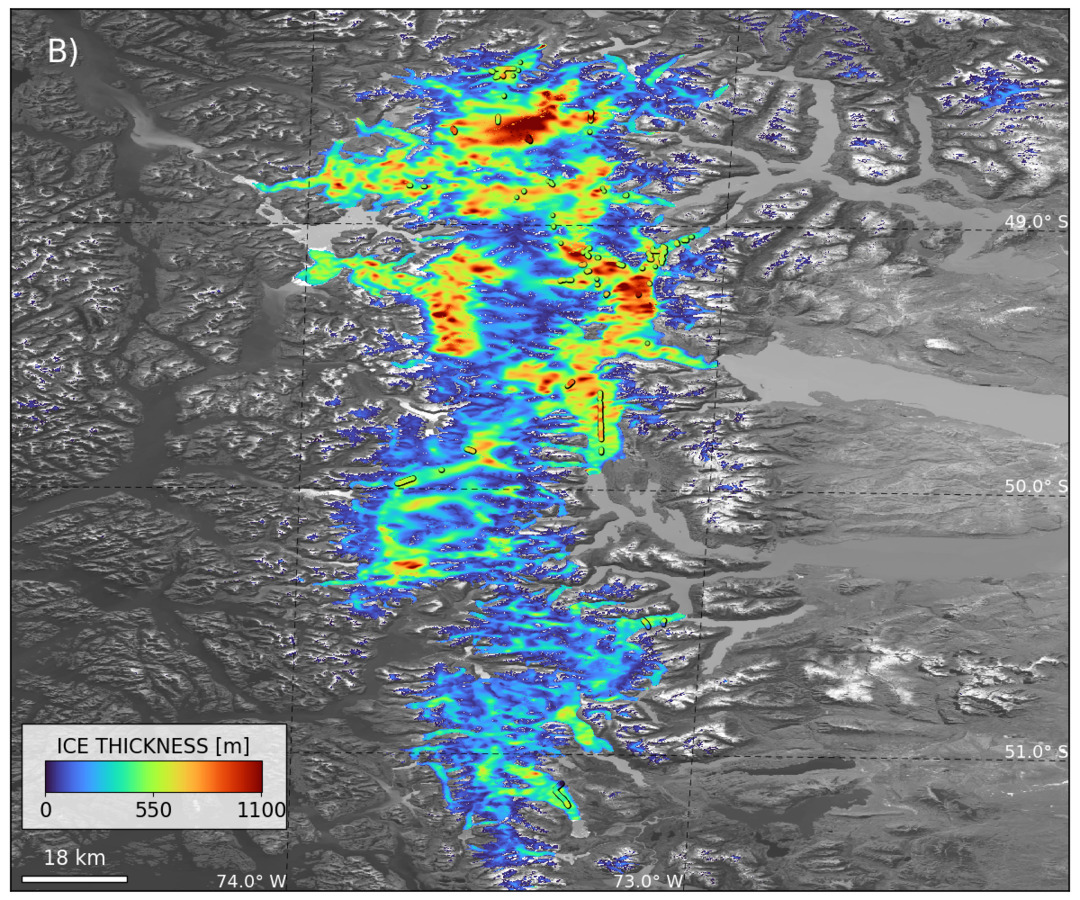}\\
    \includegraphics[width=.49\linewidth]{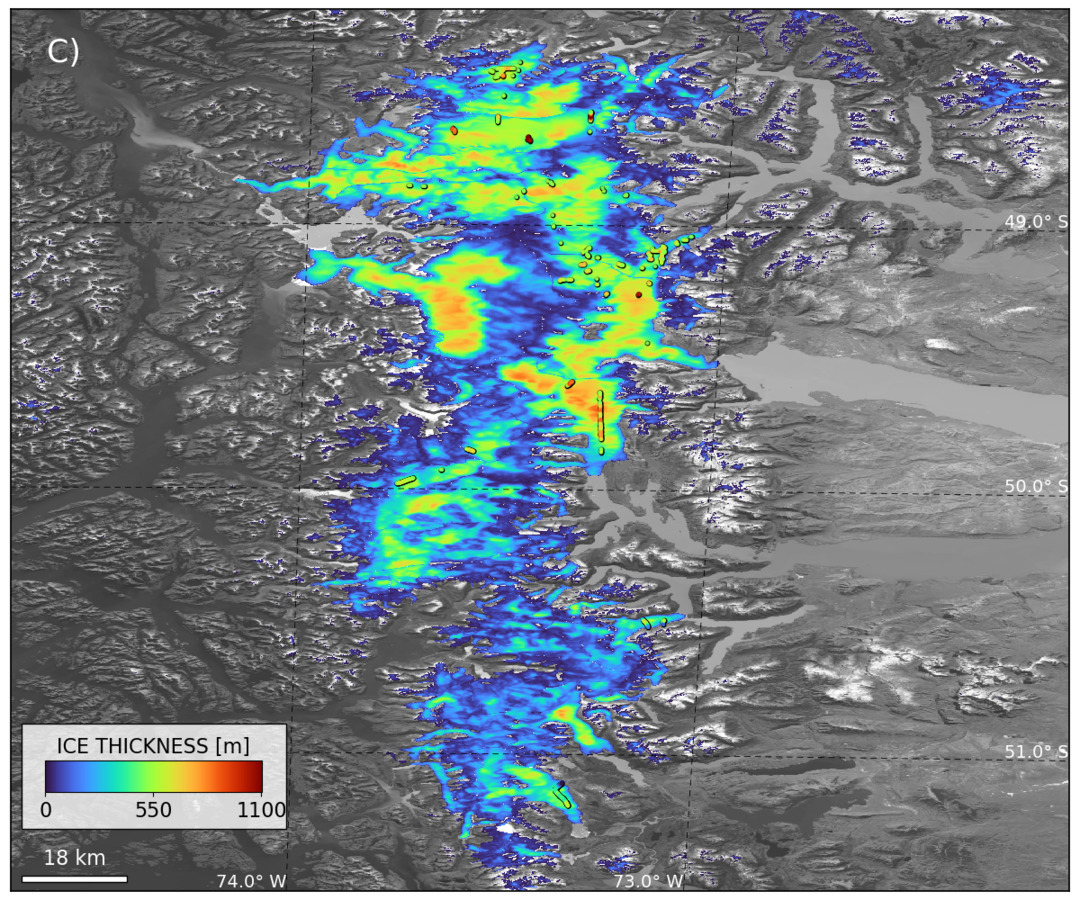}
    \includegraphics[width=.49\linewidth]{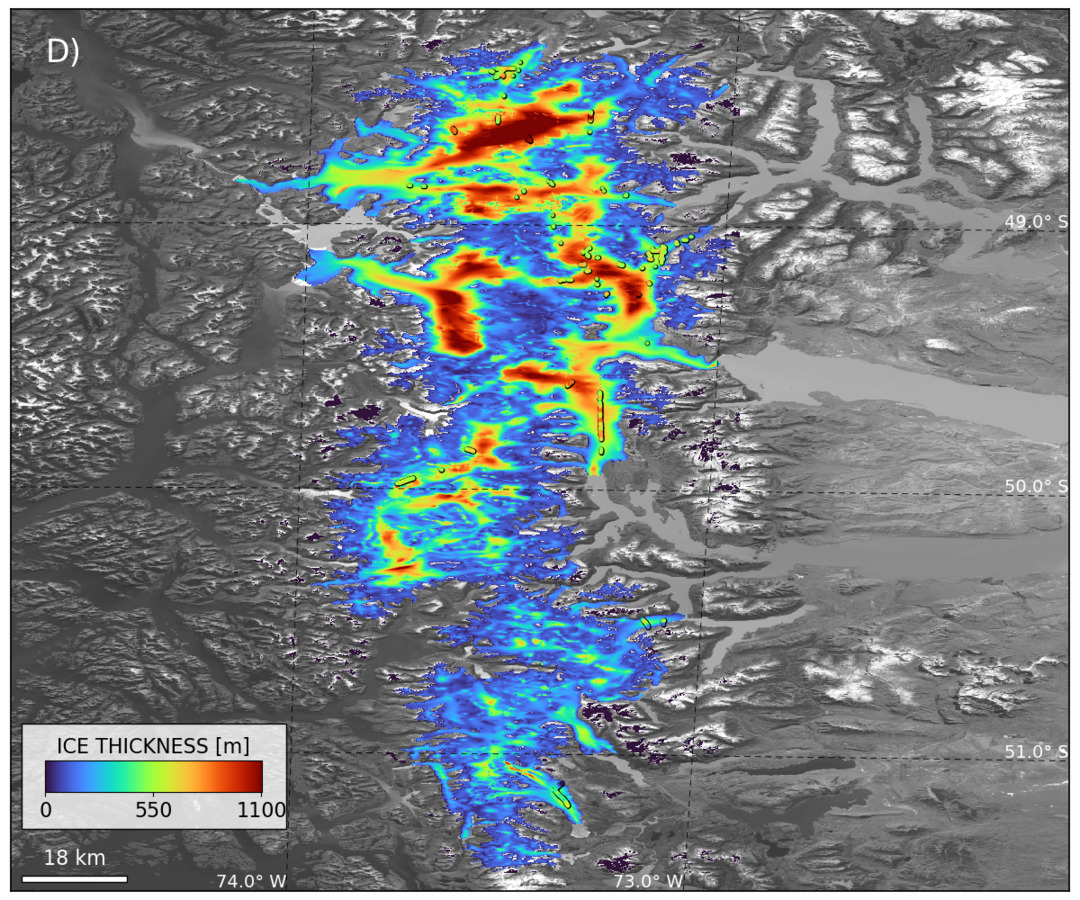}\\
    \includegraphics[width=.49\linewidth]{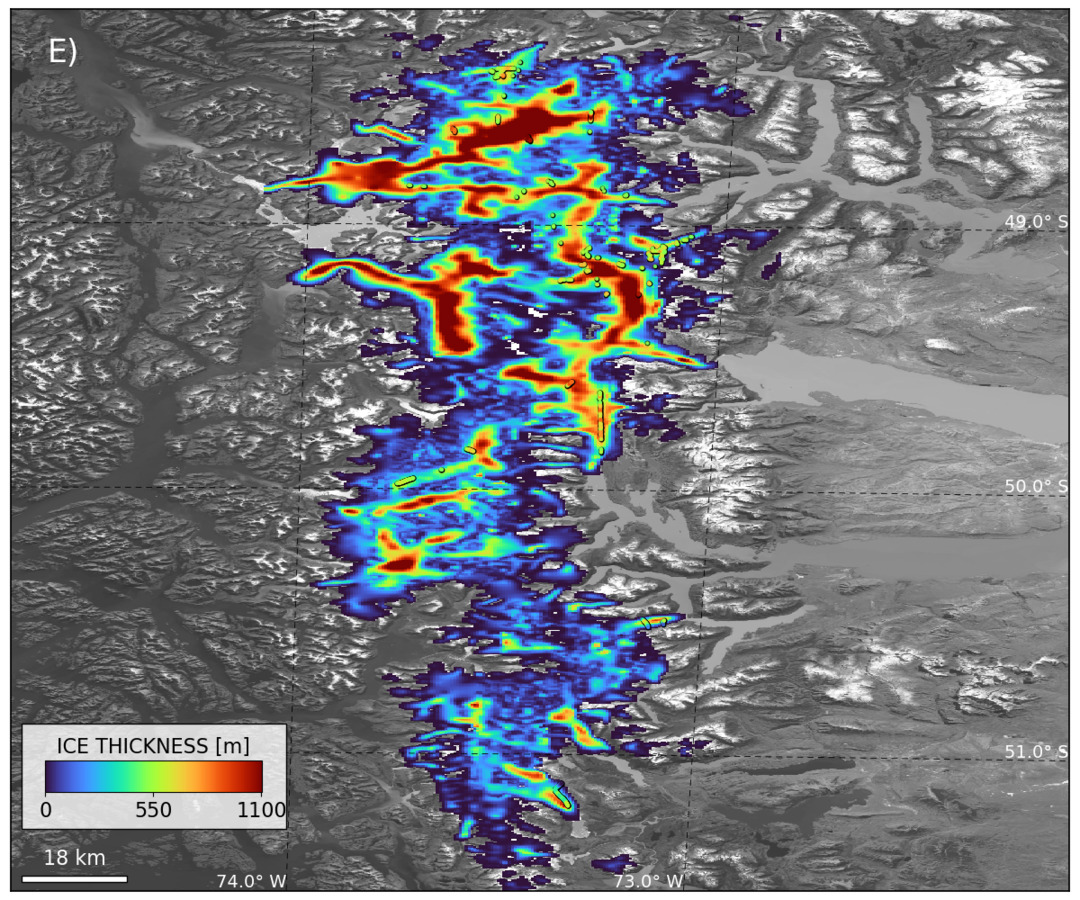}
    \label{fig:southern_andes_spi}
\end{figure}

%\begin{figure}[h!]
%    \caption{Cordillera Darwin (RGI 17). A=IceBoost v2; B=Millan et al. 2022 \cite{millan2022}; C=Farinotti et al. \cite{farinotti2019}.}
%    \includegraphics[width=.49\linewidth]{figs/17_southern_andes/fig_cordillera_darwin_iceboost.jpg}
%    \includegraphics[width=.49\linewidth]{figs/17_southern_andes/fig_cordillera_darwin_millan.jpg}\\
%    \includegraphics[width=.49\linewidth]{figs/17_southern_andes/fig_cordillera_darwin_farinotti.jpg}
%    \label{fig:southern_andes_cordillera}
%\end{figure}

\clearpage

\section{Technical validation}
%\subsection{Model inputs uncertainties}
\label{sec:feature_errors}
Sect \ref{sec:training_data_quality_control} presents the quality check done on the training dataset. All the remaining sections present the uncertainties on the model inputs and ice thickness output at deploy time, when the model is tasked to generate ice thickness maps.

% We now estimate the errors of each individual model input. 
% These errors are computed per-pixel. They will be used to calculate the error of the modeled ice thickness.

\subsubsection{Training data quality control and pre-processing}
\label{sec:training_data_quality_control}
% post-processing and policies for quality checks
IceBoost v2.0 adopts a series of filters to maximize the quality of the training set. Data time-tagged older than 2005 is removed; data registered in GlaThiDa with less than 1 meter thickness is removed; data registered outside glacier polygons is removed (unless those collected over the Antarctic peninsula); data flagged as either erroneous or limited to parts of the glacier in the different datasets are completely removed. The final combined dataset was manually checked for outliers, for every individual glacier. The consistency of crossover tracks, if present, was used as a criterion. If crossover tracks were not present but close measurements were obtained, proximal measurements were judged for consistency. Some tracks were removed for inconsistencies between tracks, some measurements featured unreasonable data. Over the Antarctic peninsula, the echograms were checked and the questionable bed picks where manually removed, using the filtered dataset from BedMachine Antarctica \cite{morlighem2022_bedmac_An}. Generally, data removal was aggressive: if data were found to be suspicious, they were removed. Most of the removed data were in the periphery of Greenland (along the southern and eastern coasts) and Antarctica, as well as in the Antarctic peninsula and over the two Patagonian icefields. Finally, measurements that did not have the complete set of valid inputs (e.g. ice velocity not available because of incomplete coverage) were also removed from the training set. Missing ice velocity was the first cause for deleting otherwise valid ground truth data. We acknowledge that despite the implemented quality control pipeline, some outliers can still be present in the final dataset.

As in the previous model version, IceBoost v2.0, we encode the training dataset by averaging both ice thickness data and the whole input feature vector in a 100x100 per-glacier pixel grid. This procedure is implemented to compensate for the different resolution at which different ground penetrating radar devices acquire the thickness data along the route, that would otherwise result in a training dataset with less entries for less resolved tracks. The downscaling pipeline reduces the training dataset from 7,069,690 data points to its final size of 378,373 data points, collected over 1,661 glaciers (less than 1\% of all existing glaciers).

\subsubsection{Elevation, slope and curvature uncertainties}
\label{sec:error_elevation}

The absolute height accuracy of the Tandem-X Edited DEM, 30m is typically $<$10 m, with relative vertical accuracy (LE90) typically under 2 meters on gentle slopes (slope $<$ 20\%), and under 4 meters on steeper slopes (slope $>$ 20\%). We set the elevation error $\Delta z$=2 m or 4 m depending on the slope value. \\
%The error on the slope is can be calculated as $\Delta slope=\frac{\Delta z}{\Delta x}$, $\Delta x \approx$30m. For gentle terrain, $\Delta z$=2m,$\Delta x \approx$30m, $\Delta slope\approx2/30\approx0.067$. For steeper terrain $\Delta z$=4m,$\Delta x \approx$=30m, $\Delta slope\approx4/30\approx0.133$. We use a slope error of 0.1, everywhere.
The error on the slope features,  calculated using central differences, $s = \frac{\Delta z}{2\Delta x}$, is calculated as $\sigma_s = \sqrt{4^2+4^2}/(2\Delta x)$, where $\sigma_z=4 m$ everywhere as an upper bound, and $\Delta x$ is the horizontal step, which can have values ranging from 50 meters to 2 kilometers depending on the slope feature considered (i.e. $s_{50}$, $s_{100}$, $s_{200}$, etc). For example, $\sigma_{s_{50}}=\sqrt{4^2+4^2}/(2 \cdot 50)=0.056$.

The curvature is calculated as the Laplacian $curv =  \nabla^2 z = \frac{\partial^2 z}{\partial x^2} + \frac{\partial^2 z}{\partial y^2}$. It is approximated using the 5-point Laplacian stencil on a 3x3 grid, $curv \sim (z_{i+1,j}+z_{i-1,j}+z_{i,j+1}+z_{i,j-1}-4z_{i,j})/\Delta x$. The error on the curvature is calculated as $\sigma_{curv} = \sqrt{20\sigma_z^2}/\Delta x^2 = 2\sqrt{5}/\Delta x^2$ (assuming $\sigma_z=4 m$ everywhere as an upper bound). For example, $\sigma_{c_{50}}=2\sqrt{5}/(50^2)$. An extra factor 100 usually appears as the curvature is reported in units of 1/100 m. 

%is proportional to the error in elevation $\Delta z$ and inversely proportional to the square of the spatial resolution $\Delta x$=30: $\Delta curv = \frac{\Delta z}{\Delta x^2}=\frac{10}{30^2}\simeq0.011\ m^{-1}$, where $\Delta z$=10m, and $\Delta x$=30 is the Tandem-X EDEM resolution. We use $\Delta curv = 0.1\ m^{-1}$. 
To summarize:
\begin{align}
    \sigma_z &= 2-4\ m \label{eq:sigma_z}\\
    \sigma_s &= 2\sqrt{2}/\Delta x \label{eq:sigma_s}\\
    \sigma_{curv} &= 2\sqrt{5}\sigma_z/\Delta x^2\ m^{-1} \label{eq:sigma_curv}
\end{align}

\subsubsection{Distances uncertainties}
Distance from ice-free pixels. The glacier outlines in RGI are derived mostly from satellite imagery: Landsat TM/ETM+ (30 m pixels),  ASTER (15 m pixels), plus other sources in regions where higher‐resolution imagery or mapping exists. The spatial resolution therefore varies. We estimate the resolution to be that of the input imagery ($\approx$15-30 m), and the error quantified in the order a pixel ($\approx$30 m). We note, however, that an additional (and significant) source of uncertainty lies in the fast changes and retreat of glaciers in many regions of the world. This can easily be the major source of uncertainty with respect to the distance to glacier margins (or nunataks therein). For this variable therefore the error is increased and set to 100 m. 

Distance from the ocean is calculated using the shorelines product (ocean/land interface in GSHHG), with a precision in the range of 50-500m \cite{wessel1996}. This variable is most important for maritime glaciers, near the ocean (rather than continental glaciers located very far from the oceans). We set the error of this product to be 100 m.

\begin{align}
    \sigma_{noice} &= 100\ m \\
    \sigma_{ocean} &= 100\ m
\end{align}

\subsubsection{Glacier length uncertainty}
The glacier length ($lmax$), calculated using the convex hull, is the as maximum distance across the glacier. The error is taken as the 5\%, to account for larger uncertainties in irregular shapes for large glaciers:

\begin{equation}
    \sigma_{lmax} = 0.05 \cdot lmax \\
\end{equation}

\subsubsection{Ice velocity uncertainty}
The uncertainty on the ice velocity is set to 10 m $yr^{-1}$ \cite{millan2019, millan2022} everywhere  except for the Greenland periphery, where it is set to 18 m $yr^{-1}$ \cite{mouginot2017}. In the Antarctic periphery and subantarctic glaciers we also use 18 m $yr^{-1}$, a conservative estimate \cite{mouginot2019}.\\

\begin{equation}
\sigma_v =
\begin{cases}
18~\text{m/yr}, & \text{if glacier } \in \{\text{Greenland, Antarctica}\} \\
10~\text{m/yr}, & \text{else.} \\
\end{cases}
\end{equation}

\subsubsection{Temperature uncertainty}
\label{sec:error_temperature}
The temperature above 2 m (t2m) is calculated by averaging N=120 monthly maps (for the 2000-2010 period). If monthly consecutive maps can be modeled to have a lag-1 autocorrelation $\rho$ (AR(1)), where $\rho$ is the positive correlation between consecutive monthly maps, the effective sample size is approximately
\begin{equation}
    N_{eff} \approx N\frac{1-\rho}{1+\rho}
\end{equation}
For non-correlated maps, $\rho$=0, $N_{eff}=N=120$. For a positive and moderately-high correlation, $\rho=0.8$, each map provides less information and the effective sample size becomes $N_{eff}=120\frac{1-0.8}{1+0.8}\approx13.3$. 

If further assuming that each monthly map has a realistic error $\sigma \approx$ 1 K, the random error of the 10 year average becomes 
\begin{equation}
    \sigma_{random} = \frac{\sigma}{N_{eff}} = \frac{1}{\sqrt{13.3}} \approx 0.3\ K
\end{equation}

We should now also consider a systematic bias $b$ due to a finite grid of the satellite product, impacting in particular regions with complex and poorly resolved mountainous terrain. We set such bias everywhere to be $b$ = 1 K. We calculate the final temperature error by accounting for both independent errors:
\begin{equation}
    \sigma_{t2m} = \sqrt{\sigma_{random}^2 + b^2} \approx \sqrt{0.3^2 + 1} \approx 1\ K.
\end{equation}
The t2m uncertainty is approximated as 1 K everywhere. Such uncertainty is likely realistic for high mountainous regions, and overestimated over flatter and well-behaving terrain.

\subsubsection{Surface mass balance uncertainty}
\label{sec:error_smb}
RACMO2.3p2 modeled surface mass balance is used in the peripheries of the Greenland (resolution 1 km) and Antarctica (resolution 2 km), both featuring complex-terrain zones, spatial gradients, and poorly resolved orographic effects. Monthly maps are averaged over 1961–1990 for Greenland and 1979–2021 for Antarctica to produce a single multi‑decadal mean map over the two ice sheets. Such a long temporal period would reduce the influence of interannual variability (much like temperature) on the mean, and it is not considered here, but leaving the dominant source of uncertainty to RACMO2.3p2 model’s systematic bias. Such error is estimated as 10\% the averaged accumulation values. Such value is inferred from RACMO2.3p2's observed difference between the coarse (27km)-to-downscaled (2 km) products in the Antarctic peninsula \cite{noel2023}, which gives an insight to such resolution-driven bias.

For all glaciers outside the ice sheet peripheries, we use the same SMB-to-elevation lapse rate method in IceBoost v1.1 \cite{maffezzoli2025}. We use regional values of $\bar{s}=dSMB/dz$ and $\bar{q}=SMB(z=0)$ for every glacier in each regions. No robust a priori rationale can be used to formulate the uncertainties on these pairs. A sensitivity test was carried out in \cite{maffezzoli2025} to assess how much modeled integrated glacier volumes would change by changing these values, and the authors found a limited sensitivity to this parametrization. We formulate the uncertainties on these two parameters to be 10\%: $\sigma_{\bar{s}}=0.1\bar{s}$, $\sigma_{\bar{q}}=0.1\bar{q}$. To summarize, in the ice sheet peripheries:
\begin{equation}
\begin{cases}
\mathrm{SMB=RACMO2.3p2} \\
\mathrm{\sigma_{SMB}} = 0.1 \cdot \mathrm{SMB}
\end{cases}
\end{equation}
For all other glaciers:

\begin{equation}
\begin{cases}
\mathrm{SMB = \bar{q} + \bar{s} z}\\
\mathrm{\sigma_{SMB}} = \mathrm{0.1\bar{q}} + \mathrm{0.1\bar{s} \mathrm{z}}
\end{cases}
\end{equation}
where z is elevation in meters; $\mathrm{\bar{s}}$ has units of $\mathrm{mm\ w.e. yr^{-1}\ m^{-1}}$; $\mathrm{\bar{q}}$ has units of $\mathrm{mm\ w.e. yr^{-1}}$; SMB has units of $\mathrm{mm\ w.e. yr^{-1}\ m^{-1}}$. The regional ($\mathrm{\bar{q}}$, $\mathrm{\bar{s}}$) values can be found \cite{maffezzoli2025} (Table A1, Appendix A).

\subsection{Ice thickness uncertainty via Monte Carlo simulations}
\label{sec:ice_thickness_uncertainties}
To estimate the errors of the modeled glacier ice thickness maps, we perform a Monte Carlo (MC) perturbation analysis. For every glacier, we collect the feature set X. Each input feature is assigned an uncertainty $\sigma$ (e.g., in elevation, slope, velocity, or mass balance, Sect. \ref{sec:feature_errors}). We then generate n=50 random realizations of the input features X, perturbing each set X as $X + \epsilon$, with $\epsilon\sim N(0, \sigma^2)$. \\

IceBoost is evaluated for each realization, $f(X + \epsilon)$, producing an ensemble of ice thickness maps. The standard deviation of this envelope, $\sigma_{MC} = \sigma(f(X + \epsilon))$ embeds two contributions: i) the aleatoric variability of predictions due to imperfect knowledge of the inputs (enforced using perturbations) and ii) the epistemic uncertainty, resulting from differences between the two learning algorithms (XGBoost and CatBoost). While the first component vanishes as input uncertainties tend to zero, the second does not and quantifies the systematic disagreement between the XGBoost and CatBoost separate models. We take $\sigma_{MC}(x,y)$ as the ice thickness error: $\sigma_h(x,y)=\sigma_{MC}(x,y)$. An example of the $\sigma_h(x,y)$ error map for the Geikie Plateau (East Greenland) is displayed in Fig. \ref{fig:geikie}. The error maps are released for each glacier. In general, higher error corresponds to higher ice thickness. The main reason for that is that higher ice thickness corresponds to regions where slope are minimum. Induced uncertainty in the slope variables (via perturbations) have a significant effect in the predicted thickness. Over mountainous terrain (higher thickness), where ice is more shallow, induced perturbations have a weaker effect, resulting in a narrower envelope of solutions, hence smaller $\sigma_{MC}(x,y)$. Such effect is further investigated in Section \ref{sect:jensen_gap}.

\subsection{Model non-linearity and Jensen Gap}
\label{sect:jensen_gap}
We aim to identify where IceBoost behaves nonlinearly and how input perturbations (i.e., uncertainties in the predictors) affect the predicted ice thickness. According to Jensen’s inequality, for a random variable $X$ and a function $f$,

\begin{align}
& \mathbb{E}[f(X)] - f(\mathbb{E}[X]) > 0 \quad \text{if $f$ is convex},\
& \mathbb{E}[f(X)] - f(\mathbb{E}[X]) < 0 \quad \text{if $f$ is concave},
\end{align}
and the quantity
\begin{align}
\mathcal{J}(f,X) = \mathbb{E}[f(X)] - f(\mathbb{E}[X])
\end{align}
is the Jensen Gap.

$\mathcal{J}$ depends jointly on the curvature of $f$ and the variance of $X$, and it vanishes for linear models. Thus, it provides a measure of how much the model's prediction can change as result of the inputs uncertainties and the model's curvature.

In our setting, $\mathbb{E}[f(X)]$ is the mean thickness obtained by applying IceBoost to the perturbed features, whereas $f(\mathbb{E}[X])$ is the thickness predicted from the unperturbed features.

Because IceBoost combines two gradient-boosted decision tree models, it is inherently nonlinear. To probe this non-linearity, we use the $n=50$ Monte Carlo perturbations described in Sect. \ref{sec:ice_thickness_uncertainties} to approximate the distribution of $X$ and to compute $\mathcal{J}(f,X)$ at each glacier pixel. The resulting Jensen Gap maps identify where - and by how much - input uncertainty would shift the mean modeled thickness. Positive values indicate locally convex model behavior (uncertainty increasing the expected thickness), while negative values indicate concavity (uncertainty decreasing it).

\begin{figure}
    \centering
    \caption{Modeled ice thickness A-C-E and corresponding Jensen Gap B-D-F. Ground truth data, if any, are overlayed on the same colorbar. A) Ruth glacier and Denali glacier system (Alaska); C) Barnes ice cap and Baffin island glaciers (Nunavut, Canada); E) Renland ice cap and glaciers in the Scoresby Sound system (Eastern Greenland). Negative (positive) Jensen Gap signals decreased modeled ice thickness as a result of uncertain input features.}
    \includegraphics[width=.49\linewidth]{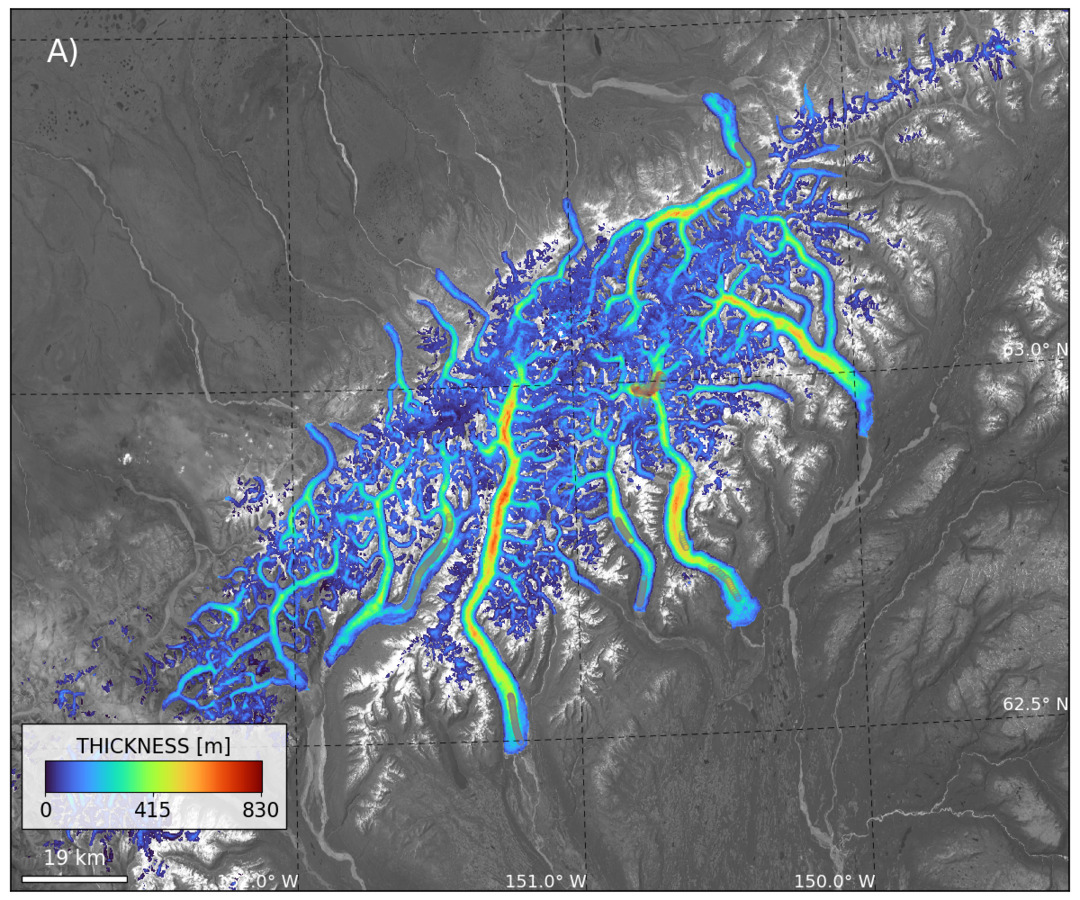}
    \includegraphics[width=.49\linewidth]{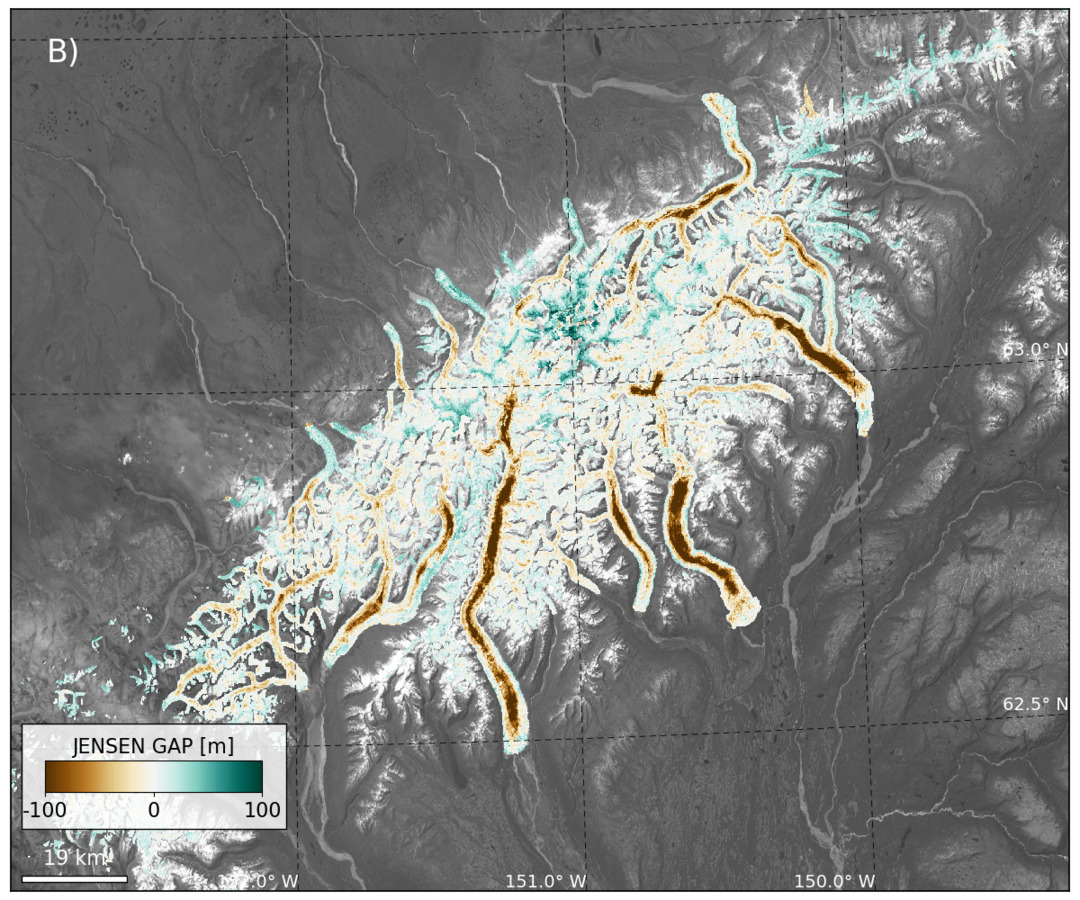}\\
    \includegraphics[width=.49\linewidth]{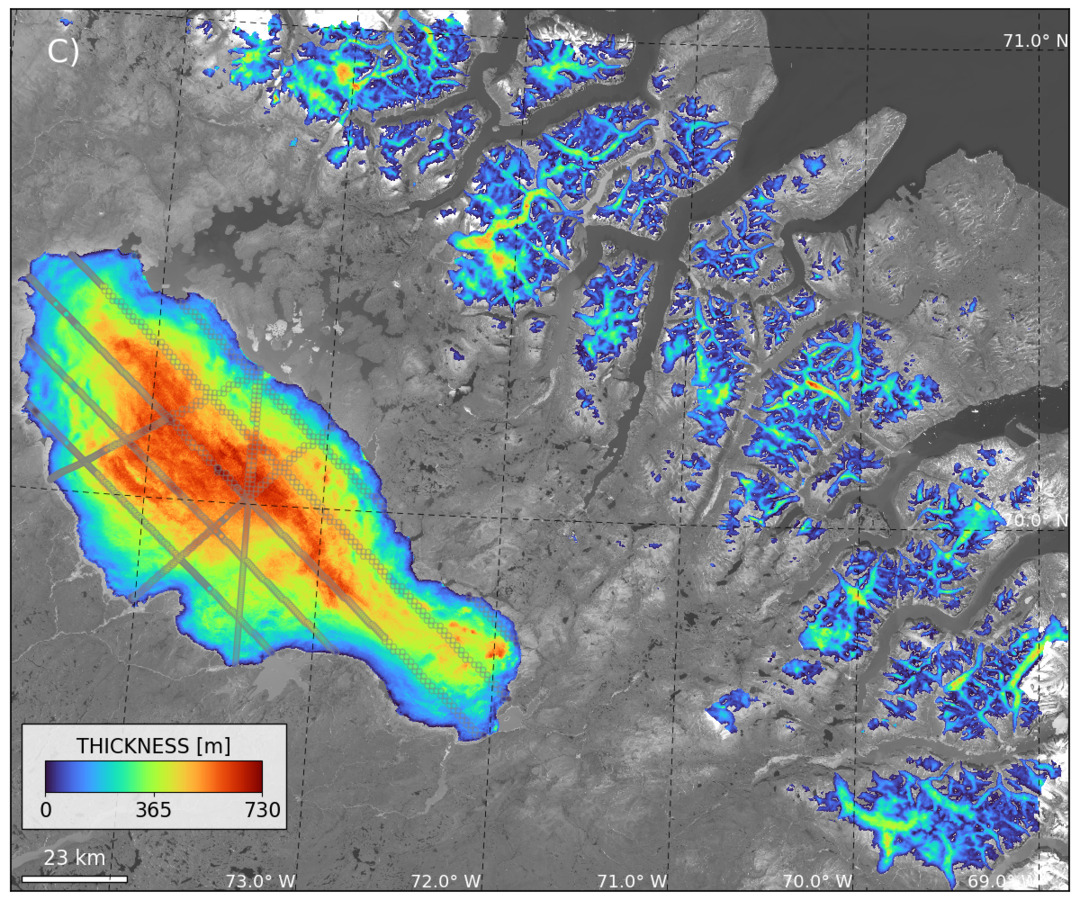}
    \includegraphics[width=.49\linewidth]{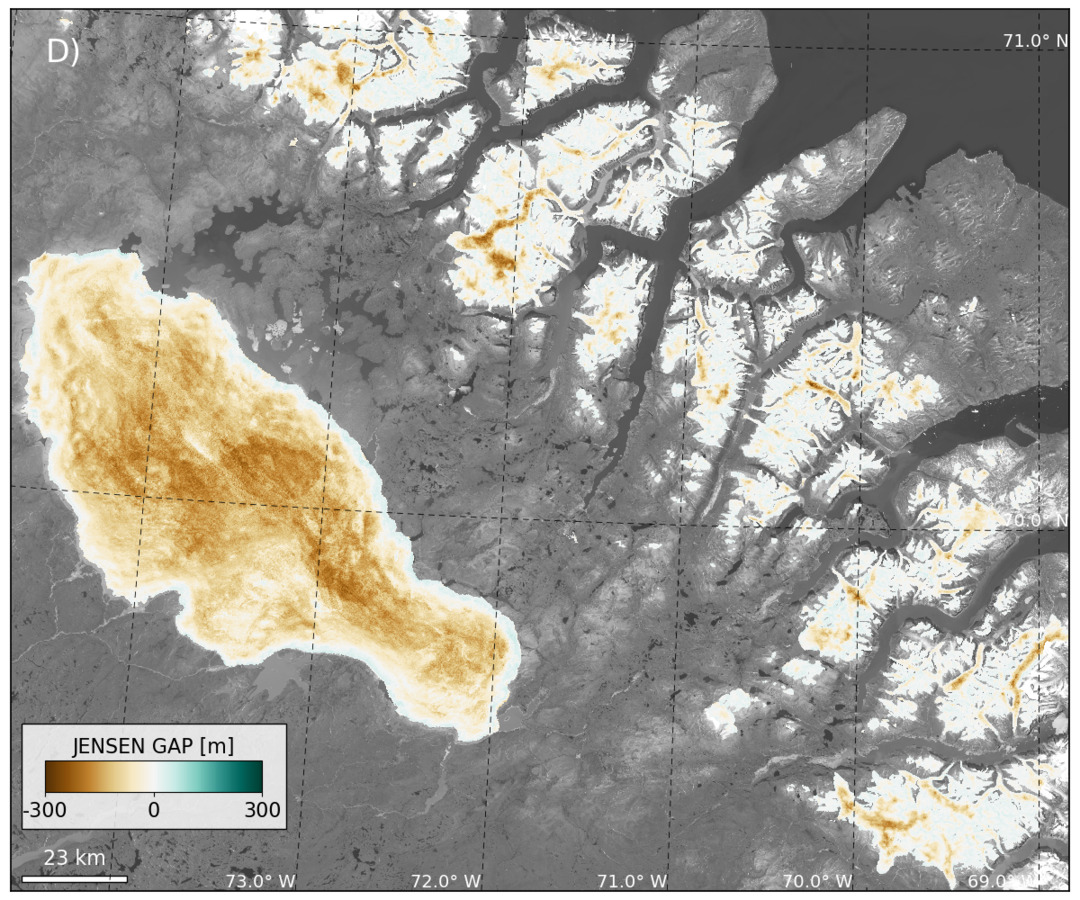}\\
    \includegraphics[width=.49\linewidth]{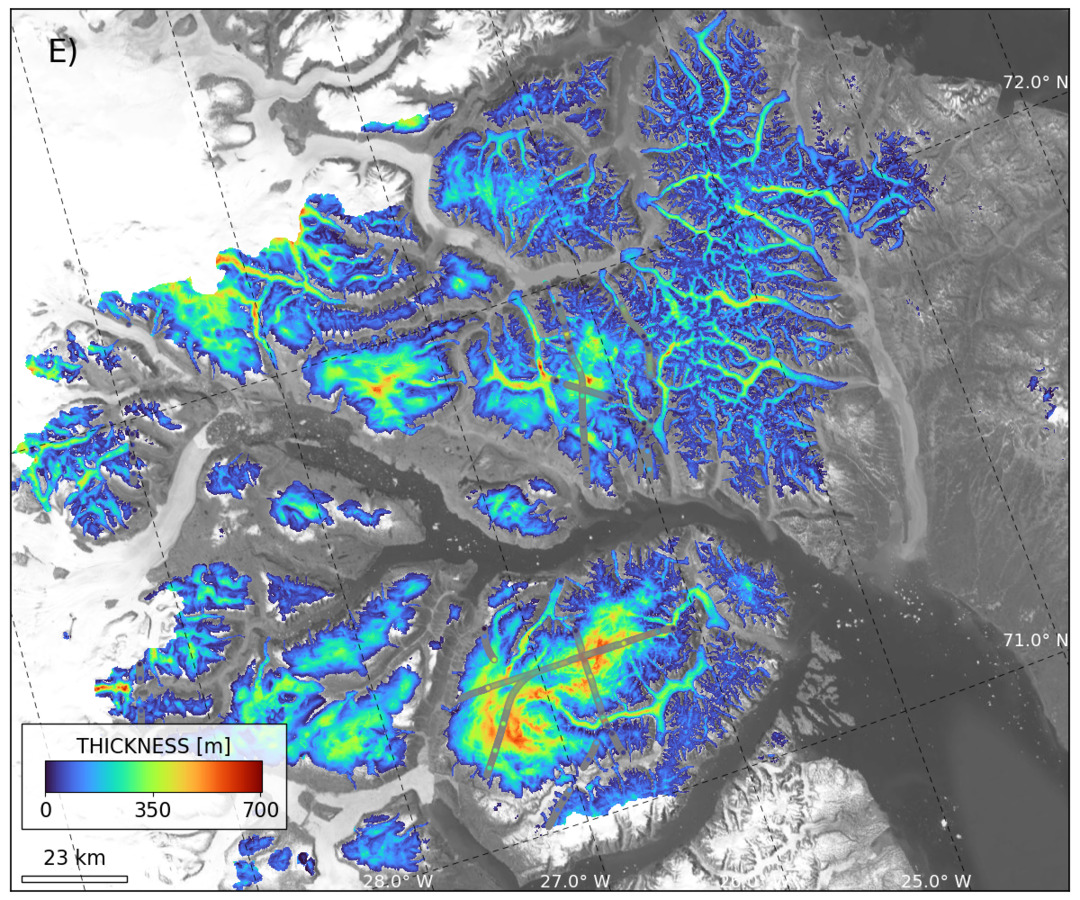}
    \includegraphics[width=.49\linewidth]{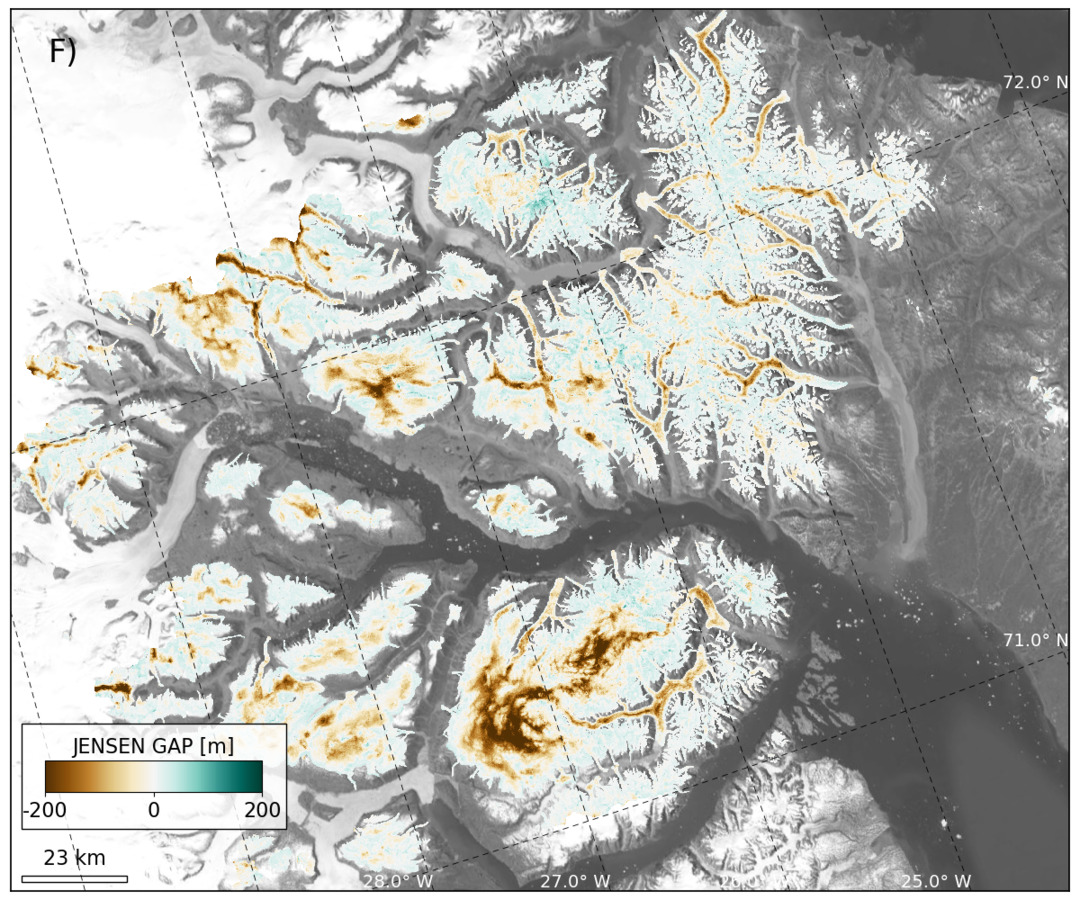}\\
    \label{fig:jensen_gap}
\end{figure}

We find that $\mathcal{J}$ is often negative (Fig. \ref{fig:jensen_gap}). These regions typically correspond to thick ice over low-slope terrain. Previous feature-importance shapely analyses (\cite{maffezzoli2025}) showed that surface slope is one of the strongest predictors of thickness. Because the model is highly sensitive to slope, perturbing the slope symmetrically has an asymmetric effect on thickness: over low-sloping terrain, a perturbation that increases the slope causes a large drop in predicted thickness. The corresponding perturbation that decreases the slope results in a much smaller gain in thickness. This concave response dominates the sign of the Jensen Gap. Although $\mathcal{J}$ reflects the integrated effect of all input uncertainties, we suggest that, where ice is thickest, slope variability is the primary driver. The Jensen Gap thus provides an empirical, spatially distributed measure of the curvature of the learned mapping $f$, and highlights where input uncertainties most strongly interact with it.

\clearpage
\section{Data records}
We release the following products:
    \begin{itemize}
        \item The training dataset (ice thickness + input features), the trained model (iceboost-v2.0) and the produced maps are all released on Zenodo at \url{https://doi.org/10.5281/zenodo.17724512}.
        \item A Web visualizer is made available at: \url{https://nmaffe.github.io/iceboost_webapp/}. Individual glacier files can also be downloaded from there.
    \end{itemize}

Each glacier .tif file contains 5 fields: ice thickness, its error, the Jensen Gap, the surface elevation (Tandem-X Edited DEM, 30m) referenced to the ellipsoid and to the geoid (EIGEN-6C4 gravity field model, 9km). The projection is UTM, except for: Greenland (rgi5, projection 3413), and Antarctica (rgi19, south of 60° S, projection 3031). Glaciers in rgi19 north of 60° S are released in UTM projection. All layers are release with horizontal resolution of 100 m, except for small glaciers, where is a finer grid is produced. The files attributes include the glacier RGI codes, their names (if any), a representative lat-lon point inside the glacier that can be useful for geographical filtering, the glacier area in $km^2$, the glacier ice volumes (total and below sea level) in $km^3$ with errors, any ground truth measurements acquired inside the glacier (latitutde, longitudes and ice thickness), and the raster spatial resolution and projection.

Note that the elevation tiles over the Jan Mayen island are taken from TandemX-EDEM v.2 (not version v.1 otherwise used), which is not yet publicly available. They are available in the tif files at 100 meters, and can be made available on their original resolution upon request to the TanDEM-X science coordination team.

\section{Data availability}
%In addition, we require a short “Data Availability” statement to repeat the main points in the Data Record, meaning some duplication is expected. We suggest repeating the lists of accession numbers, URLs, and the reference numbers for all datasets associated with the paper.
The main dataset release consists of n=215,547 and n=274,531 glacier .tif files, respectively for RGI v6.2 and RGI v7.0. Each tif file includes 5 arrays:

\begin{itemize}
    \item The modeled ice thickness.
    \item The modeled ice thickness error.
    \item The Digital Elevation Model surface elevation.
    \item The geoid elevation.
    \item The Jensen Gap.
\end{itemize}
The attributes include ice thickness measurements, ice volume, ice volume below sea level.

\section{Code Availability}
The code is hosted and maintained on GitHub: \url{https://github.com/nmaffe/iceboost/}

\section{Competing Interests}
The authors declare no competing interests.

\section{Funding}
The work has been supported by the EU Horizon Europe Marie Sklodowska-Curie Actions programme (Grant no. 101066651), project SKYNET.

\clearpage
\bibliography{sn-bibliography}% common bib file
%% if required, the content of .bbl file can be included here once bbl is generated
%%\input sn-article.bbl

\end{document}